\documentstyle[twocolumn,epsf,aps,prb]{revtex} \begin{document} %****
\title{Self-Consistent Renormalization Model of Mott Gap Collapse in the
Cuprates}

\author{R.S. Markiewicz}
\address{Physics Department, Northeastern University, Boston MA 02115, USA}
\maketitle
\begin{abstract}
A generalized antiferromagnetic approach to the Mott transition is
analyzed with special emphasis on electron doped
cuprates, where evidence for electronic phase separation is weak or
absent. Fluctuations are incorporated via a self-consistent
renormalization, thereby {\it deriving} a `nearly-antiferromagnetic Fermi 
liquid' susceptibility.  The calculation is sensitive to hot-spot effects.
Near optimal doping, an approximately electron-hole symmetrical Mott
gap collapse is found (quantum critical points).  The calculation
satisfies the Mermin-Wagner theorem (N\'eel transition at $T=0$ only --
unless interlayer coupling effects are included), and the mean-field gap
and transition temperature are replaced by pseudogap and onset temperature.
The resulting susceptibility is used to calculate the doping dependence of the
photoemission dispersion, in excellent agreement with experiment.
Discussions of interlayer coupling, doping dependence of $U$, and
extension to a three-band model are included.
\end{abstract}
%\pacs{PACS numbers~:~~74.20.Mn, 74.72.-h, 71.45.Lr, 74.50.+r }

%****
\narrowtext
%****

\section{Introduction}

Schrieffer, Wen, and Zhang\cite{SWZ} originally proposed that the magnetic
insulating phase in underdoped cuprates could be understood via a spin
density wave (SDW) approach to the Mott transition, and successfully described 
the spin wave spectrum of the undoped parent compound, which is an 
antiferromagnetic (AFM) insulator.  Kampf and Schrieffer\cite{KaSch} showed 
that precursors of the Mott transition could give rise to a {\it pseudogap} in 
the quasiparticle spectrum, between incipient upper and lower Hubbard bands 
(U/LHBs).  Attempts were quickly made to go beyond mean field theories by
incorporating fluctuation effects, but a number of problems soon arose.  While 
some calculations found evidence for pseudogaps\cite{VAT}, others did
not\cite{FLEX}.  Many calculations found evidence for instabilities -- either 
to incommensurate magnetism\cite{ShSi} or to phase separation\cite{Schulz}.

In a broader context, there is an ongoing controversy as to whether the
Mott transition can be described starting from a band structure picture,
or whether the strong correlations require a local point of view.  For
large hole dopings, a local picture seems justified by the observation
that the magnetic correlation length $\xi$ remains finite in the presence
of the pseudogap, whereas mode coupling theories would predict a
divergence in $\xi$ as temperature $T\rightarrow 0$.  However, this effect 
may be due to nanoscale phase separation physics.

Newer experiments have suggested that indeed phase separation and/or
stripe physics is present in the hole doped cuprates\cite{Tran,patch4}
down to arbitrarily small dopings\cite{Birg}, thereby validating
the early models.  It remains difficult to develop a theory
which simultaneously treats both the effects of strong fluctuations and
(nanoscale) phase separation.  A simpler alternative has recently been
proposed.  While phase separation is a significant complication For {\it
hole doping}, this instability is greatly reduced or absent in {\it 
electron-doped}
materials\cite{nparm,KLBM}, allowing a much simpler analysis.  Moreover, 
for electron doping, the band picture involving short-range AFM order
seems justified, in that the correlation length diverges for all dopings
up to the QPT.  

Remarkably, simple mean field calculations suggest an explanation for both
the electron-hole asymmetry\cite{MK1} and the anomalous properties of the
electron-doped cuprates\cite{KLBM,MK3}.  Thus, for a band structure with
electron-hole symmetry, the AFM state is stable only at half filling,
being unstable against phase separation (negative compressibility) for any
finite doping (this is also found in dynamic mean field theory\cite{ZPB}).
Introducing electron-hole asymmetry (here via a finite second-neighbor
hopping $t'$), the doped system remains unstable for any finite doping
{\it toward} the Van Hove singularity (VHS) (hole-doping for $t'<0$).
However, for doping {\it away} from the VHS, electronic phase separation
can be eliminated if the asymmetry is large enough.  Similarly, the RPA
spin wave spectra\cite{SWZ} are found to be stable in electron doped
materials, even when both upper and lower Hubbard bands cross the
Fermi level\cite{MK3}, while they are unstable against incommensurate SDWs in 
the presence of hole doping\cite{SiTe,ChuF}.  

The resulting physics is much simpler for electron doping, with the gap 
decreasing gradually as the UHB fills.  A quantum critical point (QCP) (Mott gap
closing) is found\cite{nparm} just beyond optimal doping, which can be 
understood theoretically\cite{KLBM} if the effective Hubbard $U$ parameter 
decreases with doping (e.g., due to screening).  Since the pseudogap is 
associated with short-range AFM order, it appears predominantly at the hot spots
-- the points where the Fermi surface crosses the Brillouin zone boundary for 
long-range AFM order.  As the gap shrinks, both upper and lower Hubbard bands 
cross the Fermi level, leading to two-band conduction, as observed 
experimentally\cite{NC2}.

The pseudogap closing also leads to a QCP near optimal doping in hole doped 
cuprates, consistent with predictions\cite{Varm,Cast1}.  However, its possible 
observation\cite{Tal1} is complicated by the presence of stripes, and this QCP
has been analyzed in terms of a mode-coupling theory of two dimensional CDWs,
taken as a model for stripes\cite{RM5,Cast2}.  Remarkably, for {\it both} 
electron and hole doping, the QCP occurs close to the point where hot spots 
vanish.  It will be shown why this is expected to be the case\cite{MK4}.

Thus, there appear to be {\it alternative pathways to Mott gap collapse} in the
cuprates.  Since the transition is better behaved in electron-doped cuprates, 
it makes sense to develop a model which can describe the transition for
these materials, then use the acquired insights to tackle the harder problem
of hole doping with the attendant (nanoscale) phase separation.  
Here the mean field and RPA results are extended by incorporating fluctuations 
via mode-coupling theory\cite{MuDo}, following Moriya's  self-consistent
renormalization (SCR)\cite{Mor,HasMo} procedure.  Mode coupling theories have 
been applied to charge density wave (CDW) systems\cite{LRA,RM5}, and have
led to a successful theory of weak itinerant magnetic systems\cite{Mor,HasMo}.  
They have also been used to study glass transitions\cite{glass}, and
recently extended to glasses in cuprates\cite{Cast2}.  The mode coupling
analysis is particularly convenient, being the simplest model for which
the Mermin-Wagner theorem is satisfied, and one can try to understand how
to have a Mott (pseudo)gap without a superlattice.  The present calculations 
are in general consistent with the results of Ref~\onlinecite{VAT}. The
resulting pseudogaps compare well with recent photoemission experiments,
while parallel results for hole doping provide clear evidence for
additional complications associated with stripe phases.  

The SCR theory involves a set of parameters\cite{MTU} which also arise in the 
nearly antiferromagnetic Fermi liquid (NAFL)\cite{NAFL} and spin 
fermion\cite{Chu,Chu2} models and in renormalization group (RG) calculations
of quantum phase transitions\cite{Her,Mil}.  These parameters are generally 
estimated empirically from experiments.  However, the good agreement between 
experiment and mean field theory for electron doped cuprates encourages us
to try to {\it calculate} these parameters from first principles, in
terms of a renormalized Hubbard parameter $U_{eff}$ and a mode coupling
parameter $u$.  Most of the defining integrals are dominated
by the region of hot spots.  

Some of these results have been reported previously in the discussion of the
mean-field results\cite{KLBM,MK4} and in a conference
procedings\cite{ICTP}. This paper is organized as follows.  Section II
presents an outline of the calculation, including the reduction to one
band (Appendix A), doping dependence of $U$ (Appendix B), and path
integral formalism (Appendix C).  The fluctuation-induced correction to
the susceptibilities is calculated, and is found to diverge for a
two-dimensional system, driving the N\'eel temperature $T_N$ to zero
(Mermin-Wagner theorem). Since the transition occurs when a Stoner factor
equals unity, it is controlled by the {\it real} part of the bare
susceptibility.  Hence Section III reviews the properties of $Re\chi$, 
showing that plateaus in $\chi$ as a function of doping, $\vec q$, or
$\omega$ are all controlled by the physics of hot spots.  In turn, these
plateaus provide {\it natural phase boundaries} for QCPs\cite{MK4}.  The 
resulting susceptibility has a form similar to that postulated for a
nearly antiferromagnetic Fermi liquid (NAFL), and a calculation of the
NAFL parameters (Appendix D) finds that there are {\it extra} (cutoff)
parameters, which cannot be neglected.  In
Section IV, this renormalized susceptibility is incorporated into
the lowest-order correction to the electronic self energy, allowing a 
calculation of the spectral function associated with the {\it pseudogap} ($T_N=0
$).  Excellent agreement is found with the ARPES spectra of NCCO.  Section
V shows that inclusion of interlayer hopping leads to a finite $T_N$
(Appendix E).  Results are discussed in Section VI, and Conclusions in
Section VII. 
%The Appendices include (A) a comparison of the
%Mott gaps in one and three band models; (B) an estimate of the doping
%dependence of $U$; (C) details of the self-consistent renormalization 
%calculations; (D) parameter evaluations; and (E) details of the
%interlayer coupling calculations.

\section{Outline of the Calculation}

\subsection{$t-J$ vs Mode-Coupling $t-U$ Models}

While the strongly correlated Hubbard model is often approximated by a $t-J$
model, this is not appropriate in the present analysis.  In the electron
doped cuprates, both Hubbard bands need to be accounted for, since (a)
ARPES can detect both bands (at least up to the Fermi level), and
(b) the Mott gap is found to collapse with doping, leading to an overlapping of
both bands with the Fermi level.  This situation is difficult to incorporate 
into the $t-J$ model, where one Hubbard band is neglected.  For this 
reason, an alternative procedure is used in the present paper.

While mean field calculations reproduce the low temperature properties, 
they do not account for thermal fluctuations, and hence predict that Neel
order persists to too high temperatures -- indeed, for a two-dimensional
system, Neel order can exist only at $T=0$ (Mermin-Wagner theorem)\cite{MW}.  
Even when fluctuation effects depress the onset of {\it long-range order},
{\it short-range order} can still persist up to the mean-field transition
temperature, producing a pseudogap similar to the ones found in 
one-dimensional CDWs\cite{LRA}.  In the AFM these fluctuations are associated
with spin wave excitations, in particular the Goldstone modes near $\vec Q$.  
Here, this effect is calculated, following the self consistent renormalization
(SCR) scheme of Moriya\cite{Mor,HasMo}. 

The calculation recovers the mean-field results, that antiferromagnetism
is stable for electron doping (in the presence of a second-neighbor
hopping parameter $t'<0$), but for hole doping the lowest energy state is
either an incommensurate SDW\cite{IMac} or phase separated\cite{VVG,KMWhiSc}.
[Unrestricted Hartree-Fock calculations find that for $t'=0$, the
incommensurate antiferromagnetic state has lower free energy, while for
sufficiently negative $t'$, the phase separated state is the lower free
energy state\cite{CK0}.]  While the SCR technique can be generalized to
deal with competing phases\cite{OnI}, only the antiferromagnetic fluctuations 
will be treated here. To keep the results consistent with a low-doping
$t-J$ analysis, the mode coupling parameter is chosen to reproduce the
$t-J$ expression for the spin stiffness parameter at half filling.

\subsection{Model Dispersion and Doping Dependence of $U$}

The cuprates are treated in a one-band model.  By comparison with a
3-band model (Appendix A), this can be shown to be an excellent 
approximation for the magnetic properties.  The bare electronic dispersion is 
\begin{equation}
\epsilon_k=-2t(c_x+c_y)-4t'c_xc_y, 
\label{eq:0}
\end{equation}
with $c_i=\cos{k_ia}$.  The dispersions for undoped Sr$_2$CuO$_2$Cl$_2$
SCOC and electron-doped NCCO can be fit by assuming $t=0.326eV$,
$t'/t=-0.276$, with $U$ taken as an effective doping dependent
parameter\cite{KLBM}, with $U=6t$ at half filling.  Remarkable, virtually
the same parameters are found\cite{PeAr} to describe the spin wave
spectrum\cite{RoRi} in La$_2$CuO$_4$: $t=0.34eV$, $t'/t=-0.25$, and
$U/t=6.2$.  The former values will be used here.

Many textbooks on strong correlation physics\cite{Ful,NNag} note that the 
Hubbard $U$ should be doping dependent, based on
the original results of Kanamori\cite{Kana}, but there are no satisfactory
results for the doping dependence in the cuprates.  A simple model calculation,
which gives semiquantitative agreement with experiment in NCCO\cite{nparm,KLBM},
is described in Appendix B.

\subsection{Mode Coupling Calculation:\\
Derivation of NAFL Parameters}

A naive perturbation theory breaks down due to mode coupling, and the SCR scheme
is introduced to calculate the renormalized susceptibility near the 
antiferromagnetic wave vector $\vec Q$.  The (path integral) formalism is 
reviewed in Appendix C, and only the main results are given here.
The divergence of the susceptibility is
controlled by the (inverse) Stoner factor $\delta_q=1-U\chi_0(\vec Q+\vec q,
\omega )$, where $\chi_0$ is the bare magnetic susceptibility.  Within the SCR, 
$\delta_0$ is renormalized to $\delta>0$, so there is no divergence, but when 
$\delta =0^+$ there are strong fluctuations (pseudogap regime).  The physics is 
controlled by the dispersion of $\delta$ near $\vec Q$, 
\begin{equation}
\delta_q(\omega )=\delta +Aq^2+A_zq_z^2-B\omega^2-iC\omega,
\label{eq:0B18}
\end{equation}
Eq.~\ref{eq:B18}; the important parameters $A$, $B$, and $C$ are 
evaluated in Appendix D (B is small and can generally be ignored).  This
leads to a susceptibility 
\begin{equation}
\chi (\vec q,\omega )={\chi_Q\over 1+\xi^2[(\vec q-\vec Q)^2+a_z
(q_z-Q_z)^2]-\omega^2/\Delta^2-i\omega /\omega_{sf}},
\label{eq:15}
\end{equation}
with coefficients given by Eqs.~\ref{eq:B36}-\ref{eq:B39} in terms of $A$, $B$, 
and $C$.  Here and below $\vec q$ and $\vec Q$ are treated as
two-dimensional vectors, while $q_z$ is introduced explicitly, and
$a_z=A_z/A$.  Interlayer coupling will be ignored, $a_z=0$, until Section
V and Appendix E, where it is considered as a possible mechanism for
generating a finite N\'eel temperature. The similarity of Eq.~\ref{eq:15}
to the corresponding result for CDW's\cite{RM5} should be noted -- the SCR
is a form of mode coupling theory.  

Equation~\ref{eq:15} is of nearly antiferromagnetic Fermi liquid 
(NAFL)\cite{NAFL} form, and the same parameters also arise in a
renormalization group calculation of QCP's\cite{Her,Mil}.  Thus, 
evaluation of the parameters $A$, $B$, and $C$ would amount to a {\it
microscopic derivation of NAFL theory}.  The present work evaluates these
parameters in terms of two interaction parameters, the Hubbard $U$ and 
a mode coupling parameter $u$.  It is found that $U$ has an important
doping dependence, estimated in Appendix B, which is consistent with
experiment. However, an attempt to directly calculate the mode coupling
parameter $u$ fails, giving anomalously small values -- this problem has
been noted previously, although there is debate about whether $u$ 
diverges\cite{Chu} or vanishes\cite{Her,Mil}.  Here, the $t-J$ model is
used to fix the value of $u$.  Finally, it must be noted that the SCR
theory is a model of weak ferro- or antiferromagnetism\cite{Mor}, and is
here found to underestimate the Mott gap splitting near half filling.
This most likely is due to imperfect self-consistency: the parameters are
estimated from the bare susceptibility $\chi_0$, while the opening of the
pseudogap leads to significant modifications of $\chi$.

Despite these limitations, the resulting model is well behaved, and in good
qualitative agreement with experiment, suggesting that a full derivation can
ultimately be carried out along these lines.  There are a number of
{\it deviations} from the simplest NAFL theory.  In Section III, it will be 
shown that the theory contains two additional {\it cutoff} parameters, $q_c$ 
and $\omega_c^-$, which cannot be neglected (or sent to infinity).  One 
consequence of this is that the $A$ parameter develops a significant temperature
dependence, particularly in the electron-doping regime.  

Equation~\ref{eq:15}, with $Im\chi\sim\omega$, only holds in the presence of hot
spots\cite{Mil}.  Hot spots complicate the evaluation of the A, B, and C 
parameters, since the expressions contain integrals which are formally 
divergent near the hot spots.  For the present band structure, hot spots exist 
only when the chemical potential $\mu$ is in the range $4t'\le\mu\le 0$, or 
for doping $0.25>x>-0.19$ (electron dopings are considered as negative).  
%%nonanalytic

The following section demonstrates how hot spots control the
properties of the magnetic susceptibility (plateau formation), and
summarizes the evaluation of the
SCR parameters.  The integrals are more singular at the end points of the hot
spot range -- the H and C points.

\section{Hot Spot Plateaus and Generic QCPs}

\subsection{Plateaus in Doping Dependence}

\subsubsection{The Pseudo-VHS}

\begin{figure}
\leavevmode
   \epsfxsize=0.38\textwidth\epsfbox{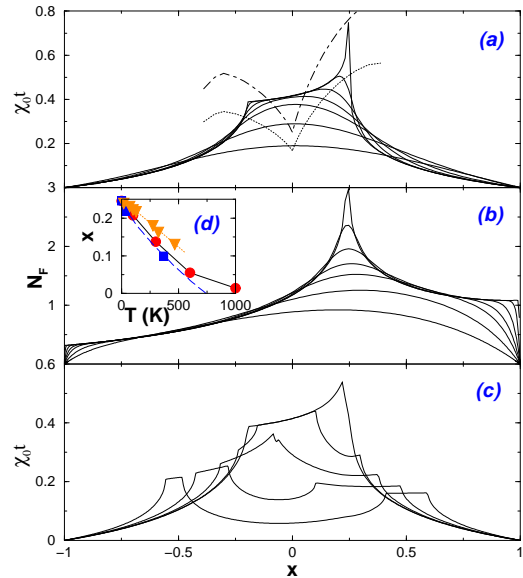}
\vskip1.5cm
\caption{(a) Susceptibility $\chi_0$ at $\vec Q$ as a function of doping for 
several temperatures.  From highest to lowest curves near $x=0.1$, the
temperatures are $T$ = 1, 100, 300, 600, 1000, 2000, and 4000 K. Dotted line =
$1/U_{eff}$, dot-dashed line = $1.5/U_{eff}$.  (b) Density of 
states $N_F$ for the same temperatures.  (c) Susceptibility $\chi_0$ at $\vec Q$
as a function of doping for several frequencies at $T=1K$: $\omega$ = 0.01, 0.1,
0.3, 0.6, 1.0 eV.  (d) Pseudo-VHS (peak of $\chi_0$) as a function of 
temperature $T_V$ (circles) or scaled frequency $T_c^-=\omega_c^-/\pi$ 
(squares); triangles = $T_{incomm}$.}
\label{fig:0a}
\end{figure}

The present analysis is based on a self-consistent renormalization scheme:
in two dimensions, fluctuations prevent the establishment of long range AFM
order.  Hence, the relevant quantity on which the study is based is the
bare magnetic susceptibility,
\begin{equation}
\chi_0(\vec q,\omega) =-\sum_{\vec k}{f(\epsilon_{\vec k})-f(\epsilon_{\vec k+
\vec q})\over\epsilon_{\vec k}-\epsilon_{\vec k+\vec q}+\omega+i\delta},
\label{eq:0a}
\end{equation}
where $\delta$ is a positive infinitesimal.  This susceptibility has been 
analyzed in a number of papers, but generally only $Im(\chi )$ is explored in
detail (e.g., Refs.~\onlinecite{BCT,SiZLL,LaSt}), whereas the Stoner criterion
involves $Re(\chi )$, which was studied in Ref.~\onlinecite{OPfeut}.  The 
extended discussion which follows is intended to bring out salient features for
the computation of the NAFL parameters.  The doping dependence of $\chi_0(
\vec Q,\omega)$ is illustrated in Fig.~\ref{fig:0a}a, where $\vec Q=(\pi ,\pi )
$. The susceptibility has a remarkable doping dependence, with the large peak at
the Van Hove singularity (VHS) shifting\cite{OPfeut} to half filling with 
increasing temperature $T$. The peak position of this `pseudo-VHS' defines a 
temperature $T_V(x)$, Fig.~\ref{fig:0a}d (circles).  This behavior can readily 
be understood from the form of $\chi_0(\vec Q,0)$, Eq.~\ref{eq:0a}.  The
denominator $\epsilon_{\vec k}-\epsilon_{\vec k+\vec Q}=-4t(c_x+c_y)$, is {\it 
independent of $t'$}, and hence has a stronger divergence than the density
of states (dos).  Indeed, this divergence matches the strong VHS found for
$t'=0$ (perfect nesting), and like that VHS falls {\it at half filling},
$x=0$.  There is one crucial difference -- at low temperatures, this
divergence is {\it cut off} by the Fermi functions, which leave the
integrand non zero in a wedge which intercepts the zone diagonal (where
the denominator vanishes) only at isolated points: the hot spots.  Hence,
the residual divergence at low $T$ is still dominated by the conventional
VHS.  However, {\it at finite $T$}, excitations along the zone diagonal
become allowed, leading to a stronger divergence of $\chi_0(\vec Q,0)$
near $x=0$. 

The strong temperature dependence of the pseudo-VHS is in strong contrast to the
density of states, $N_F$, Fig.~\ref{fig:0a}b, and also with the pairing
correlations\cite{OPfeut}.  The denominator of the pairing susceptibility 
involves the {\it sum} of the energies, $\epsilon_{\vec k}+\epsilon_{\vec k+\vec
Q}=-8t'c_xc_y$, rather than their difference (as in Eq.~\ref{eq:0a}), and hence 
always peaks at the ordinary VHS.

The difference between nesting and pairing susceptibilities has a
fundamental significance.  By mixing electron and hole-like excitations,
the superconducting gap is always pinned to the Fermi level, and can open
up a full gap at any doping.  On the other hand, a nesting gap need not be
centered on the Fermi surface, and is constrained to obey Luttinger's
theorem, conserving the net number of carriers in the resultant Fermi
surface.  Hence, the only way a nesting instability (such as
antiferromagnetism) can open a full gap at the Fermi level is for the
instability to migrate with increased coupling strength to integer filling
of a superlattice zone (e.g., half filling of the normal state).

Since the susceptibility has such a distinct {\it temperature} dependence from
the density of states, one might ask how the {\it frequency} dependence
compares.  This is illustrated in Fig.~\ref{fig:0a}c at low temperature (1K).
While the frequency introduces additional sharp features  and has
an overall very distinct appearence from the T-dependence, nevertheless the main
peak also shifts from the VHS toward lower doping with increasing $\omega$
-- in fact, the shift is almost the same when comparing $\hbar\omega$ and
$\pi k_BT$, Fig.~\ref{fig:0a}d.  The dashed line in Fig.~\ref{fig:0a}d is
$T_c^-=\hbar\omega_c^-/\pi k_B$, with\cite{BCT} 
\begin{equation}
\omega_c^-={4t(\hat\mu -\tau)\over 1-\tau},
\label{eq:00e}
\end{equation}
with $\tau=2t'/t$ and $\hat\mu=\mu /2t$. 
The proportionality of frequency and temperature dependences holds only
in the hole doped regime: temperature shifts the susceptibility peak only to
half filling, $x=0$, while frequency will shift the peak beyond half filling 
($x<0$).

Also in contrast to $N_F$, the susceptibility has (at low $T$) a plateau shape,
with sharp falloff in intensity beyond the plateau edges on both electron and 
hole doping sides of half filling.  This shape is characteristic of hot spot 
physics.  Hot spots are those points where the Fermi surface (FS) intersects the
replica FS shifted by $\vec Q$.  They are located at $c_x=-c_y=c_{x0}$, with
\begin{equation}
c_{x0}=\cos{ak_{x0}}=\sqrt{\mu\over 4t'},
\label{eq:C1}
\end{equation}
and equivalent points.  The edges of the plateau are those points at which the 
overlap terminates (hot spots cease to exist).  For the band structure of 
Eq.~\ref{eq:0} these points occur at chemical potential $\mu$ = $4t'\equiv\mu_v
$ (the VHS) and 0, or at dopings $x$ = 0.25, -0.19 (taking electron dopings as 
negative).  Since these two end points play an important role, it is convenient 
to label them, and they are here called `hot' hot spot and `cold' hot spot
(or H-point and C-point) for the hole and electron-doped termination
points, respectively.  It will be demonstrated below that at each
doping, the hot spots also lead to a susceptibility plateau in momentum
space, around $\vec Q$, collapsing to a logarithmic (square root)
divergence at the H- (C-)point.  The $H$-point is the VHS, and hence also
involves a conventional ETT. The physics is simpler near the $C$-point,
where the topology hardly changes but the FS and $\vec Q$-FS become
decoupled (it is therefore a form of Kohn anomaly\cite{OPfeut}).  

\subsubsection{Mean Field Mott Transition}
%Natural Phase Boundaries of QCP's}

\begin{figure}
\leavevmode
   \epsfxsize=0.33\textwidth\epsfbox{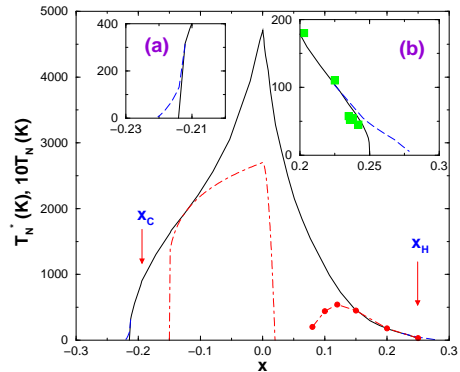}
\vskip0.5cm
\caption{Mean field magnetic transition temperatures determined from Stoner 
criterion using $U_{eff}$ of Fig.~\protect\ref{fig:0a}.  Solid line: 
commensurate (at $\vec Q$); long dashed line: incommensurate. Dot-dashed
line = 10$T_N$, where $T_N$ is the onset of long range AFM order, from 
[\protect\onlinecite{AIP}] and [\protect\onlinecite{Ich}] (with filled
circles). Insets = blowups near C- and H-points.  Squares in inset b =
pseudogap data of [\protect\onlinecite{Kras}].} 
\label{fig:0d}
\end{figure}

For the parameter values expected in the cuprates, these susceptibility
plateaus control the physics of the Mott gap collapse\cite{MK4}.  As a 
function of doping, the mean field Mott gap is found to close at a
doping just beyond the edge of the plateau, {\it for both electron and
hole doping}, Fig.~\ref{fig:0d}.  The solid and long dashed lines are the
commensurate and incommensurate mean field Mott transition temperatures
$T^*(x)$ calculated using the estimated $U_{eff}(x)$, dotted line in
Fig.~\ref{fig:0a}.  For electron doping, there is a double transition, 
first from commensurate to incommensurate antiferromagnetic order at the
plateau edge, then to the loss of any magnetic order at a slightly higher
doping (inset a).  For hole doping, the dominant antiferromagnetic order
is incommensurate for all dopings, but the difference in $T_N$ becomes
significant only near the H-point (inset b).  When fluctuations are
included (below), it is found that the N\'eel transition is shifted to
zero temperature, while a pseudogap first appears near the mean field
$T_N$.  Interlayer coupling can then restore a finite $T_N$, Section V.  
For the real cuprates, the terminations of the Mott gaps are preempted by
superconducting transitions, close to the critical regime.  

While the present local calculation (minimizing the free energy at a fixed
doping) finds incommensurate magnetic order for hole doping, some
global calculations (comparing free energies over a wider doping range) 
find that this is precluded by a phase separation instability.

The structure in the low temperature susceptibility, Fig.~\ref{fig:0a}, with its
largest peak at the H-point on the hole doped side, is in striking contrast to 
the calculated doping dependence of the N\'eel transition, Fig.~\ref{fig:0d},
which has a broad plateau on the electron-doped side, but falls off more 
quickly with hole doping, showing no sign of a peak near the VHS.  
This contrast can be accounted for by two effects.  First, the shift of 
spectral weight with temperature of the pseudo-VHS, noted in 
Fig.~\ref{fig:0a}, would tend to produce a symmetric falloff of $T_N$
with either electron or hole doping. But the dos peak at the VHS leads to
better screening of $U_{eff}$ for hole doping, thereby further depressing
$T_N$.  The experimentally observed\cite{AIP} $T_N$ (dot-dashed line)
shows an even stronger falloff with hole doping, perhaps due to phase
separation.  Since stripes can frustrate magnetic order, the figure also
includes the magnetic ordering temperature of quasi-static stripe arrays,
from Nd-substituted LSCO\cite{Ich}, which is taken as a lower bound
for the N\'eel ordering transition in the absence of stripes.  The mean
field calculation provides an approximate envelope of the resulting data,
but overestimates the transition temperatures by a factor of 10.  Note
that in the hole doped regime, there is good agreement between the mean
field transition and the pseudogap\cite{MK4} (squares in
Fig.~\ref{fig:0a}b = data of Krasnov\cite{Kras}, assuming $2\Delta
=4.6T^*$).  Calculation of the N\'eel transition beyond the mean field
level will be discussed in Section V.  

\subsection{Plateaus in Momentum Space}

\subsubsection{Plateaus}

\begin{figure}
\leavevmode
   \epsfxsize=0.33\textwidth\epsfbox{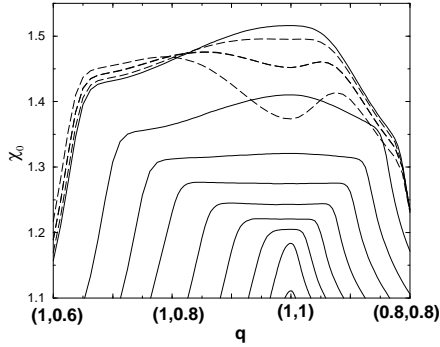}
\vskip0.5cm
\caption{Susceptibility $\chi_0$ near $\vec Q$ for a variety of dopings at 
$T=100K$.  From highest to lowest solid curves near $S\equiv\vec Q$, the 
chemical potentials are $\mu$ = -0.35, -0.30, -0.25, -0.20, -0.15, -0.10, 
-0.055, -0.02, and 0 eV.  For the dashed curves (top to bottom), $\mu$ = 
-0.352, -0.355, and -0.359eV.}
\label{fig:41}
\end{figure}
\begin{figure}
\leavevmode
   \epsfxsize=0.33\textwidth\epsfbox{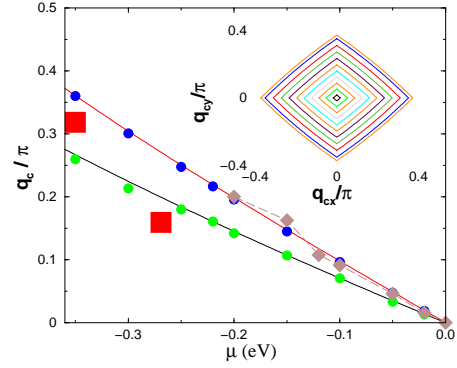}
\vskip0.5cm
\caption{Plateau width $q_c$, comparing Eq.\protect\ref{eq:0e} (solid lines)
and the measured widths (circles) from Fig.~\protect\ref{fig:41}.  Upper
curve along $[q_c,0]$ direction, lower along $[q_c,q_c]/\sqrt{2}$ direction.
Squares = experimental inverse correlation lengths $\xi^{-1}$ from 
Ref.~\protect\onlinecite{BoRe}; diamonds = $T^*_A/5000K$.  Inset = plateau 
boundary for a series of chemical potentials $\mu$ from 0 (smallest) to 
-0.359eV (largest).}
\label{fig:41a}
\end{figure}
\begin{figure}
\leavevmode
   \epsfxsize=0.33\textwidth\epsfbox{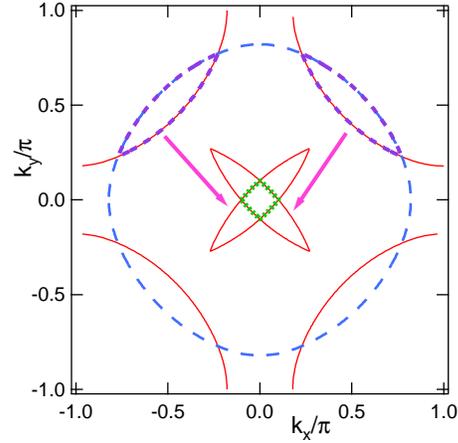}
\vskip0.5cm
\caption{Illustrating origin of plateaus (dotted line) from crossed
hole pockets (short dashed lines).}
\label{fig:41aa}
\end{figure}
\begin{figure}
\leavevmode
   \epsfxsize=0.33\textwidth\epsfbox{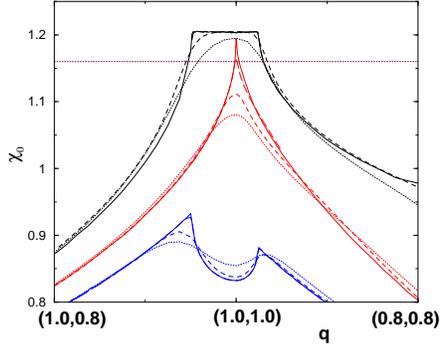}
\vskip0.5cm
\caption{Susceptibility $\chi_0$ near $\vec Q$ for several dopings near the
C-point.  Upper group at $\mu$ = -0.05eV, middle at $\mu$ = 0 (C-point), and
bottom at $\mu$ = +0.05eV.  Temperatures are $T=200K$ (dotted lines), 100K (short
dashed lines), 10K (long dashed lines), 1K (solid lines).  Horizontal line =
$U_{eff}(\mu =0)$.}
\label{fig:42}
\end{figure}

In analyzing either thermal fluctuations or the quantum fluctuations associated 
with QCPs, it is necessary to understand the susceptibility near the AFM
vector $\vec Q$. At each doping, hot spot physics leads to a plateau in momentum
space, centered on $\vec Q$. Figure~\ref{fig:41} shows how $\chi_0$
varies near $\vec Q$ at a low temperature (100K) for a series of different
dopings.  Results near $T=0$ are presented in 
Ref.~\onlinecite{ICTP}.  For all dopings there is a plateau in $q$.  The
width of the plateau at $T=0$ can be readily determined\cite{MK4}: in any
direction, it is the minimum $q$ needed to shift the replica FS so that
the hot spots are eliminated.  This can be found from the dispersion,
Eq.~\ref{eq:0}, by substituting $\vec k\rightarrow (\vec Q+\vec q)/2$, or
\begin{equation}
-2t(\hat s_x+ \hat s_y)-4t'\hat s_x\hat s_y=\mu,
\label{eq:0e}
\end{equation}
with $\hat s_i=\sin{(q_ia/2)}$.  As shown in Fig.~\ref{fig:41a}, this formula
agrees with the (anisotropic) plateau width measured from
Fig.~\ref{fig:41} (circles). The inset shows the shape of the plateau as a
function of doping.  The diamond shape of the plateau, Eq.~\ref{eq:0e}, is
related to the profile of the hole pockets formed by the overlap of the
shifted and unshifted FSs.  Specifically, the plateau is the region of
overlap of the two hole pockets, shifted to have a common center, as
illustrated in Fig.~\ref{fig:41aa}. The remaining parts of the pockets
also show up, as ridges\cite{RMrec} in the susceptibility, radiating from
the corners of the diamond (similar to the peaks in the $\mu =0.05eV$ data
in Fig.~\ref{fig:42}, below).  As noted by B\'enard, et al.\cite{BCT}, the
susceptibility in two-dimensions acts as a FS caliper.  The plateau width
leads to a natural limit on the magnetic correlation length, $\xi_c\sim
1/q_c$, in agreement with experimental data\cite{BoRe} (squares in
Fig.~\ref{fig:41a}), as noted previously\cite{LaSt,OPfeut}.  


The plateaus in $q$ help in understanding 
the doping dependence of the susceptibility near $\vec Q$, Fig.~\ref{fig:0a}a. 
At each doping on the plateau (in $x$) there is a plateau in $q$ centered at 
$\vec Q$, with the width of the plateau decreasing to zero as $x\rightarrow x_C
$, Fig.~\ref{fig:42}.  The critical points $q=q_c$ are precisely those points 
at which the $\vec Q+\vec q$-shifted-FS no longer overlaps the original FS.  

\subsubsection{Cusps}

For electron-doping beyond the C-point ($\mu>0$), the plateau ends and the 
susceptibility displays split peaks away from $\vec Q$, Fig.~\ref{fig:42}, 
with a dip in between.  The change in character of $\chi_0$ means that
{\it $\mu =0$ is a QCP}.  (The corresponding QCP at the H-point was
analyzed in Ref.~\onlinecite{OPfeut}.)  However, the {\it magnitude} of
$\chi_0$ also changes rapidly near $\mu=0$, so there should be a
transition to a non-magnetic phase near the same doping\cite{MK4}, as 
discussed in the previous subsection (note the line depicting $U(\mu =0)$
in Fig.~\ref{fig:42}).

\begin{figure}
\leavevmode
   \epsfxsize=0.33\textwidth\epsfbox{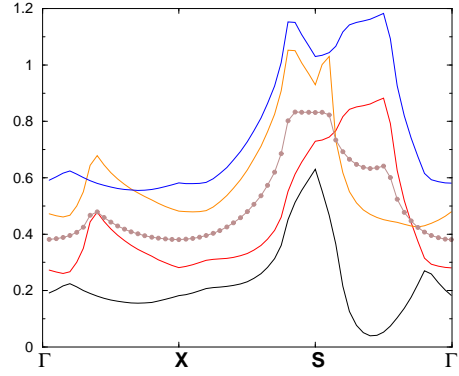}
\vskip0.5cm
\caption{Bare susceptibility $\chi_0$, for $x=0$, (beaded line); various solid
lines show contributions of individual quadrants, shifted vertically by 0.1 for
clarity. The special points in the Brillouin zone are $\Gamma =(0,0)$, $X=(\pi,
0)$, and $S=\vec Q$.  (T=100K.)}
\label{fig:2}
\end{figure}

%%nonanalytic
The origin of these $\mu> 0$ {\it cusps} can be readily understood from
Fig.~\ref{fig:2}.  Here, the contributions of individual quadrants to the
$\chi$ integral are plotted separately, with each quadrant containing two
hot spots.  It can be seen (Appendix D) that the dispersion at {\it each} hot 
spot contains a cusp, near which the dispersion is linear in $|q|$. However,
when adding the contributions of the 8 hot spots, the linear terms cancel,
leaving a quadratic dispersion.  For $\mu>0$ and $q=0$ there are no hot
spots -- the Fermi surfaces in $\vec k$ and $\vec k+\vec Q$ do not
intersect.  However, translating one Fermi surface by $\vec q$ will lead to an
intersection, with corresponding hot spot, beyond some threshold $\vec q_c$.
Since only one or two hot spots are restored for a given direction of 
$\vec q_c$, the linear terms do not cancel, leading to a linear in $q$
dispersion for $q>q_c$.  Exactly at $\mu=0$, $\chi_0$ has a $\sqrt{q}$ cutoff
as $q\rightarrow Q$ (Appendix D3).  
%This non-analyticity seems to be a common feature near a
%QCP\cite{BelKirV}, and will be discussed further in Section VI.  

\begin{figure}
\leavevmode
   \epsfxsize=0.33\textwidth\epsfbox{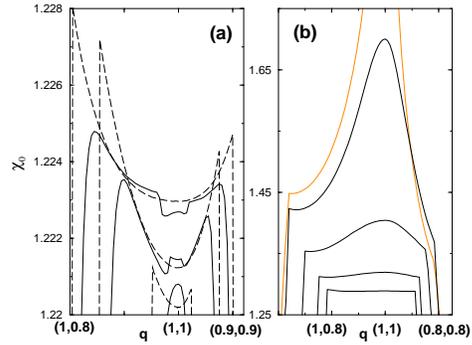}
\vskip0.5cm
\caption{(a): Expanded view of susceptibility $\chi_0$ on the plateaus near 
$\vec Q$ for a variety of dopings at $T=100K$ (solid curves) or 1K (dashed 
curves).  From highest to lowest curves near $\vec Q$, the chemical potentials 
are $\mu$ = -0.20, -0.15, and -0.05 eV (for both solid and dashed curves). 
All curves except $\mu=-0.20 eV$ have been shifted vertically to fit within the 
expanded frame. (b): Similar plateaus for the hole doped materials ($T=1K$),
with (from highest to lowest) $\mu$ = -0.359, -0.35, -0.3, -0.25, and -0.22 eV.}
\label{fig:44}
\end{figure}

Technically, similar cusps also arise at the plateau edges\cite{MK4} 
for electron doping, $0>\mu >-0.22eV$. The tops of the plateaus are not 
completely flat, Fig.~\ref{fig:44}a: the highest susceptibility is shifted away 
from $\vec Q$, and the sharp steps near $\vec Q$ are again hot spot effects, 
this time associated with the loss of hot spots at large values of $\vec q-\vec 
Q$.  However, these effects are much weaker than those associated with $\mu >0$ 
($\Delta\chi /\chi\le 0.5\%$ -- compare the vertical scales of 
Figs.~\ref{fig:42},~\ref{fig:44}).  Thus near the mean-field transition any
structure on the plateaus is smeared out by thermal broadening.  Even at $T=0$, 
these features are likely to be negligible compared to dispersion in $U$ which
arises from renormalization effects\cite{AGGAH}.

\subsubsection{Curvature (A)}

\begin{figure}
\leavevmode
   \epsfxsize=0.33\textwidth\epsfbox{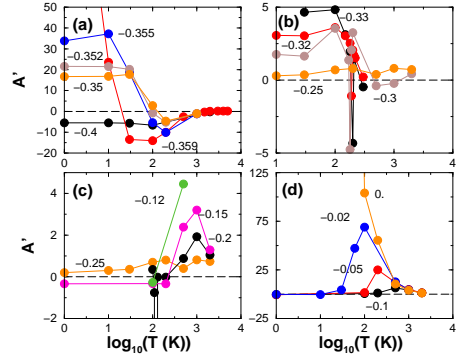}
\vskip0.5cm
\caption{Temperature dependence of $A'$ for several dopings. }
\label{fig:45}
\end{figure}

The plateau is a region of anomalously small local curvature $\hat A =A/U$ 
(Eq.~\ref{eq:0B18}) of the susceptibility, $\chi_0(\vec Q+\vec q)=\chi_Q-\hat 
Aq^2$, where $A$ is an important NAFL parameter. Clearly, at $T=100K$ the 
curvature $A$ has gone negative near the $H$-point, Fig.~\ref{fig:41}.  At
even lower temperatures, it reverts to positive values, Fig.~\ref{fig:44}b.  
The temperature dependence of the normalized parameter $A'=(\pi /a)^2(U/t)A$ 
is illustrated in Fig.~\ref{fig:45} at several dopings.  The temperature 
dependence is dominated by divergences at both H- and C-points.  The divergence 
at the H-point Fig.~\ref{fig:45}a is the well-known logarithmic VHS.  However, 
at finite temperatures spectral weight is shifted away from the VHS and $A$ 
turns negative, only recovering a positive sign above $T\simeq 2000K$. The
temperature at which $A$ turns negative can be defined as $T_{incomm}$:
$A<0$ for $T>T_{incomm}$.  From Fig.~\ref{fig:0a}d, $T_{incomm}$ is
comparable to but larger than $T_V$ (for $x\le 0.06$ $A$ remains
positive).  This in fact explains the origin of $T_{incomm}$.
Figure~\ref{fig:45}a demonstrates that $A$ is negative at $T\rightarrow 0$
beyond the H-point ($\mu =-0.4eV$).  Thus, increasing $T$ above $T_V$
produces the same susceptibility crossover.  A similar crossover was
discussed by Sachdev, et al.\cite{SCS}, except that they assumed that in
the high temperature phase the AFM fluctuations remained centered on the
commensurate $\vec Q$, whereas here $A$ is negative. At sufficiently high
temperatures $A$ again becomes positive for all dopings -- i.e., the
leading singularity of $\chi_0$ is always at $\vec Q$.  

At the C-point, the collapse of the plateau width translates into a
divergence of the curvature at $\vec Q$ ($\hat A\rightarrow\infty$).  
This divergence of the high-temperature susceptibility is cut off at low $T$, 
Fig.~\ref{fig:45}d, when the thermal smearing becomes smaller than the
plateau width.  For smller $T$, $A$ is controlled by the curvature on the
plateau.  The temperature at which $A$ has a peak, defined as $T^*_A$, is
plotted as diamonds in Fig,~\ref{fig:41a} (the peak is only found for
$x\le 0$).  Rather surprisingly, $T^*_A$ scales with the plateau width $q_c
$, even though the dynamic exponent is $z=2$.  Further, the maximum slope scales
approximately as $A_{max}\sim T_A^{*-1.5}$, which follows from the fact that $A
\sim T^{-1.5}$ at the C-point.  

At intermediate doping, Fig.~\ref{fig:45}b,c, $A$ is generally a
scaled-down version of the behavior near the two end points, with a
crossover near $\mu =-0.25eV$, where the T-dependence is weak.  Also for
intermediate temperatures, there can be fine structure on the plateau
(e.g., solid lines in Fig.~\ref{fig:44}a) which can lead to wild swings in
$A(T)$.  However, at these dopings they are not relevant, since the
susceptibility peaks are away from $\vec Q$, and this fine structure 
is not generally reported in Fig.~\ref{fig:45}.

%\begin{figure}
%\leavevmode
%   \epsfxsize=0.33\textwidth\epsfbox{fls.eps}
%\vskip0.5cm
%\caption{Doping dependence of $A'$ for several temperatures. (Inset = same data
%on an expanded scale.)}
%\label{fig:43}      
%\end{figure}  

\subsection{Plateaus in Frequency}

\begin{figure}
\leavevmode
   \epsfxsize=0.33\textwidth\epsfbox{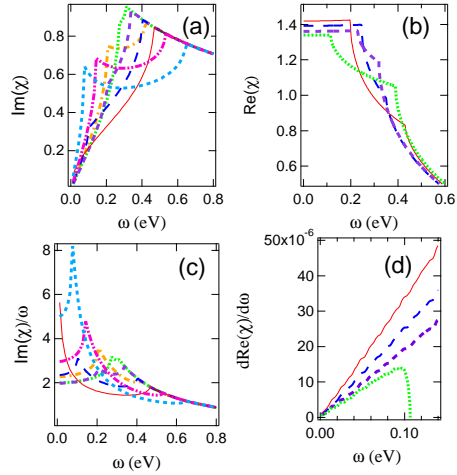}
\vskip0.5cm
\caption{(a) $Im\chi (\vec Q,\omega)$, (b) $Re\chi (\vec Q,\omega)$, (c) 
$Im\chi /\omega\equiv\hat C$, and (d) $dRe\chi (\vec Q,\omega)/d\omega$,
for (a,c): $\mu$ = 0 (solid line), -0.05 (long dashed line), -0.10
(dashed line), -0.15 (dotted line), -0.20 (dot-dashed line), -0.25
(dot-dot-dashed line), and -0.30eV (short dashed line); (b,d): $x$ = 0
(solid line), 0.04 (long dashed line), 0.10 (dashed line), and 0.15
(dotted line).}
\label{fig:44b}
\end{figure}

Figure~\ref{fig:44b} illustrates $Im\chi (\vec Q,\omega)$, $Re\chi (\vec Q,
\omega)$, and $Im\chi /\omega\equiv\hat C$.  At $T=0$, the imaginary part
of the susceptibility $\chi (\vec Q,\omega )$ can be calculated
analytically\cite{BCT}:
\begin{eqnarray}
Im(\chi (\vec Q,\omega ))=\sum_{\vec k}(f(\epsilon_{\vec k})-
f(\epsilon_{\vec k+\vec Q}))\delta (\epsilon_{\vec k+\vec Q}-\epsilon_{\vec k}
-\omega)
\nonumber \\
={F(\theta_1,\tilde k)-F(\theta_2,\tilde k)\over 4t},
\label{eq:C8aa}
\end{eqnarray}
where $F(\theta ,x)$ is an elliptic integral, $\tilde k=\sqrt{1-(\omega /8t)^2}
$, and $sin(\theta_i)=sin(\phi_i)/\tilde k$, with
\begin{equation}
\cos^2{(\phi_1)}=\cases{c^2_-&if $\omega\le\omega_c^-$\cr
                        \hat\omega /2&if $\omega >\omega_c^-$\cr},
\label{eq:C8bb}
\end{equation}
\begin{equation}
\cos^2{(\phi_2)}=\cases{c^2_+&if $\omega\le\omega_0$\cr
                        1&if $\omega >\omega_0$\cr},
\label{eq:C8cc}
\end{equation}
with $\hat\mu =\mu /2t$, $\hat\omega =\omega /4t$, $c^2_{\pm}=a_{\pm}+\sqrt{a_
{\pm}^2-\hat\omega^2}$, and $a_{\pm}=1-(\hat\mu\pm\hat\omega )/\tau$.
The real part $Re\chi$ can be found from the Kramers-Kronig result,
\begin{equation}
Re\chi (\vec Q,\omega)={1\over\pi }\int_0^{\infty}{Im\chi (\vec Q,\omega')
\omega'd\omega'\over\omega^{'2}-\omega^2}.
\label{eq:C8g}
\end{equation}

\begin{figure}
\leavevmode
   \epsfxsize=0.33\textwidth\epsfbox{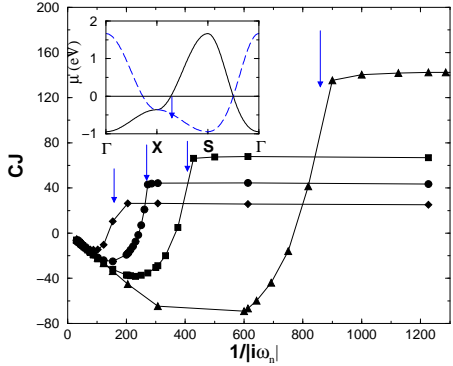}
\vskip0.5cm
\caption{$\hat C$ calculated for several values of $\mu$: $\mu$ =
-0.355 (diamonds), -0.357 (circles), -0.358 (squares), -0.359eV (triangles) 
[$\mu_v$ = -0.3599eV].  Inset: Band dispersion $\epsilon_{\vec k}$ (solid line) 
$\epsilon_{\vec k+\vec Q}$ (dashed line), for $\mu =0$.  Arrow = $\omega_c^-$.}
\label{fig:11}
\end{figure}

\begin{figure}
\leavevmode
   \epsfxsize=0.33\textwidth\epsfbox{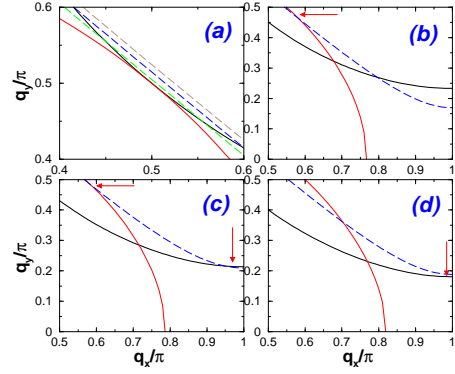}
\vskip0.5cm
\caption{Origins of critical cutoffs.  Thick solid line = FS; thin solid line =
Q-shifted FS; dashed lines = Eq.~\protect\ref{eq:C8d}, for several values of 
$\omega$.  Chemical potential $\mu$ = (a) 0, (b) -0.1, (c) -0.14, (d) -0.2eV.
horizontal arrows indicate $\omega_0$, vertical arrows $\omega_c^-$.}
\label{fig:44a}
\end{figure}

Thus, there are also plateaus in the frequency dependence of $Re\chi$.
Furthermore, hot spots generate an imaginary part of the susceptibility linear 
in frequency, which also approximates a plateau, particularly near the
H-point, Fig.~\ref{fig:11}.  The origin of this plateau and of the
critical frequencies $\omega_c^-$, Eq.~\ref{eq:00e}, and 
\begin{equation}
\omega_0={8t\over\tau}[\sqrt{1-\hat\mu\tau}-1]
\label{eq:C8e}
\end{equation}
can be understood from Fig.~\ref{fig:44a}.  The thick (thin) solid lines 
represent the original (Q-shifted) Fermi surfaces, while the dashed lines
represent
\begin{equation}
\omega =\epsilon_{\vec k+\vec Q}-\epsilon_{\vec k},
\label{eq:C8d}
\end{equation}
for various values of $\omega$.  Equation~\ref{eq:C8d} gives the points at which
the denominator of $\chi_0(\vec Q,\omega)$, Eq.~\ref{eq:0a}, vanishes. Thus at
$T=0$, $Im(\chi_0(\vec Q,\omega))$ is proportional to the length of the dashed
line lying between the original and Q-shifted FSs (i.e., where $f(\epsilon_{\vec
k})-f(\epsilon_{\vec k+\vec q})=\pm 1$).  Since the two FSs meet at an
angle, forming a wedge, $Im(\chi_0(\vec Q,\omega ))\sim\omega$. 

From Fig.~\ref{fig:44a}, the critical frequencies (denoted by arrows) are
points where the $\omega$ dependence of this length changes abruptly, leading 
to a sharp change in $Im\chi$.  Thus, near the H-point, the plateau width
is $\omega_c^-$ (inset, Fig.~\ref{fig:11}), while near the C-point it is
$\omega_0$. The vertical arrows in Fig.~\ref{fig:44a} indicate
$\omega_c^-$, where the dashed line 
(Eq.~\ref{eq:C8d}) intersects the FS at the zone boundary, while the horizontal 
arrows are\cite{BCT} $\omega_0$, where the dashed line ceases to intersect the 
Q-shifted FS.  There is a crossover at $\mu_c\simeq -0.14eV$: for $\mu >\mu_c$, 
$\omega_0<\omega_c^-$ while for $\mu <\mu_c$, $\omega_0>\omega_c^-$.  Combining 
Eqs.~\ref{eq:00e},\ref{eq:C8e}, $\omega_0=\omega_c^-$ at $\mu_c=[1-z(2-\sqrt{z}
)^2]2t/\tau =-0.1384eV$, with $z=1-\tau$.  For $\omega >min\{\omega_0,\omega_c^
-\}$, $Im(\chi_0(\vec Q,\omega))\sim\omega^{1/2}$, so $C\sim 1/\omega^{1/2}$ 
-- i.e., the susceptibility is no longer on the plateau. 

The height of the plateau $C$ is an important parameter of the SCR
model.  It can be represented as another frequency $\omega_1=1/C$, with
$C=U\hat C(\omega =0)$.  From Eq.~\ref{eq:B21} of Appendix C2, $C$ can be
found explicitly
\begin{equation}
C={1\over 2\pi Js_{x0}^2(1+\tau c_{x0})}={1\over\omega_1}
\label{eq:C8}
\end{equation}                                                          
(with $J=4t^2/U$, $s_{x0}^2=1-c_{x0}^2$). 
Defining a width parameter $\alpha_{\omega}=min\{\alpha_{\omega}^-,\alpha_
{\omega}^0\}$, with $\alpha_{\omega}^0=\omega_0/\omega_1$, then
\begin{equation}
\omega_1/\omega_c^-={2\pi t(1-\tau )\over U}[{1+\tau c_{x0}\over -\tau}]
\equiv{1\over\alpha_{\omega}^-}.
\label{eq:C8b}
\end{equation}
This latter is in good agreement with the numerical results (arrows in 
Fig.~\ref{fig:44a}) and is similar to the result found by Onufrieva and 
Pfeuty\cite{OPfeut}, using a hyperbolic band approximation valid near a VHS, 
$\omega_1/\omega_c^-=2\pi t(1-\tau )/U$.  
%These plateaus are further discussed in Appendix B2,3.

Because of the dynamic scaling $\omega\sim q^z$, this crossover is also
reflected in the behavior on the plateau {\it in $\vec q$},
Fig.~\ref{fig:44}: for $\mu >-0.14eV$, the plateau has a negative
curvature, which can almost be scaled between different dopings, while for
$\mu <-0.14eV$, the plateau starts to fill in, ultimately developing a
peak at $\vec Q$.  (See also Fig. 3a in Ref.~\onlinecite{ICTP}.)
Note that the plateau width collapses in frequency at
both the $H$- and $C$-points, while the collapse in wave number
($q_c\rightarrow 0$) is only present near the $C$-point.

\subsection{Parameter Evaluation for Mode Coupling Theory}

\begin{figure}
\leavevmode
   \epsfxsize=0.33\textwidth\epsfbox{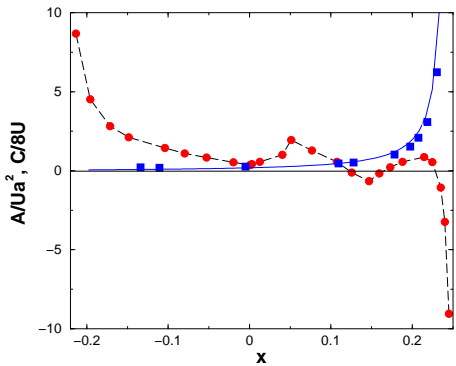}
\vskip0.5cm
\caption{Calculated values of $A$ (circles) and $C$ (squares), for $U=6t$.  
Solid line = Eq.~\protect\ref{eq:C8}}
\label{fig:10}
\end{figure}

\begin{figure}
\leavevmode
   \epsfxsize=0.33\textwidth\epsfbox{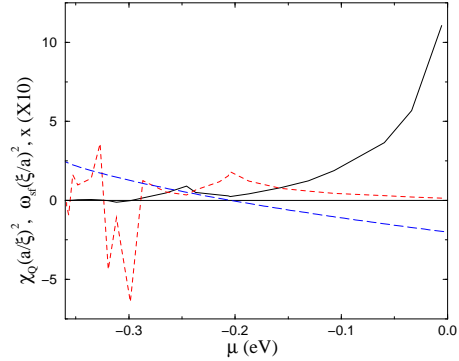}
\vskip0.5cm
\caption{Calculated values of $\chi_{\vec Q}/\xi^2$ (solid line) and $\omega_
{sf}\xi^2$ (short dashed line), assuming $U=6t$.  Long dashed line = doping 
$x(\mu )$ ($\times 10$).}
\label{fig:9}
\end{figure}

The evaluation of the SCR parameters $A$ and $C$ was discussed above.  The
collapse of the $\vec q$ and/or $\omega$ plateau widths near the H- and
C-points leads to the introduction of additional parameters $q_c$ and
$\alpha_{\omega}$.  The narrow width of the $\vec q$-plateau, particularly
for electron doping, leads to an additional complication not included in 
the conventional SCR analysis: the curvature of the bare susceptibility near $
\vec Q=(\pi ,\pi )$ (the $S$-point of the BZ) is strongly temperature dependent,
and for some dopings may even change sign.  
In principle, it is not difficult to incorporate an $A(T)$ into 
the analysis near the mean-field N\'eel temperature $T_N^*$ (pseudogap onset).  
But for the present 2D system, long range N\'eel order only sets in at $T_N=0$, 
and for $T<<T_N^*$, a self consistent value of $A$ should be found, by taking 
into account the effect of the pseudogap in modifying the electronic dispersion 
and hence $\chi$.  For the present, this complication is ignored, and in the 
following section $A$ is taken as $A=A(T_N^*)$, where $T_N^*$ is the
magnetic pseudogap onset, the temperature where $\chi_0(\vec Q)U_{eff}=1$, 
using the effective $U_{eff}$ found earlier\cite{KLBM} (Appendix B).  This 
should be the most important $A$ for controlling the pseudogap, and
moreover at lower temperatures the band renormalization should strongly
modify $A(T)$.  With this choice, the resulting $A(\mu )$ is plotted in
Fig.~\ref{fig:10}, along with the $C$ parameter, evaluated at $T=0$.  Note
that for electron doping, this choice of $A$ is always positive and varies
smoothly with doping, diverging at the C-point.  By contrast, for hole
doping $A$ is often negative, again illustrating the instability of the
uniform AFM phase.  [In principle, a positive $A$ can be found by taking
an incommensurate nesting vector; for this paper, only commensurate
nesting is considered; the negative $A$ will suggest when electronic
inhomogeneity may be important.] Given $A$ and $C$, Fig.~\ref{fig:9} shows
the calculated values of $\chi_{\vec Q}$ and $\omega_{sf}$, normalized to
$\xi^2$. 

In the following, the present results are applied to understanding the ARPES
spectra of electron doped cuprates, concentrating on the four dopings analyzed
by Armitage, et al.\cite{nparm}.  For convenience, Table I summarizes the
parameters for these dopings.  From the mean-field analyses\cite{KLBM}, the 
effective Hubbard parameters were found to be $U_{eff}/t$ = 6 ($x=0$), 5 ($x=-0
.04$), 3 ($x=-0.10$), and 2.5 ($x=-0.15$).  [These numbers differ somewhat
from those of Ref.~\onlinecite{KLBM}, since a second neighbor hopping,
$t''$, was included in the latter analysis, to give the best fit of the
Fermi surfaces.]  The Stoner factor has a quantum correction, $\eta$,
Eq.~\ref{eq:B35}, which tends to suppress the AFM transition; 
hence a smaller renormalization of $U$ is required.  This is reflected in Table 
1: for $x$ = -0.1, -0.15, there are two rows, the upper row using the 
mean-field $U$ parameters, the lower with the quantum correction.  Note
that the $U$'s are enhanced by essentially the quantum correction factor. 
These values will be used in the subsequent analysis.

The SCR analysis also involves a mode coupling parameter $u$, evaluated
in Appendix D6. As found previously, direct evaluation of this parameter is 
unsatisfactory -- the results of Table 1 being anomalously small due to
the flatness of the susceptibility plateau ($\partial\chi
/\partial\omega\sim 0$), Fig.~\ref{fig:44b}d.  Below, an empirical way to
estimate $u$ is suggested.
\vskip 0.1in
\begin{tabular}{||c|c|c|c|c|c|c|c|c||}        %\hline
\multicolumn{9}{c}{{\bf Table I: Electron Doped Cuprates}}\\
            \hline\hline
x& $U/t$&$A/a^2$&$\omega_1$(eV)&$\alpha_{\omega}$&$q_ca$&$\eta$&$T^*_A(K)$&$u^
{-1}$ (eV) \\
     %\hline
    \hline\hline
0&6&0.696&0.345&0.583&0.635&1.29&1020&760         \\     \hline
-0.04&5&1.16&0.540&0.455&0.518&1.25&850&3200    \\        \hline
-0.10&3&1.34&1.32&0.176&0.342&1.19&500&2700      \\     \hline
''&3.5&1.56&1.13&0.206&''&1.24&''&2300       \\     \hline
-0.15&2.5&1.75&2.16&0.054&0.172&1.08&56&4000      \\     \hline
''&2.9&2.03&1.86&0.062&''&1.15&''&3500\\     \hline
\end{tabular}
\vskip 0.1in
It is convenient to compare the present results with parameters estimated for
the SCR model\cite{MTU} from experimental data for (optimally) hole-doped
cuprates.  The parameters are defined as $T_0=Aq_B^2/2\pi C$, $T_A=Aq_B^2/2\chi_
0$, $y_0=\delta_0(T=0)/Aq_B^2$, and $y_1\simeq 12a^2u/\pi^3AC$.  The results
are listed in Table II, where the first line gives the hole-doped results
estimated in Ref.~\onlinecite{MTU}.  Moriya, et al.\cite{MTU} took $q_B^2=1/4\pi
a^2$ ($q_Ba=0.282$), while for Table II it is assumed that $q_B=q_c$.  A
key difference is that Moriya, et al.\cite{MTU} assume the system is in
the paramagnetic phase ($y_0>0$) at and above optimal (hole) doping, while
in the present work $y_0<0$, and the system is paramagnetic due to the
Mermin-Wagner theorem, with the Mott gap appearing as a pseudogap.  The
small magnitude of $y_0$ is suggestive of a system pinned close to a QCP.
Finally, the parameter $y_1$ is estimated using the value $u^{-1}
=0.256eV$ (below), and not the anomalous values of Table 1.
\vskip 0.1in
\begin{tabular}{||c|c|c|c|c||}        %\hline
\multicolumn{5}{c}{{\bf Table II: SCR Parameters}}\\
            \hline\hline
x& $T_0$ (K)&$T_A$ (K)&$y_0$&$y_1$ \\
     %\hline
    \hline\hline
$\sim0.2$&1600-4000&3000-10000&0.01-0.02&3         \\     \hline
0.0&180&1150&-5.27&0.75            \\      \hline
-0.04&310&1300&-3.31&0.7        \\        \hline
-0.10&380&670&-1.23&1.5         \\     \hline
-0.15&200&220&-0.31&1.85          \\     \hline
\end{tabular}

\section{ARPES Spectra}

\subsection {SCR Transition and Correlation Length}

\begin{figure}
\leavevmode
   \epsfxsize=0.33\textwidth\epsfbox{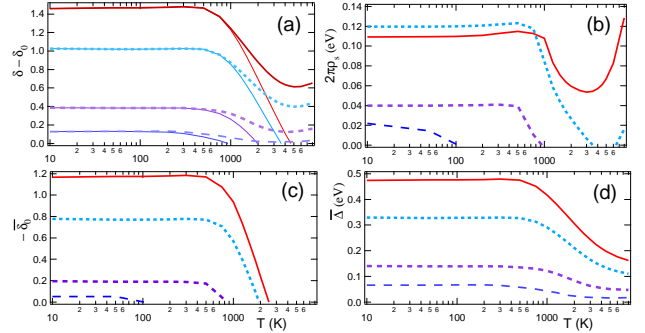}
\vskip0.5cm
\caption{(a) $\delta -\delta_0$ (thin solid lines = $-\delta_0$); (b) $\rho_s$ 
calculated from Eqs.~\protect\ref{eq:B40},~\protect\ref{eq:B34f}; (c) $-\bar
\delta_0$; (d) $\bar\Delta$, Eq.~\protect\ref{eq:20a}.
In all the plots, the solid curves correspond to $x=0.0$, dotted lines:
$x=-0.04$, short dashed lines: $x=-0.10$, long dashed lines: $x=-0.15$.}
\label{fig:11a1}
\end{figure}

Given the above parameters, the doping dependence of the MF and SCR
transitions is compared in Fig.~\ref{fig:11a1} for the four electron
dopings studied in 
Refs.~\onlinecite{nparm},~\onlinecite{KLBM}.  The MF transition occurs when the 
bare Stoner factor $\delta_0=1-\chi_{\vec Q0}U$ becomes negative, 
Fig.~\ref{fig:11a1}a.  However, in SCR the renormalized Stoner factor $\delta$
stays positive, so there is no $T>0$ phase transition (Mermin-Wagner theorem), 
although $\delta -\delta_0$ has a strong increase near the temperature
where $\delta_0$ changes sign.  There is still a zero-T N\'eel transition,
controlled by the quantum corrected Stoner factor, $\bar\delta_0=\eta
-\chi_{\vec Q0}U$. 
%This quantum correction accounts 
%for part of the observed doping dependence of $U(x)$ -- indeed, the value of 
%$U_{eff}$ had to be adjusted for the $x=-0.10$ and -0.15 data, as indicated in 
%the added rows in Table I.  
From Fig.~\ref{fig:11a1}c, it can be seen that at $x=-0.15$, the system is
close to a QCP, $\bar\delta_0(T=0)\rightarrow 0$.  This QCP is controlled
by the Stoner criterion of the zero-T antiferromagnet.  While there is no
long range order, there is still a Mott (pseudo)gap, controlled by {\it
short-range} order, Fig.~\ref{fig:11a1}d.  This will be discussed further
below.

In the renormalized classical regime, the vanishing of $\delta$ as $T\rightarrow
0$ is controlled by a correlation length, which can be written as\cite{CHN}
\begin{equation}      
\xi =\xi_0e^{2\pi\rho_s/k_BT}
\label{eq:16}
\end{equation}
with spin stiffness $\rho_s$ given by Eq.~\ref{eq:B40}.  The prefactor $\xi_0$ 
%can take two different forms, depending on the frequency cutoff $\alpha_{\omega}
%/C$.  When this is {\it less} than $T$, the prefactor is a constant $\xi_0=q_c^{
%-1}$, but in the more probable case $T<\alpha_{\omega}/C$ it 
is $T$-dependent, $\xi_0=\sqrt{Ae/2CT}$.  This T-dependence agrees with the {\it
one-loop} $\sigma$-model results\cite{KoCha} rather than the more accurate 
two-loop results\cite{CHN,Tkhsh}.  This difference is presumably a deficiency of
the present model in not using fully self consistent parameters.  Note that 
$\rho_s$ is found to be nearly $T$-independent below the pseudogap onset.
Equation~\ref{eq:16} is used to fix the value of $u$.  From Eq.~\ref{eq:B40}, 
$\rho_s\propto u^{-1}$, while the $\sigma$-model calculations\cite{CHN,KoCha} 
give $\rho_s=JS^2$.  Equating these two expressions for $x=0$, $T=0$ gives 
$u^{-1}=0.256eV$, which is assumed for all dopings.  The calculated values
of $\rho_s$ are illustrated in Fig.~\ref{fig:11a1}c, based on
Eqs.~\ref{eq:B34b},~\ref{eq:B40}.  

\subsection{General Results}

\begin{figure}
\leavevmode
   \epsfxsize=0.33\textwidth\epsfbox{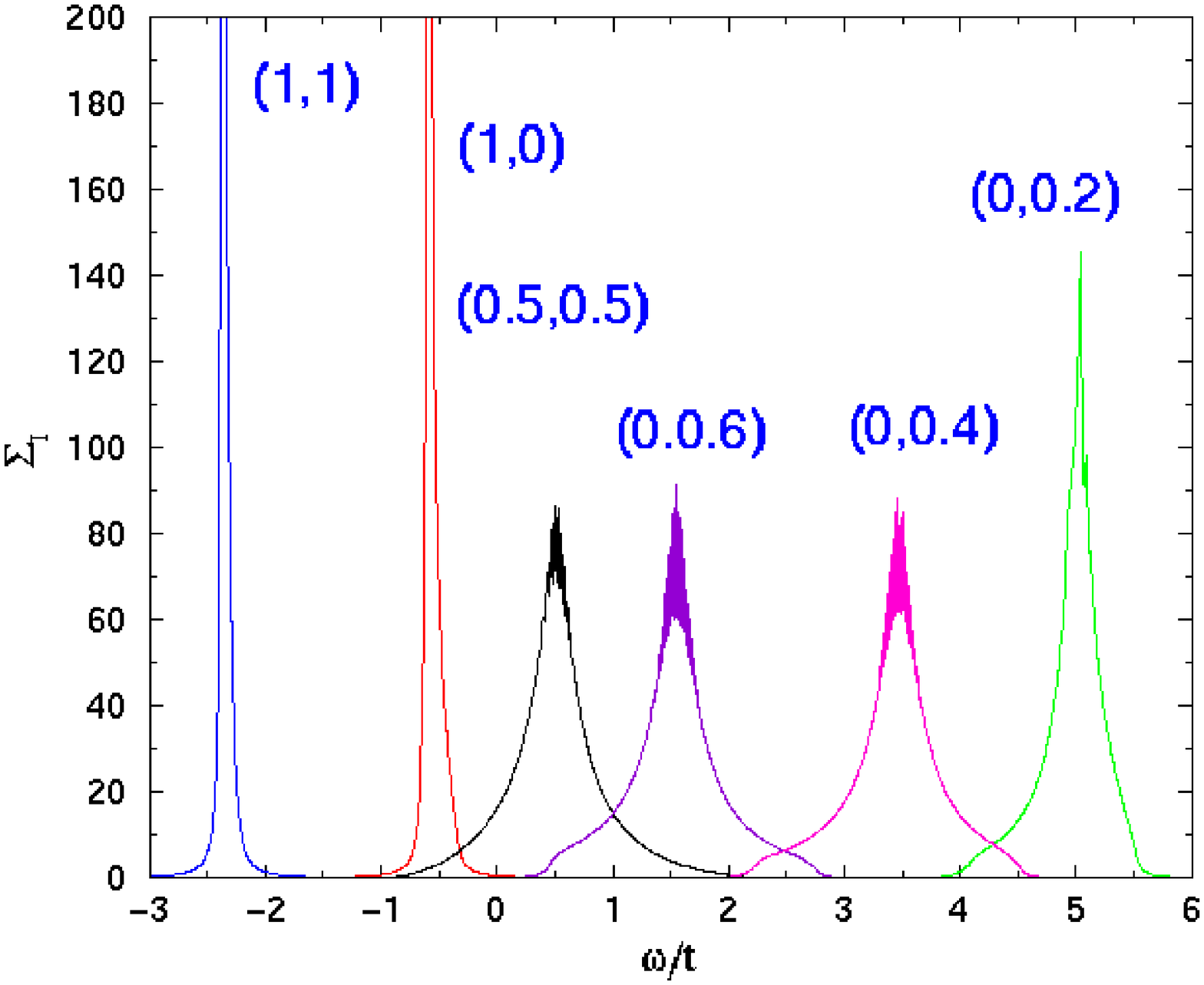}
\vskip0.5cm
\caption{Imaginary part of the self energy, Eq.~\protect\ref{eq:18}, assuming
$1/C=0.05t$, $\delta =0.002$, $\alpha_\omega =1$, $T=100K$.  The branches are 
labelled $(k_x,k_y)$, in units of $\pi$.}
\label{fig:6}
\end{figure}

Given the susceptibility, Eq.~\ref{eq:15}, the self energy can be calculated as
\begin{eqnarray}
\Sigma (\vec k,i\omega_n)={g^2\chi_0\over\beta V}\sum_{\vec q, i\omega_m}G_0(
\vec k+\vec q,i\omega_n+i\omega_m)D_0(\vec q,i\omega_m)
\nonumber \\
={g^2\chi_0\over V}\sum_{\vec q}\int_{-\alpha_{\omega}/C}^{\alpha_{\omega}/C}
{d\epsilon\over\pi}{n(\epsilon )+f(\xi_{\vec k+\vec q})\over i\omega_n+
\epsilon -\xi_{\vec k+\vec q}}{C\epsilon\over (\delta +Aq^{'2})^2+(C\epsilon 
)^2},
\label{eq:17}
\end{eqnarray}
with bare Green's function $G_0(\vec k,i\omega_n)=1/(i\omega_n-\xi_{\vec k})$ 
and magnetic propagator $D_0$, Eq.~\ref{eq:B25}; for the form of the 
integral, see the discussion near Eq.~\ref{eq:B32}.  In addition, $\chi_0= 
\chi_0(\vec Q,0)$, $\vec q=\vec Q+\vec q'$, $n$ is the Bose function, and
\begin{equation}
g^2\chi_0=U^2\chi_0(U\chi_0(\vec Q,i\omega_n)+{1\over 1+U\chi_0(\vec Q,i\omega_n
)})\simeq{3U\over 2}
\label{eq:17b}
\end{equation}
(Ref.~\onlinecite{BSW2}). The last form is an approximation based on the 
empirical substitution $\chi_0\rightarrow\simeq 1/U$ in the pseudogap 
regime.  After analytical 
continuation, the imaginary part of the retarded self energy is
\begin{eqnarray}
Im\Sigma^R(\vec k,\omega )
={-g^2\chi_0\over V}\sum_{\vec q}\int_{-\alpha_{\omega}/C}^{\alpha_{\omega}/C}
d\epsilon [n(\epsilon )+f(\xi_{\vec k+\vec q})]\times
\nonumber \\
\times\delta(\omega +\epsilon -\xi_{\vec k+\vec q}){C\epsilon 
\over (\delta +Aq^{'2})^2+(C\epsilon )^2}.
\label{eq:18}
\end{eqnarray}
The resulting self energy is plotted in Fig.~\ref{fig:6} for $T=100K$.  
(The weak oscillations seen in some branches of $\Sigma_I$ are an artifact due
to an insufficient density of points in the numerical integration.)  Note
that $Im\Sigma$ has the form of a broadened $\delta$-function peaked at $\omega
=\xi_{\vec k+\vec Q}$.  If it were a $\delta$-function, $Im\Sigma =-\pi\bar
\Delta^2\delta (\omega -\xi_{\vec k+\vec Q})$, then
\begin{equation}
Re\Sigma^R(\vec k,\omega )
={1\over\pi}\int_{-\infty}^{\infty}d\epsilon {Im\Sigma^R(\vec k,\epsilon )\over
\epsilon -\omega }={\bar\Delta^2\over\omega -\xi_{\vec k+\vec Q}},
\label{eq:19}
\end{equation}
so away from the $\delta$-function
\begin{equation}                           
G(\vec k,\omega )={1\over \omega -\xi_{\vec k}-Re\Sigma^R(\vec k,\omega )}=
{\omega -\xi_{\vec k+\vec Q}\over (\omega -\xi_{\vec k})(\omega -\xi_{\vec k+
\vec Q})-\bar\Delta^2}.
\label{eq:20}
\end{equation}
This is exactly the Green's function of the mean field 
calculation\cite{SWZ,MK3}, with the substitution
$\Delta\rightarrow\bar\Delta$, where $\bar\Delta$ can be evaluated by integrating 
\begin{eqnarray}
\bar\Delta^2=-{1\over\pi}\int_{-\infty}^{\infty}d\omega Im\Sigma^R(\vec k,\omega
 )
\nonumber \\
%={2g^2\chi_0a^2\rho_s\over \pi A}.
={U\over 8u}(\delta -\delta_0),
\label{eq:20a}
\end{eqnarray}
Fig.~\ref{fig:11a1}d.  This result is due to the Bose term $n(\epsilon )$
in the square bracket of Eq.~\ref{eq:18}, the Fermi function $f$ making no
contribution.  This leads to $\bar\Delta$ being independent of $\vec k$.

Equations~\ref{eq:20},\ref{eq:20a} constitute an important result: the 
connection between the Mott gap and short-range magnetic order.  Recalling
that $\Delta = U<M_i>$, or $\Delta^2=U^2<S_i>^2$, where $<M_i>=(-1)^i<S_i>$ 
is the staggered magnetization, then, in the spirit of an alloy analogy,
a {\it short-range order parameter} can be defined as
\begin{eqnarray}
\bar\Delta_{SR}^2(i\omega )={-g^2\over 4\beta}\int_0^{\beta}\sum_{<i,j>}<S_{i+}
(\tau )S_{j-}(0)>e^{i\omega\tau}d\tau
\nonumber \\
={-g^2\over 4\beta}\sum_k(c_x+c_y)\chi_{+-}(k,i\omega )
\simeq {g^2\over 2\beta}\sum_k\chi_{+-}(k,0)
\label{eq:20b}
\end{eqnarray}
which is equivalent to Eq.~\ref{eq:20a}.  (In the last equality in 
Eq.~\ref{eq:20b} the limit $i\omega\rightarrow 0$ is an adiabatic 
approximation\cite{Mor}, while the approximation is made that $\chi$ peaks near
$\vec Q$.)  Thus, {\it as long as there is short-range magnetic order ($\bar
\Delta$ or $\rho_s$ non-zero), there will be a Mott (pseudo)gap.}  
Equation~\ref{eq:20b} was also derived by Schmalian, et 
al.\cite{SPS}, but they did not discuss its significance.
%Finally, Eq.~\ref{eq:20a} can be rewritten by approximating $\rho_s$ 
%(Eq.~\ref{eq:B40a}) by $\rho^a_s$ (Eq.~\ref{eq:B40}):
%\begin{equation}
%\bar\Delta^2={U\over 8u}(\delta -\delta_0),
%\label{eq:20c}
%\end{equation}
%where $\bar\delta_0$ is the quantum corrected Stoner factor, {\it which controls
%the $T=0$ QCP}.  That is, the onset of short range order is determined by the
%{\it same} Stoner criterion that controls the T=0 termination of the AFM
%ordered phase.  This confirms the claim made in Ref.~\onlinecite{KLBM} that the
%observed Mott gap collapse constitutes a QCP.

\subsection{Application to the Cuprates}

\begin{figure}
\leavevmode
   \epsfxsize=0.33\textwidth\epsfbox{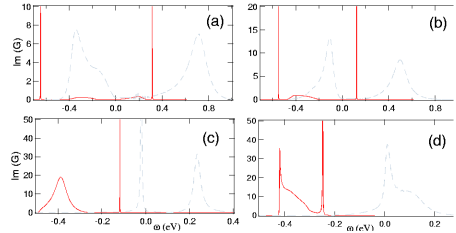}
\vskip0.5cm
\caption{Spectral functions for (a) $x=0$, (b) $x=-0.04$, (c) $x=-0.10$, and (d)
$x=-0.15$, at $T=100K$.  Solid lines at $(\pi ,0)$, 
%short dashed lines at $(0.6\pi ,0)$, 
and long dashed lines at $(\pi /2,\pi /2)$.}
\label{fig:7a1}
\end{figure}

\begin{figure}
\leavevmode
   \epsfxsize=0.33\textwidth\epsfbox{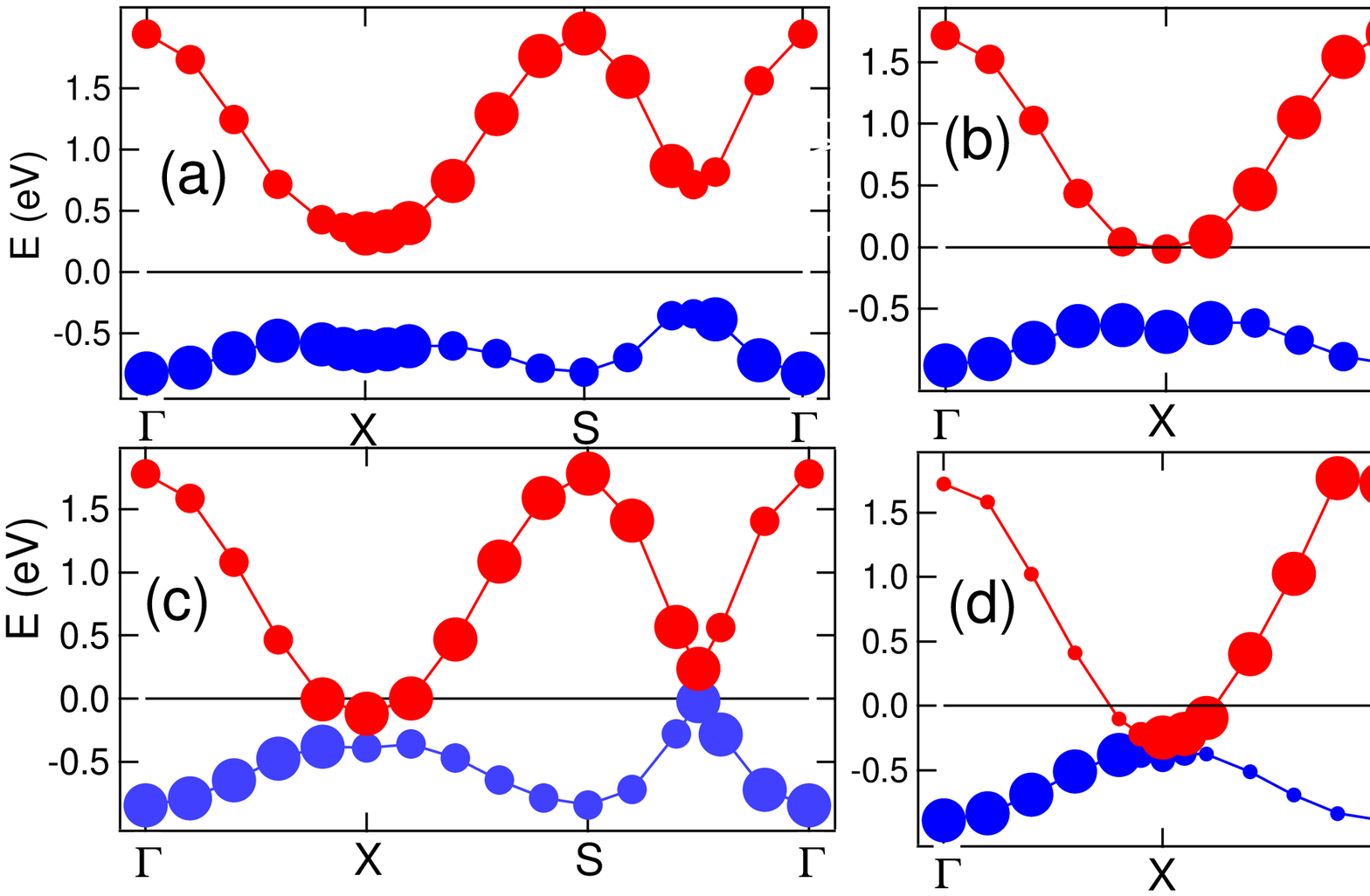}
\vskip0.5cm
\caption{Dispersion relations for electron doped materials, calculated at $T=100
K$: (a) $x=0$ ($U/t=6$), (b) $x=-0.04$ ($U/t=5$), (c) $x=-0.10$ ($U/t=3.5$), and
(d) $x=-0.15$ ($U/t=2.9$).  Weaker features are denoted by smaller circles; for
$x=-0.15$ all shadow features are extremely weak.}
\label{fig:7e}
\end{figure}

\begin{figure}
\leavevmode
   \epsfxsize=0.33\textwidth\epsfbox{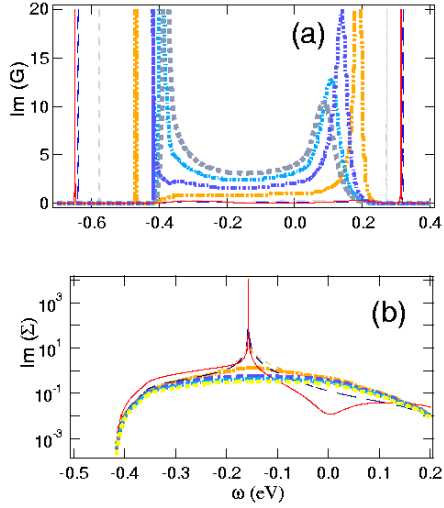}
\vskip0.5cm
\caption{Temperature dependence of (a) spectral function and (b) imaginary part
of self energy, for $x=0.0$ at $(\pi ,0)$.  Temperatures are 100, 500, 1000,
2000, 3000, 4000, and 5000K.}
\label{fig:7ee}
\end{figure}

\begin{figure}
\leavevmode
   \epsfxsize=0.33\textwidth\epsfbox{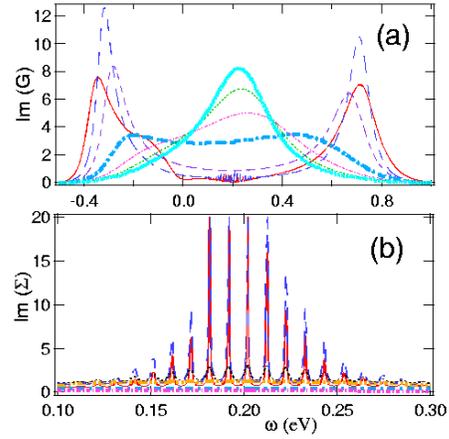}
\vskip0.5cm
\caption{Temperature dependence of (a) spectral function and (b) imaginary part
of self energy, for $x=0.0$ at $(\pi /2,\pi /2)$.  Temperatures are 100, 500, 
1000, 2000, 3000, 4000, and 5000K.}
\label{fig:7ef}
\end{figure}

\begin{figure}
\leavevmode
   \epsfxsize=0.33\textwidth\epsfbox{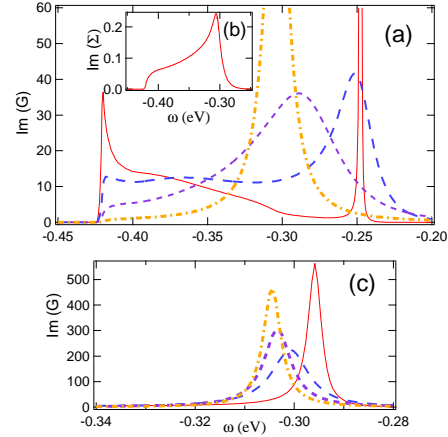}
\vskip0.5cm
\caption{Temperature dependence of spectral function 
for $x=-0.15$ at $(\pi ,0)$, for $U/t$ = 2.9 (a) and 2.5 (c).  Temperatures are 
100 (solid line), 500 (long-dashed line), 1000 (short-dashed line), and 2000K
dot-dashed line).  (b): imaginary part of self energy at $T$ = 100K,
$U/t$ = 2.9.}
\label{fig:7eg}
\end{figure}

Using the correct $Im\Sigma^R$ from Eq.~\ref{eq:18}, and the calculated 
parameter values from Table I, ARPES spectra are calculated for electron-doped 
cuprates, at the four dopings for which detailed data are 
available\cite{nparm}.  Figure~\ref{fig:7a1} shows typical calculated
spectra for several $\vec k$-points in the a-b plane.  Broadened Hubbard bands 
are found, which gradually smear out at high temperatures as $\delta$ increases 
($\xi$ decreases). There is a well defined pseudogap, with two peaks in the 
spectral function at a given $\vec k$.  It should be stressed that since
there is no interlayer coupling, long range antiferromagnetic order exists
only at $T=0K$.  The resulting dispersions are shown in 
Fig.~\ref{fig:7e}.  Figures~\ref{fig:7ee}-~\ref{fig:7eg} illustrate the 
temperature dependence of $Im(G)$ and $Im(\Sigma)$, for two dopings, $x=0$ and 
-0.15.  The broadening of the peaks can be understood from 
Eq.~\ref{eq:18}: particle-hole excitations are present within a range
$\pm\alpha_{\omega}/C$ of $\xi_{\vec k+\vec q}$.  Away from this
particle-hole continuum the main peaks are sharp, while they broaden when
they enter the continuum.  


Note that the Mott gap collapse is anisotropic: for the undoped case, the nodal 
gap collapses between 2-3000K, while a gap persists near $(\pi ,0)$ above
5000K.  $Im(\Sigma )$ has striking oscillatory structure, particularly
near $(\pi /2,\pi /2)$, which produces a similar weak structure in $Im(G)$ at 
low T.  [Similar, weaker oscillations are present near $(\pi ,0)$, which can be
better seen in Fig. 4c of Ref.~\onlinecite{ICTP}.]  In addition, there is a very
intense, strongly T-dependent peak in $Im(\Sigma )$ exactly at $\xi_{\vec k+
\vec q}$ (Fig.~\ref{fig:7ee}b -- also present but not shown in 
Fig.~\ref{fig:7ef}b).  It is the divergence of this peak as $T\rightarrow 0$
which signals the AFM transition.  At low temperatures, the peak positions in 
$Im(G)$ have a temperature dependence consistent with the collapse of the
Mott gap -- e.g., the LHB shifts to higher energies (toward midgap) at
higher temperatures.  Some experiments on hole doped cuprates find the
{\it opposite} dependence\cite{KiRo}, which can possibly be understood as
a localization or phase separation effect.

\begin{figure}
\leavevmode
   \epsfxsize=0.33\textwidth\epsfbox{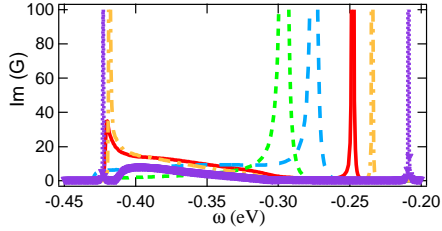}
\vskip0.5cm
\caption{$U$-dependence of spectral functions for $x=-0.15$ at $T=100K$ near the
$T=0$ QCP, for $U/t$ = 2.5 (short dashed line), 2.7 (long dashed line), 2.9
(solid line), 3.0 (dot-dashed line), and 3.2 (dotted line).}
\label{fig:7db}
\end{figure}

In contrast, for $x=-0.15$, Fig.~\ref{fig:7ef}, the splittings are absent near 
$(\pi /2,\pi /2)$, and vanish near $(\pi ,0)$ by $\sim$500K, and the lines 
actually {\it sharpen} on warming.  If the effective $U$ is reduced to $2.5t$, 
no splitting is found, but the peak position and broadening have an anomalous 
T dependence.  Clearly, the system is very close to a QCP.  Figure~\ref{fig:7db}
shows in more detail how the spectrum evolves with $U$ near this point.

\begin{figure}
\leavevmode
   \epsfxsize=0.33\textwidth\epsfbox{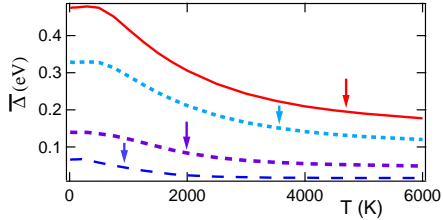}
\vskip0.5cm
\caption{Temperature dependence of gap $\bar\Delta$ for (from highest 
to lowest) $x=0$, -0.04, -0.10, and -0.15.  Arrows show mean field
transition temperature $T_N$.}
\label{fig:7da}
\end{figure}
Thus, the SCR calculation agrees with the mean-field results\cite{KLBM}, if
the mean-field gaps and transition temperatures are interpreted as the opening
of a pseudogap at finite T, with the long-range AFM appearing only at T=0.  A
direct comparison of the transition temperatures is presented in 
Fig.~\ref{fig:11a1}d, presented on a linear T scale in Fig.~\ref{fig:7da}.
Moreover, the overall dispersions, Fig.~\ref{fig:7e} are in quite good
agreement with the mean field results\cite{KLBM} and experiments\cite{nparm}.
%For the doped materials, agreement with the present SCR results is quite 
%satisfactory, but the model predicts too small a gap at half filling. 
%In part, the disagreement may arise because 
This agreement is somewhat surprising, since the model is not fully 
self-consistent.  For instance, the parameter $C$ involves Landau damping of 
the spin waves by electron-hole excitations, and hence depends on the 
electronic dispersion near the Fermi level.  Thus, the opening of the pseudogap
should have a strong influence on $C$, which is not accounted for.  
%Given the
%general good agreement with experiment, it should be possible to improve the 
%dispersion with relatively modest changes in the parameters; for instance, 
%in Fig.~\ref{fig:7f}, only the parameter $u$ is changed, but the observed 
%gap is reproduced.
%\begin{figure}
%\leavevmode
%   \epsfxsize=0.33\textwidth\epsfbox{arpee314n.eps}
%\vskip0.5cm
%\caption{Modified dispersion relation for the half filled case, at $T=100K$, 
%assuming same parameters as Fig.~\protect\ref{fig:7e}a, except
%$u^{-1}=4.47eV$.}
%\label{fig:7f}
%\end{figure}

\begin{figure}
\leavevmode
   \epsfxsize=0.33\textwidth\epsfbox{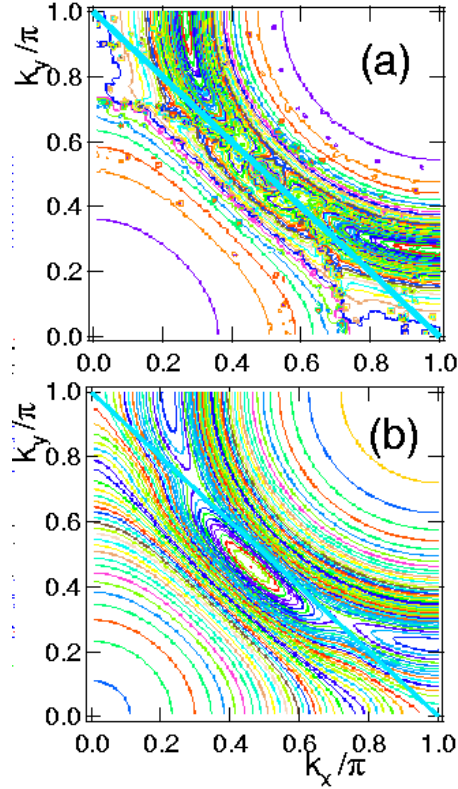}
\vskip0.5cm
\caption{Fermi surface map for $x=-0.04$ (a) and $-0.10$ (b).}
\label{fig:7g}
\end{figure}

Finally, Fig.~\ref{fig:7g} displays Fermi surface map for $x=-0.04$ and 
$-0.10$. The pseudogap along the zone diagonal associated with the hot-spot
scattering\cite{HlR} is clearly seen.  These should be compared with the
mean-field\cite{KLBM} and experimental\cite{nparm} results.  One
interesting aside: in the mean field calculation, with sharply defined
bands, it was necessary to include a $t''$ parameter to reproduce the
experimental hole pocket near the zone diagonal.  In the SCR calculation
the spectral function peaks are considerably broader, and no $t''$
parameter is needed.
%\begin{figure}
%\leavevmode
%   \epsfxsize=0.33\textwidth\epsfbox{arpee43.eps}
%\vskip0.5cm
%\caption{Constant energy maps for $x=-0.04$, $E$ = -0.167 (a), -0.108
%(b), -0.05 (c), and 0 eV (d).}
%\label{fig:7g2}
%\end{figure}

\section{Three Dimensional N\'eel Order}

In the physical cuprates, the interlayer hopping has an anomalous dispersion,
generally written as $t_z=t_{z0}(c_x-c_y)^2$.  This formula holds for
bilayer splitting, and in general when the CuO$_2$ planes are stacked {\it
uniformly}.  However, as explained in Appendix E, many of the cuprates,
including NCCO, have a {\it staggered layering}, with the Cu in one
CuO$_2$ plane laying above a vacancy in the neighboring CuO$_2$ sheet.
This leads to a {\it magnetic frustration}: the Cu in one sheet has four
nearest neighbors in the adjacent sheet, two with spin up, two with spin
down.  This frustration is reflected in a more complicated dispersion of $t_z$:
\begin{equation}
t_z=t_{z0}(c_x-c_y)^2\cos{k_xa\over 2}\cos{k_ya\over 2},
\label{eq:n21}
\end{equation}
which vanishes at $(\pi ,0)$ and $(0,\pi )$, and leads to a greatly
reduced interlayer coupling. (Effects of AFM frustration associated 
with layering have been discussed in Ref.~\onlinecite{PiL}.)

The consequences of both uniform and staggered stacking are explored in
Appendix E.  If the c-axis resistivity is coherent, it can be used to
estimate the interlayer hopping $t_{z0}$.  It is found that the value of
$t_{z0}$ needed to produce a given resistivity anisotropy is approximately
5 times smaller for uniform stacking, to account for the frustration in
the staggered stacking.  With the corresponding $t_{z0}$'s determined from
resistivity, both forms of interlayer coupling give rise to comparable
interlayer coupling, and hence a finite N\'eel temperature.  While the
optimal $Q$-vector depends on doping, at half filling both forms predict
$\vec Q=(\pi ,\pi ,0)$, consistent with experiment in La$_2$CuO$_4$.
Even for quite strong anisotropy, this mechanism can account for the
observed $T_N$s (in fact, tends to overestimate $T_N$), without the
necessity of invoking additional mechanisms, such as a
Kosterlitz-Thouless transition, with the reduced spin dimensionality
caused by spin-orbit coupling effects\cite{Ding,SinT,KAASh,KK}.

Within mode coupling theory\cite{STeW} (Appendix E), the N\'eel
temperature is found from the gap equation (Eqs.~\ref{eq:D1},~\ref{eq:B34e})
\begin{equation}
\chi_0(T)U=\eta +{3uTa^2\ln{({T\over T_{3D}})}\over\pi A},
\label{eq:21}
\end{equation}
where $T_{3D}\sim t_z^2$ is defined below Eq.~\ref{eq:B34i}.  It is found
that $T_{3D}$ is approximately constant, independent of doping in the
electron-doped regime.  Apart from a small numerical factor, 
Eq.~\ref{eq:21} differs from the isotropic three-dimensional result by
the logarithmic factor, which diverges ($T_N\rightarrow 0$) as
$t_z\rightarrow 0$.  

Equation~\ref{eq:21} can be rewritten in a suggestive form.  Approximating 
$\rho_s$ by $\rho_s^a=A(\chi_0U-\eta )/12ua^2$ (Eq.~\ref{eq:B40}), then, 
using Eq.~\ref{eq:16}, the N\'eel transition occurs when
\begin{equation}
J_z[{\xi(T_N)\over\xi_0(T_N)}]^2=\Gamma T_N
\label{eq:22a}
\end{equation}
where $J_z=J(t_{z0}/t)^2$, $J=4t^2/U$, and $\Gamma=4t_{z0}^2/UT_{3D}$.  A
very similar form was proposed earlier\cite{BiGGS}.

\begin{figure}
\leavevmode   
   \epsfxsize=0.33\textwidth\epsfbox{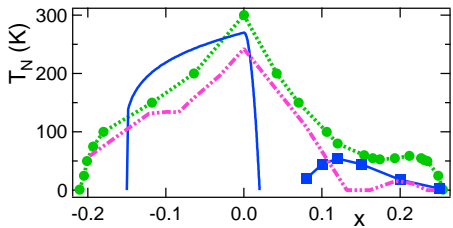}
\vskip0.5cm
\caption{Comparison of experimental N\'eel temperatures for NCCO and LSCO,
(solid line), and for the stripe (magnetic) ordering transitions observed
in Nd-substituted LSCO [\protect\onlinecite{Ich}] (solid line with
squares) with the model of interlayer coupling with staggered stacking and
$t_{z0}=t/10\sim 30meV$, plotted as $T_N/10$ (dot-dot-dash line). Also
included is the approximate expression, Eq.~\protect\ref{eq:22b} (dotted
line with circles).  (Note that there is a range of hole doping for which
$A$ is found to be negative; in this range $T_N$ was arbitrarily assumed
to vanish in the staggered model, $T_N=0$.) } 
\label{fig:nD7} 
\end{figure}

Figure~\ref{fig:nD7} compares the calculated value of $T_N$ with the
experimental values.  
%Since stripes complicate the physics of the hole
%doped regime, the figure also includes the magnetic ordering temperature
%of quasi-static stripe arrays, from Nd-substituted LSCO\cite{Ich}.  
While
the overall doping dependence is comparable, the calculated $T_N$ is about
an order of magnitude higher.  The calculation is for staggered stacking,
with $t_z$ adjusted to reproduce the observed resistivity anisotropy, but
Appendix E shows that the overestimate is generic: the coefficient of the
logarithm needs to be larger to reduce $T_N$.  Also shown in
Fig.~\ref{fig:nD7} (dotted line) is a simplified model, which assumes that
\begin{equation}
T_0^*={\pi A\over 3ua^2\ln ({T\over T_{3D}})}
\label{eq:22b}
\end{equation}
is doping independent, $T_0^*=1200K$.  This model reproduces qualitatively
the shape of the numerical calculation, but with a magnitude comparable to
experiment.  The magnitude of $T_N$ could be matched almost quantitatively
if $U_{eff}$ also has a significant temperature dependence, as discussed
in Appendix E . The overall doping dependence is also comparable
to experiment.  The agreement could be further improved by using a smaller
value of $t'$, which would shrink the doping range over which N\'eel order
occurs.  

\section{Discussion}

\subsection{Magnon Bose Condensation}

\begin{figure}
\leavevmode   
   \epsfxsize=0.33\textwidth\epsfbox{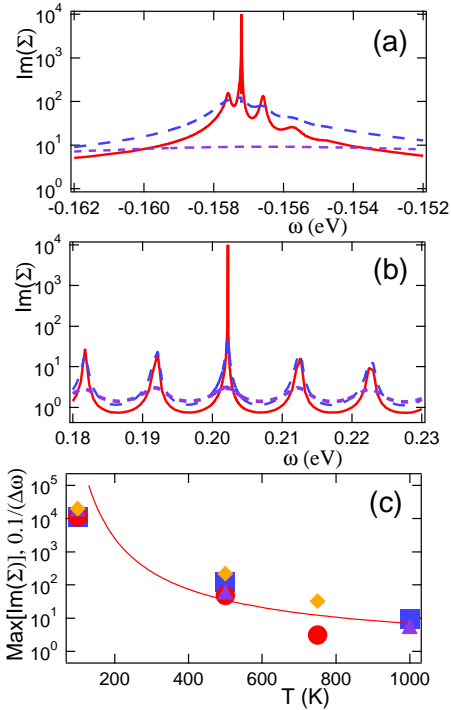}
\vskip0.5cm
\caption{(a,b): Blowups of $Im(\Sigma )$ for $x=0$ at $(\pi ,0)$ (a) and
$(\pi /2,\pi /2)$ (b) at $T$ = 100K (solid lines), 500K (long dashed
lines), and 1000K (a) or 750K (b) (short dashed lines). (c) Maximum of
$Im(\Sigma )$ vs $T$ for $(\pi ,0)$ (squares) and $(\pi /2,\pi /2)$
(circles); 0.1/(full width at half maximum) for $(\pi ,0)$ (triangles) and
$(\pi /2,\pi /2)$ (diamonds); solid line = corresponding $\xi (T)$,
Eq.~\protect\ref{eq:16}.} 
\label{fig:nD9} 
\end{figure}

Figure~\ref{fig:nD9} shows the sharp peak which arises in $Im \Sigma$ at
low T.  The growth is exponential, approximately matching that of the
coherence length, Eq.~\ref{eq:16}.  (Note that it requires a fine mesh in
the integral of Eq.~\ref{eq:18} to capture this growth.)  This peak arises
exactly at the incipient magnetic zone boundary, and turns into true Bragg
scattering at the transition to long range order: the increase in peak
height is almost exactly compensated by a decrease in the width of the
peak.  A simple physical explanation is that the SDW transition can be
interpreted as a {\it Bose condensation of the zone boundary magnons}.
Then the Mermin-Wagner theorem reduces to the fact that in a
two-dimensional system, Bose particles can only condense at $T=0$. 
A similar explanation for the transition has been presented
earlier\cite{VAT}.

\subsection{SCR Calculation of NAFL Parameters}

The SCR is perhaps the simplest model in which fluctuation effects are 
included to satisfy the Mermin-Wagner theorem, allowing one to ask questions
such as, how does a Mott gap appear if there is no long-range order to 
generate a smaller Brillouin zone?  It can be seen that the Mott gap is
really a pseudogap, and is relatively insensitive to the appearence of
long-range N\'eel order.  However, it must be kept in mind that the
original SCR\cite{Mor} is intended to describe {\it weak} itinerant
ferromagnets and AFM's, and will have to be extended to account for the
strong renormalization of the electronic bands.

Given these limitations, the present paper attempts to use the SCR to calculate
the parameters, $A$, $C$, etc., for an NAFL-like model of the low energy 
physics.  The principal findings include:

$\bullet$ the program {\it fails} for hole doped cuprates, since the parameter
$A$ is negative, signalling an instability of the commensurate AFM phase 
against either incommensurate SDW phases or nanoscale phase separation.

$\bullet$ For electron doped cuprates, considerable progress can be made, 
but the model still has problems, in particular a poorly defined spin-wave
interaction parameter $u$.  Here $u$ is estimated by comparison with
sigma model results near half filling, but its possible doping dependence 
is unknown.

$\bullet$  The present theory differs from conventional NAFL theory
by the inclusion of two cutoff parameters, $q_c$ and $\omega_c^-$.  These
cutoffs shrink to zero at either the H- or C-points, and in particular
cause the $A$ parameter to have a strong temperature dependence in the
electron-doping regime.  

$\bullet$ The combined SCR and mean-field\cite{KLBM}
results are in excellent agreement with the ARPES data\cite{nparm} on the
doping dependence of the Mott transition in the electron-doped cuprates.

$\bullet$ While the collapse of the Mott gap and the termination of the hot spot
regime are in principle independent QCPs, in practice they fall very close to 
each other.  The present calculations and Ref.~\onlinecite{MK4} explain
why this is so.

These points are discussed in more detail in the following subsections.

\subsection{Mott Transition vs Slater Antiferromagnetism}

In reviewing the history of magnetism, Anderson recently pointed
out\cite{PWA} that the theories fall into two diametrically opposed
classes: band theory and atomic models, and the latter are typically more
successful.  In particular, for strongly correlated Mott insulators Mott's
model of nearly localized spins seems like a better starting point than
Slater's theory of spin density wave antiferromagnetism.  The main problem
is that the Slater theory ties the Mott gap to antiferromagnetic order,
predicting too high a N\'eel temperature.

While a localized picture may be a more convenient starting point near
half filling, nevertheless it should be possible to develop a (perhaps
more complicated) picture based on band theory.  There is a general
desideratum to be able to extend band structure calculations to all
materials; moreover, this is important in the present instance because the
doping dependence of $U$ suggests that there is a crossover from strong to
intermediate correlations, which may be better handled by working
throughout in a band structure formalism.  The present mode coupling
calculation seems to be an appropriate starting point.  The Mermin-Wagner
theorem leads to a decoupling of the Mott gap and N\'eel order.  How this
might be extended to three-dimensional Mott insulators remains unclear.

Local physics should show up in a band structure calculation as a very
narrow band width.  In a mean field SDW calculation, the width of the
Hubbard bands, $\sim t^2/U\sim J$, is not small.

\subsection{Stoner Criterion and Crossover from Small to Large FS}

In the Hartree-Fock model, the crossover from small to large Fermi surface 
is {\it coincident} with Mott gap collapse, and comparison with experiment
suggests that this correctly describes the situation in electron doped NCCO.
The SCR theory confirms this result and offers additional insights.
Fluctuations preclude long-range order at finite temperatures (in the
absence of interlayer coupling), so the physics is that of a {\it
zero temperature} AFM {\it QCP}, controlled by a Stoner criterion: the QCP
occurs when $\chi U$ is too small.  At finite temperatures, the Mott gap
is replaced by a pseudogap, and the FS crossover still coincides with the
collapse of the Mott pseudogap, which occurs at the  QCP.  Whether a
similar crossover can occur in the absence of Mott gap collapse remains to
be seen.

\subsection{Hole Doped Cuprates}

The present calculations make two predictions for hole-doped cuprates: (1)
there are strong indications for instability against phase separation and
stripe physics; but (2) despite this, the termination of strong magnetic
fluctuations should be approximately electron-hole symmetric\cite{MK4} -- 
associated with the symmetric susceptibility plateau in
Fig.~\ref{fig:0a}a.

\subsubsection{Stripes}

The situation in hole doped cuprates is complicated by the presence of
stripes.  At all levels, mean-field, RPA, SCR, striking differences
between hole and electron doping are clearly revealed.  All techniques
provide strong evidence for the instability of the AFM state for hole
doping, while it remains stable under electron doping.  In earlier
Hartree-Fock and RPA calculations, evidence for the instability of the
hole doped phase was found {\it in the ordered magnetic phase}, within the
smaller magnetic Brillouin zone: e.g., the spin wave dispersion is
unstable.  In the SCR approach, there is no phase transition at finite
temperatures, but {\it even in the paramagnetic phase} there is evidence
for the instability.  A detailed analysis of the real part of the bare 
susceptibility (as appropriate for a Stoner criterion) provides evidence for 
instability of a commensurate magnetic phase at $\vec Q$ (negative curvature 
$A$).

\subsubsection{Pseudogap}

In hole doped cuprates, ARPES finds two features which are commonly
referred to as pseudogaps -- a `hump' feature found near $(\pi ,0)$ at
higher binding energy than the main, superconducting `peak', and the
`leading edge gap', a loss of spectral weight in the immediate vicinity of
the Fermi level.  This latter feature is not explained by the present
calculation; it may be associated with the onset of strong
superconducting fluctuations\cite{SPS,pg2}.  Alternatively, such a gap
has been found in a dynamical cluster expansion calculation of the Hubbard
model\cite{MNess}.

On the other hand, the `hump' feature can be consistently interpreted as
the collapse of the Mott pseudogap\cite{SPS,MK4}.  While ARPES only sees
the feature below the Fermi level associated with the lower Hubbard band
(LHB), tunneling\cite{pg2,Miya,Kr} finds an approximately symmetrical peak
feature above the Fermi level, associated with the UHB -- as if the
pseudogap were pinned to the Fermi level.  This has led to a number of
alternative models for the pseudogap, in terms of superconducting or
charge density wave (CDW) fluctuations (the latter possibly related to
stripe physics).  Recently\cite{MK4} it was found that the same mean field
model of Mott gap collapse can approximately explain the data (see
Fig.~\ref{fig:0d}b). In this model, the lower pseudogap peak is the VHS
of the LHB, while the upper peak is due to the leading edge of the UHB.
As the Mott gap collapses, the two features merge.  
A careful tunneling study of the `hump' features could look for the
predicted asymmetry of the features about the Fermi level.  Such a study
should best be done in single layer Bi$_2$Sr$_2$CuO$_6$,\cite{AYYA} where
complications due to bilayer splitting are absent. 

The above interpretation requires that for hole doping also the Mott gap
must collapse slightly above optimal doping.  This is consistent with
recent experimental observations of a QCP\cite{Tal1}.  The early SCR
results of Table II point in the same direction: the very small
and positive values found for the Stoner parameter $y_0$ in optimally
and overdoped cuprates suggests the proximity to a QCP near optimal doping.
Moreover, the model predicts that at the QCP, where the pseudogap just
closes, the Fermi level is {\it exactly at the VHS} (H-point).
This result had been found experimentally in some lightly overdoped
cuprates\cite{Surv,Tal2}. 

While the doping of Mott gap collapse is approximately electron-hole
symmetric, some significant differences remain.  Thus, for electron doping
the magnetic correlation length remains large up to the QCP, while the 
correlation length is only a few lattice constants on the hole doped
side.  This may be related to competing order -- indeed, in Nd substituted
LSCO, long-range incommensurate magnetic order is found\cite{Tran} up to
$x=0.2$.  Alternatively, Schmalian, et al.\cite{SPS} were able to
reproduce a hump-like pseudogap with small correlation length, by a
careful summation of the full diagrammatic perturbation series.
[Note that near a VHS, all competing electron-hole instabilities -- SDW,
CDW, flux phase, shear (`Pomeranchuk') instability -- will lead to similar
pseudogaps near $(\pi ,0)$, and indeed in the presence of strong
fluctuations, all will contribute in a comparable fashion,
$\Delta^{*2}\sim\sum_i\Delta_i^2$.]

\subsection{VHS}

\subsubsection{Electron-Hole Asymmetry}

To study the role of the VHS in the Mott transition and high-$T_c$
superconductivity, one would ideally like to study a system in which one
could turn the VHS on or off.  Electron vs hole doping of the cuprates
would appear to approach this ideal.  In switching from electron to hole
doping (from NCCO to LSCO) $T_c$ increases by less than a factor of two,
and apparently remains d-wave, while the normal state properties change
drastically, with nanoscale phase separation on the hole doped side only.  

On the other hand, the one hole vs one electron systems should be much
more similar: in either case, nearest neighbor hopping is frustrated by
breaking local antiferromagnetic order.  Hence in both cases, the low
energy states will be magnetic polarons.  Indeed, the electron-doped
polarons may be more localized, since there is no interpolaron attraction 
(i.e., tendency to phase separation).  The ARPES spectra for low-electron
doping ($x$=-0.04) show an additional pseudogap at the Fermi level, which
may be related to localization.

\subsubsection{Temperature Dependent VHS}

As noted by Onufrieva and Pfeuty\cite{OPfeut}, the VHSs
associated with the susceptibilities (and hence with charge or spin nesting) are
{\it different} from those associated with the density of states (and
superconductivity).  Thus, whereas superconductivity will occur at the same
optimal doping for all temperatures, the doping of maximal nesting instability
{\it is a strong function of temperature}.

This contrasting behavior of nesting vs pairing {\it susceptibilities} is
related to a characteristic difference in the nature of the two {\it
instabilities}.  A superconducting instability has an intrinsic
electron-hole symmetry, which means that the gap is tied to the Fermi
level, and a full (s- or d- wave) gap can be opened at any doping level.
On the other hand, a nesting gap is dispersive, and only part of it lies
at the Fermi level (except in special cases).  Furthermore, a
(superlattice) Luttinger's theorem must be obeyed, requiring the presence
of residual Fermi surface pockets.  Stated differently, a full nesting gap 
can only open at integer filling, so {\it as the interaction strength
increases, any nesting instability must migrate to integral doping}
(e.g., half filling in the original band structure).  This same VHS
migration is mirrored in the T-dependence of the magnetic (or charge)
susceptibility.

\subsubsection{VHS Transitions}

We have seen that the doping-dependent $U_{eff}$ gives rise to a Mott gap
collapse near the edges of the susceptibility plateau in Fig.~\ref{fig:0a}.
If $U_{eff}$ is smaller (dot-dashed line: $U_{eff}$ reduced by 2/3), more
complicated behavior should arise.  Due to the peak in $\chi$ near the
H-point, there could be a reentrant transition, with one magnetic order
near half filling, and a second near the VHS.  For an even smaller
$U_{eff}$ (or replacing $U_{eff}\rightarrow J$)\cite{OPfeut}, the transition 
near $x=0$ can be eliminated, leaving a spin density wave transition near
the VHS.  In principle there could even be a phase separation {\it between
two AFM phases}: an insulating phase near half filling and a metallic
phase near the VHS.

\subsection{Future Directions}

It must be stressed, however, that the present theory is not fully 
self-contained.  There are three significant limitations.  First, an
improved calculation of the doping dependence of the Hubbard $U$ parameter is 
a desideratum, perhaps along the lines of earlier
calculations\cite{Esi,ChAT,BSW2}.  Since all of these calculations lead to
different doping dependences for $U(x)$, the actual doping dependence must
be regarded as an unresolved issue.  

Second, several unsuccessful attempts have been made to calculate the quartic 
interaction parameter $u$. A simple {\it ansatz} for $u(x)$ is introduced,
based on consistency with the $t-J$ model, which leads to good results. A
deeper understanding of why this works, and whether $u$ is doping
dependent, is desirable.

Finally, the susceptibility was calculated with the bare electronic bands,
but when the bands are renormalized to first order in $\chi$ a pseudogap opens.
This gap should be self consistently incorporated into the calculation of
$\chi$, as in the FLEX and spin fermion approaches; it is expected to have a 
profound effect on the temperature dependence of the parameters (especially $A$)
and on the residual density of states in the pseudogap, particularly near $x=0$.

\section{Conclusions}

The main results of this rather long paper are briefly summarized:

\par\noindent $\bullet$
Fluctuation effects were added to the mean field Hubbard model via a
mode coupling calculation, which allowed satisfying of the Mermin-Wagner
theorem ($T_N=0$).  It was found that the mean-field gap $\Delta_{mf}$ and 
N\'eel temperature $T_N^{mf}$ evolved into a pseudogap $\Delta_{ps}\sim
\Delta_{mf}$ and an onset temperature $T^*\sim T_N^{mf}$ (as is familiar
from the related CDW results).  

\par\noindent $\bullet$
The resulting dispersions and Fermi surfaces are in excellent agreement
with photoemission experiments on electron-doped cuprates\cite{nparm},
while the pseudogap seems consistent with ARPES and tunneling results in
hole doped cuprates\cite{MK4}.  It is interesting to note that a recent
$t-t'-t''-J$ model calculation seems consistent with the first doped carriers
forming weakly interacting quasiparticles in pockets of the respective
upper or lower Hubbard bands, for either electron or hole doping\cite{LHN}.

\par\noindent $\bullet$
The zero-temperature N\'eel transition is controlled by a Stoner-like 
criterion, hence is sensitive to the bare susceptibility and in turn to
the Fermi surface geometry (hot spots).  This lead to an approximately
electron-hole symmetric QCP near optimal doping (termination of hot spot
regime), at which both zero temperature N\'eel transition and pseudogap
transition simultaneously terminate.

\par\noindent $\bullet$
The model leads to a NAFL-type susceptibility, and the calculation of the
NAFL parameters has been reduced to a calculation of the coupling
parameters $U$ and $u$, the former having a significant doping (and
possibly temperature) dependence. At present, $U(x)$ is estimated from
experiment, and the mode coupling $u$ via consistency with the $t-J$ model.
(A small portion of the renormalization of $U$ arises from quantum
corrections to the Stoner criterion.)

\par\noindent $\bullet$
Whereas the antiferromagnetic state at $\vec Q$ is stable to electron
doping, hole doping leads to an incommensurability, which is interpreted
as an indication of instability to phase separation (as found in the mean 
field calculations).  This asymmetry follows from the properties of the
VHS.

\par\noindent $\bullet$
Finally, a striking {\it temperature/frequency dependence} of the
susceptibility peak, from the VHS at low $T$ to half filling at high $T$,
found earlier\cite{OPfeut}, is interpreted in terms of Luttinger's
theorem: if the coupling is strong enough to open a full gap, the gap must
fall at half filling.

Acknowledgments:  
Part of this work was done while I was on sabbatical at the 
Instituto de Ciencia de Materiales de Madrid, CSIC, Cantoblanco, E-28049 
Madrid, Spain, supported by the Spanish Ministerio de Educaci\'on through
grant SAB2000-0034.  I thank my hosts, Maria Vozmediano and Paco Guinea,
for a very stimulating visit, for numerous discussions, and for correcting
an error in the original calculation.

I thank Walter Harrison for stimulating conversations on
calculating the interlayer coupling.

\appendix
\section{Three Band Model}

A major simplification of the present calculation is to treat the cuprates in a
one-band model.  This is consistent with the Zhang-Rice picture\cite{ZR}, 
although the approximation is less drastic for electron doping, since the upper
Hubbard band is already predominantly copper-like.  Nevertheless, the model also
describes the doping dependence of the `lower Hubbard band', which is really a
charge transfer, predominantly oxygen-like band.  Here an explanation for why 
this simplification works is suggested.  

\begin{figure}
\leavevmode
   \epsfxsize=0.33\textwidth\epsfbox{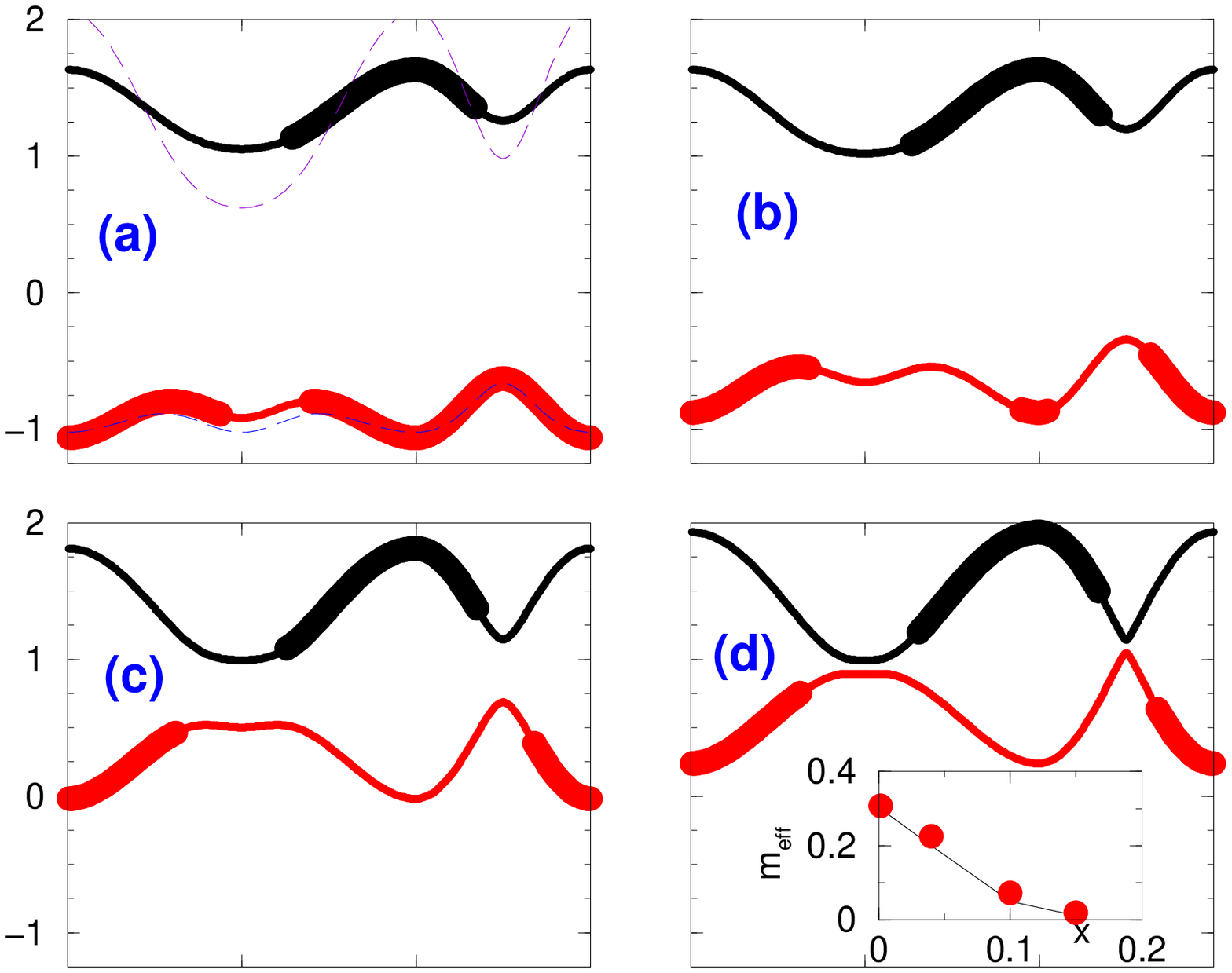}
\vskip0.5cm 
\caption{Dispersion of two antibonding bands in three-band model, assuming
$m_Q$ = 0.3 (a), 0.2 (b), 0.05 (c), and 0.01 (d). Inset to d: effective 
magnetization $m_{eff}=mU/6t$ for the three-band (solid line) and one-band 
(filled circles) models.  The one-band result has been multiplied by 3/4 to 
better agree with the three-band results.}
\label{fig:5}
\end{figure}

Even without carrying out self-consistent calculations, the nature of the
Mott transition can be understood by introducing a doping dependent gap.  
The energy bands can be calculated from the hamiltonian matrix
\begin{eqnarray}
H=\sum_j\Delta d^{\dagger}_jd_j
+\sum_{<i,j>}t_{CuO}[d^{\dagger}_jp_i+(c.c.)] \nonumber \\
+\sum_{<j,j'>}t_{OO}[p^{\dagger}_{j}p_{j'}+(c.c.)]
%\nonumber \\
%+Un_{j\uparrow}n_{j\downarrow}\bigr),
\label{eq:100}
\end{eqnarray}
where $\Delta$ is the difference in on-site energy between copper and oxygen,
$t_{CuO}$ is the copper-oxygen hopping parameter, and $t_{OO}$ the 
oxygen-oxygen hopping parameter.  For good agreement with the doping
dependence of the one band model, it is necessary to properly incorporate 
the Hartree correction to the self energy, $\Delta =\Delta_0+\Sigma_
H$, $\Sigma_H=Un_{\downarrow}$ (for up spins), and $n_{\downarrow}=n/2-m_Q$, 
with $n$ the average electron energy.  In Figs.~\ref{fig:5}, $\Delta_0=0$ is 
assumed. 

The band dispersion is extremely similar to that found in the one band model, 
Fig. 3 of Ref.\onlinecite{KLBM}, even though the lower band crosses over from 
the Zhang-Rice (hybridized copper-oxygen band) at half filling to a more copper 
like lower Hubbard band with increasing electron doping.  In addition, the 
effective magnetizations are proportional (inset, Fig.~\ref{fig:5}d), although 
the one-band model overestimates the magnetization by $~1/3$.  This can be 
understood: in the three-band model, the shape of the Hubbard bands is fixed by
the combined effects of the magnetic instability and hybridization with the
oxygen band.  In the one band model, only the former effect is present, 
necessitating a larger value of $m$ to produce the same net splitting.

This remarkable agreement between one and three-band models goes well beyond the
Zhang-Rice model.  That model is restricted to the LHB in a small range of
doping near half filling; the present results compare both LHB and UHB over the
full range of electron doping. The result is nontrivial -- in the three band 
model, the bonding and non-bonding bands are also split into upper and lower 
Hubbard bands.  This degree of agreement comes about because the parameter 
$\Delta$ is approximated by the two components (magnetic and nonmagnetic) of the
Hartree term.  In turn, this suggests that in the absence of magnetic effects 
the Cu and O energies are nearly degenerate -- as found in early LDA band 
structure calculations (see discussion in Ref.~\onlinecite{Hyb}).

\section{Charge Susceptibility and $U_{eff}$}

\begin{figure}
\leavevmode
   \epsfxsize=0.33\textwidth\epsfbox{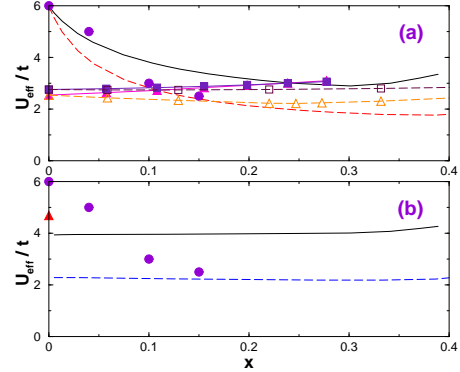}
\vskip0.5cm
\caption{Calculated $U_{eff}$ assuming (a) simple screening or (b) full vertex
correction of Chen, et al. [\protect\onlinecite{ChAT}].  In both cases, a bare 
$U=6.75t$ was assumed.  Solid lines = electron doping; long dashed lines = hole 
doping; triangles (squares) in a = paramagnetic screening of $U$, at $T$ =
1K (2000K); triangle in b = undoped; circles = data of 
Ref.~[\protect\onlinecite{nparm}].}
\label{fig:111}
\end{figure}

The proper choice of vertex corrections is an unresolved issue in the
analysis of the Hubbard model.  It is known to be of critical importance
for generating a pseudogap\cite{MVT}.  Here, by comparing simple
mean-field and SCR models to experiment, it is shown that the net effect
of vertex corrections is to make the coupling $U$ effectively doping (and
possibly temperature) dependent.  Kanamori\cite{Kana} 
showed that the effective Hubbard $U$ should decrease with doping, as an 
electron can hop around, and hence avoid, a second electron.  In the limit of a 
nearly empty (or full) band, this should lead to a correction of the form $U_
{eff}\sim U/(1+U/W)$, where $W=8t$ is the bandwidth.  It was 
found\cite{ChAT,BSW2} that Monte Carlo calculations of the susceptibility of a 
doped Mott insulator were approximately equal to the RPA susceptibility with 
suitable $U_{eff}$, and Chen, et al.\cite{ChAT} suggested the explicit form 
$U_{eff}=U/(1+<P>U)$, with P given by a vertex correction to the susceptibility 
and $<\cdot\cdot\cdot >$ an average over $\vec q$, at zero frequency.  
Figure~\ref{fig:111}b presents a calculation for $U_{eff}$ based on Chen, et al.
However, whereas Chen, et al. performed the average in the paramagnetic
phase, using bare Green's functions, here the dressed Green's functions
appropriate to the N\'eel phase are used, to approximately incorporate the
effect of this gap.  This makes little difference, since $P$ is dominated
by the intraband terms, and hence remains finite at half filling. Explicitly,
\begin{equation}
P=-{1\over N}\sum_{i,j,k}\hat U_{i,j}(k,k+q)\tilde F_{i,j}(k,k+q),
\label{eq:13a}
\end{equation}
\begin{equation}
\tilde F_{i,j}(k,k')={1-f_k^i-f_{k'}^j\over E_i(\vec k)+E_j(\vec k')-\omega-i
\delta},
\label{eq:13b}
\end{equation}
\begin{equation}
E_{\pm}(\vec k)={1\over 2}(\epsilon_k+\epsilon_{k+q}\pm E_0),
\label{eq:13c}
\end{equation}
\begin{equation}
E_0=\sqrt{(\epsilon_k-\epsilon_{k+q})^2+4\Delta^2},
\label{eq:13d}
\end{equation}
\begin{equation}
\hat U_{i,j}(k,k')={1\over 4}(1+iA_k)(1+jA_{k'})+ijB_kB_{k'},
\label{eq:13e}
\end{equation}
with $i,j$ summed over $+,-$, $\Delta$ the AFM gap, and
$A_k=(\epsilon_k-\epsilon_{k+Q})/E_{0k}$, $B_k=\Delta /E_{0k}$.
In agreement with Chen, et al., the calculation finds $U$ to be
renormalized by a factor of $~2$ at finite doping, but does not recover a 
large $U$ near half filling, although different results are found
depending on whether $x=0$ from the start (triangle) or whether $x\rightarrow 0$
from the hole or electron doping sides.  

For modelling purposes, it is useful to have a $U_{eff}$ which evolves
smoothly from a large value at half filling to a reduced, Kanemori value
at finite doping.  A simple toy model consists of taking the RPA
screening of a charge response.  There should be a close connection
between the Kanemori mechanism and screening.  Screening involves creation
of a correlation hole about a given charge, while Kanemori's $U_{eff}$
involves the ability of a second charge to move around the first, while
avoiding double occupancy.  Near half filling, the second charge must move
in the correlation hole.  Approximating\cite{KLBM} the vertex correction
by the RPA screening of the charge susceptibility,
\begin{equation}
U_{eff}={U\over 1+<\chi >U},
\label{eq:12c} 
\end{equation}
it is possible to reproduce\cite{KLBM} the experimentally observed\cite{nparm} 
doping dependence, while matching the calculation of Chen, et al. away
from half filling, Fig.~\ref{fig:111}a.  

In this calculation, issues of self-consistency are also important.  To
minimize screening at half filling, it is necessary to reproduce the gap
in the susceptibility.  Hence, the susceptibility in Eq.~\ref{eq:12c} is
approximated by the charge susceptibility in the AFM state, $\bar
\chi^{00}_0$ from Eq. 2.24 of Ref.~\onlinecite{SWZ}, evaluated with the
bare $U=6.75t$.  (In principle, at finite doping there is a coupling to
the longitudinal magnetic susceptibility\cite{ChuF}, but this is neglected
for simplicity.)  The importance of using the AFM susceptibility is
illustrated in Fig.~\ref{fig:111}a: the solid and dashed lines show
$U_{eff}$ calculated using the charge susceptibility in the N\'eel state,
while the corresponding lines with triangles use the paramagnetic
susceptibility at low $T$.  The latter calculation finds a nearly doping
independent, but small $U_{eff}$; the former reproduces a large, weakly
screened $U$ near half filling.  Such a difference is expected in terms of
screening: when there is no gap at half filling, the enhanced susceptibility 
should be better able to screen $U$, resulting in a smaller $U_{eff}$.
This suggests that $U_{eff}$ should have an important {\it temperature
dependence} as the gap decreases -- which in turn will cause the gap to
close at a lower temperature.  Figure~\ref{fig:111}a also shows that there
is a weak temperature dependence of the screening.  The calculations
suggest that the large values of $U$ found in the cuprates are
characteristic mainly of the half filled regime and relatively low
temperatures.
A similar but larger screening effect was recently reported by Esirgen, et
al.\cite{Esi}.  

This procedure is still not fully self consistent.  If there is a
large difference between the bare $U$ and the screened $U_{eff}$, the gap
in $\chi$ should depend on the actual $U_{eff}$.  However, since
$U_{eff}\simeq U$ at half filling, any simple improvement will not
significantly change the overall doping dependence.  This is the same
kind of lack of self-consistency found for the SCR approach, and will 
be here neglected.

%Since the magnetization depends on the effective $U$, a better
%approximation might be to evaluate all susceptibilities in terms of
%$U_{eff}$, finding $U$ from Eq.~\ref{eq:12c} as
%\begin{equation}
%U={U_{eff}\over 1-<\chi (U_{eff})>U_{eff}}.
%\label{eq:12d}
%\end{equation}
%XA particularly appealing choice would then be to use the RPA charge 
%Xsusceptibility $\chi^{00}_{RPA}=\chi^{00}_0/(1+U_{eff}\chi^{00}_0)$,
%Xwhich leads to 
%X\begin{equation}
%XU_{eff}={U\over 1+U_{eff}<\chi^{00}_0(U_{eff})>}.
%X\label{eq:12e}
%X\end{equation}
%(Note that this empirical approach is rather loose about replacing
%susceptibilities by their averages.)  To apply Eq.~\ref{eq:12d}, $U_{eff}$
%is adjusted at each doping to get the same bare $U$ value.  

%These results are also shown in Fig.~\ref{fig:111}a.  In both cases, the
%intraband susceptibility vanishes at half filling and the interband term
%is small, while screening is enhanced in the doped material, leading to
%much better agreement with experiment\cite{KLBM}.  Since the differences
%between the two expressions are relatively small, and the simpler 
%non-self-consistent version agrees better with the vortex correction at
%large doping, the simpler model for $U_{eff}$ is used in the text.  

\section{Path Integral Calculation}

\subsection{Formalism}

The partition function of the Hubbard model can be written as a path 
integral\cite{NNag}:
\begin{equation}
Z=\int DC^{\dagger}DCexp\bigl[-\int_0^\beta d\tau L\bigr],
\label{eq:B1}
\end{equation}
with Lagrangian
\begin{equation}
L=\sum_{i,\sigma}C^{\dagger}_{i\sigma}(\partial_{\tau}-\mu )C_{i\sigma}-
\sum_{i,j,\sigma}C^{\dagger}_{i\sigma}t_{i,j}C_{j\sigma}+U\sum_{i}n_{i\uparrow}
n_{i\downarrow}.
\label{eq:B2}
\end{equation}
The quartic term can be decoupled by a Hubbard-Stratonovich transformation
\begin{equation}
Z=\int D\phi DC^{\dagger}DCexp\bigl[-\int_0^\beta d\tau L(\phi ,C^{\dagger},C)
\bigr],                      
\label{eq:B3}
\end{equation}
\begin{eqnarray}
L(\phi ,C^{\dagger},C)=\sum_{i,\sigma}C^{\dagger}_{i\sigma}(\partial_{\tau}-\mu )C_{i\sigma}-        
\sum_{i,j,\sigma}C^{\dagger}_{i\sigma}t_{i,j}C_{j\sigma}
\nonumber \\
+{U\over 4}\sum_{i}
\phi_i^2+{U\over 2}\sum_{i}\phi_i(n_{i\uparrow}-n_{i\downarrow})
\label{eq:B4}
\end{eqnarray}
(neglecting a term involving $n_{i\uparrow}+n_{i\downarrow}$).  Integrating out 
the Fermion fields leaves an effective action in the field $\phi$:
\begin{equation}
Z=\int D\phi e^{-S_{eff}},
\label{eq:B5}
\end{equation}
\begin{equation}
-S_{eff}={U\over 4}\int_0^\beta d\tau \sum_{i}\bigl[\phi_i^2-Tr\ ln\bigl(
\partial_{\tau}-\mu-t_{i,j}+\sigma{U\over 2}\phi_i\bigr)\bigr].
\label{eq:B6}
\end{equation}
Fourier transforming in space and (imaginary) time, the trace term can be 
rewritten
\begin{equation}                                                              
Tr\ ln\bigl(-G_0^{-1}+V\bigr)=Tr\ ln\bigl(-G_0^{-1})-\sum_{n=1}^{\infty}{1 \over
n}Tr\bigl(G_0V\bigr)^n,
\label{eq:B7}
\end{equation}
with
\begin{equation}
G_0(\vec k,i\omega_n)={\delta_{\sigma,\sigma'}\delta_{n,m}\delta_{\vec k,\vec 
k'}\over i\omega_n-\xi_{\vec k}},
\label{eq:B8}
\end{equation}
\begin{equation}
V={\delta_{\sigma,\sigma'}\over\sqrt{\beta N_0}}{\sigma U\over 2}\phi (\vec k-
\vec k',i\omega_n-i\omega_m),
\label{eq:B9}
\end{equation}
with $\xi_{\vec k}=\epsilon_{\vec k}-\mu$.  In Eq.\ref{eq:B7}, the odd terms in
$n$ average to zero, so expanding the action to fourth order in $\phi$ yields
\begin{eqnarray}
S={1\over 2}\sum_{\vec q,i\omega_n}\Pi_2(\vec q,i\omega_n)\phi(\vec q,i\omega_n)
\phi(-\vec q,-i\omega_n)
\nonumber \\
+{1\over 4(\beta N_0)^2}\sum{}^{'}\Pi_4(\vec q_i,i\omega_i)\phi(\vec q_1,i\omega
_1)\phi(\vec q_2,i\omega_2)\times
\nonumber \\
\times\phi(\vec q_3,i\omega_3)\phi(\vec q_4,i\omega_4),
\label{eq:B10}
\end{eqnarray}
where the prime in the second sum means summing over all $\vec q_i$, $\omega_i$,
such that $\sum_{i=1}^4\vec q_i=0$, $\sum_{i=1}^4\omega_i=0$, 
\begin{equation}
\Pi_2(\vec q,i\omega_n)={U\over 2}[1-U\chi_0(\vec q,i\omega_n)],
\label{eq:B11}
\end{equation}
\begin{eqnarray}
\Pi_4(\vec q,i\omega_n)={U^4\over 8}\sum_{\vec k,i\epsilon_n}
G_0(\vec k,i\epsilon_n)G_0(\vec k+\vec q_1,i\epsilon_n+i\omega_1)\times
\nonumber \\
\times G_0(\vec k+\vec q_1+\vec q_2,i\epsilon_n+i\omega_1+i\omega_2)
G_0(\vec k-\vec q_4,i\epsilon_n-i\omega_4),
\label{eq:B12}
\end{eqnarray}
and $\chi_0$ is the dynamic susceptibility
\begin{eqnarray}                                                                
\chi_0(\vec q,i\omega_n)={-1\over \beta N_0}\sum_{\vec k,i\epsilon_n}
G_0(\vec k,i\epsilon_n)G_0(\vec k+\vec q,i\epsilon_n+i\omega_n)
\nonumber \\
={1\over N_0}\sum_{\vec k}{f(\xi_{\vec k+\vec q})-f(\xi_{\vec k})\over i\omega_n
+\xi_{\vec k}-\xi_{\vec k+\vec q}}.
\label{eq:B13}
\end{eqnarray}
%The latter expression for $\chi$ holds in the high temperature limit when all
%the $\phi_i$ correspond to fluctuating fields.

The mean field solution corresponds to assuming $\phi_i=\phi_0e^{i\vec Q\cdot
\vec R_i}$ and finding the saddle point solution of Eq.~\ref{eq:B6}.  Including 
interactions by summing bubble or ladder diagrams\cite{SWZ} leads to the RPA 
susceptibility (see Eq.~\ref{eq:B14} below), from which the spin wave spectra 
are calculated\cite{MK3}.  Fluctuations about the mean field solution are
described by Eq.~\ref{eq:B10}.  Due to the Mermin-Wagner theorem, these
fluctuations are in the high temperature limit $T>T_N$.  A naive perturbational
analysis diverges (as demonstrated below), so a self-consistent analysis is
necessary.  Following the self-consistent renormalization (SCR) model of 
Moriya\cite{Mor,HasMo}, the exact dynamical susceptibility can be written as
\begin{equation}
\chi (\vec q,i\omega_n)={\chi_0(\vec q,i\omega_n)\over 
1-U\chi_0(\vec q,i\omega_n)+\lambda_U(\vec q,i\omega_n)},
\label{eq:B14}
\end{equation}
with the RPA susceptibility given by Eq.~\ref{eq:B14} with $\lambda_U=0$.

Solving Eq.~\ref{eq:B14} requires an equation for $\lambda_U$.  An 
approximate solution is found by replacing $\lambda_U(\vec q,i\omega_n)$ by a
constant
\begin{equation}                                                             
\lambda\equiv\lambda_U(0,0)=\chi_0({\partial^2\Delta F(M,T)\over\partial M^2}
)_{M=0},
\label{eq:B15}
\end{equation}
where the total free energy is written as $F(M,T)=F_{HF}(M,T)+\Delta F(M,T)$,
with $F_{HF}$ the Hartree-Fock free energy and
\begin{eqnarray}
\Delta F(M,T)=-T\sum_{\vec q,n}\int_0^UdU[\chi (\vec q,i\omega_n)
-\chi_0(\vec q,i\omega_n)]
\nonumber \\
=T\sum_{\vec q,n}[ln(1-U\chi_0(\vec q,i\omega_n)+\lambda )+U
\chi_0(\vec q,i\omega_n)].
\label{eq:B16}
\end{eqnarray}

Equation~\ref{eq:B16} can be solved by expanding about the expected ordered
state.  The ordered states are found from the zeroes of the denominator of the
dynamical susceptibility, Eq.~\ref{eq:B14}.  For the present case, the largest
bare susceptibility corresponds to antiferromagnetic order, $\vec q=\vec Q$. 
Then, defining
\begin{equation}
\delta =1-U\chi_0(\vec Q,0)+\lambda ,
\label{eq:B17}
\end{equation}
and $\delta_0=\delta -\lambda$, it will be shown that $\delta\ge 0$, and
$\delta\rightarrow 0$ as $T\rightarrow 0$ -- that is, there is no finite
temperature phase transition (the Mermin-Wagner theorem is satisfied).

\subsection{SCR Analysis}

Following the conventional analysis, Eq.~\ref{eq:B16} is expanded in
terms of the small parameters $\omega$ and $\vec q'\equiv\vec q-\vec Q$
(analytically continuing $i\omega_n\rightarrow\omega +i\epsilon$): 
\begin{equation}
1-U\chi_0(\vec q,\omega )+\lambda =
\delta +Aq'^2-B\omega^2-iC\omega,
\label{eq:B18}
\end{equation}
where the expansion coefficients are
\begin{equation}
A=-Ua^2\sum_{\vec k}[{f'(\epsilon_{\vec k})\over (\epsilon_{\vec k+\vec Q}-
\epsilon_{\vec k})^2}+{f(\epsilon_{\vec k})-f(\epsilon_{\vec k+\vec Q})\over
(\epsilon_{\vec k+\vec Q}-\epsilon_{\vec k})^3}],
\label{eq:B19}
\end{equation}
\begin{eqnarray}
A={Ua^2\over 2}\sum_{\vec k}[({f(\epsilon_{\vec k})-f(\epsilon_{\vec k+\vec Q})
\over (\epsilon_{\vec k+\vec Q}-\epsilon_{\vec k})})({1\over 8}+{32t'c_y^2s_x^2
\over (\epsilon_{\vec k+\vec Q}-\epsilon_{\vec k})^2})
\nonumber \\
-f'(\epsilon_{\vec k})({1\over 16}+{2t'\over (\epsilon_{\vec k+\vec Q}-
\epsilon_{\vec k})}+{32t^{'2}c_y^2s_x^2\over (\epsilon_{\vec k+\vec Q}-
\epsilon_{\vec k})^2})
\nonumber \\
-f''(\epsilon_{\vec k})(t'(1-c_xc_y)+{2s_x^2(t^2+4t^{'2}
c_y^2)\over (\epsilon_{\vec k+\vec Q}-\epsilon_{\vec k})})],
\label{eq:B19b}
\end{eqnarray}
\begin{equation}
B=U\sum_{\vec k}[{f(\epsilon_{\vec k})-f(\epsilon_{\vec k+\vec Q})\over 
(\epsilon_{\vec k+\vec Q}-\epsilon_{\vec k})^3}],
\label{eq:B20}
\end{equation}
\begin{equation}
C=-2\pi U\sum_{\vec k}[f'(\epsilon_{\vec k})\delta (\epsilon_{\vec k+\vec Q}-
\epsilon_{\vec k})]. 
\label{eq:B21}
\end{equation}
These are similar to results found previously\cite{HasMo}, specialized to the
dispersion of Eq.~\ref{eq:0}.  However, with this dispersion the $A$ and $B$ 
integrals formally diverge.  Hence a more careful analysis is needed, 
presented in Appendix D.  For now, $A$, $B$, and $C$ will be treated as 
parameters.

\subsubsection{Free Energy}

Here, $\lambda$ is determined by minimizing the free energy, including
quartic, $~\Pi_4$, corrections\cite{NNag}.  With the variational estimate
\begin{equation}
F=F_0+{1\over\beta}<S-S_0>_{S_0},
\label{eq:B22}
\end{equation}
the action $S$ (Eq.~\ref{eq:B10}) becomes
\begin{eqnarray}
S={U\over 4}\sum_{\vec q,i\omega_n}(\delta_0+Aq'^2+C|\omega_n|)
\phi(\vec q,i\omega_n)\phi(-\vec q,-i\omega_n)
\nonumber \\
+\tilde u\int_0^{\beta}d\tau \int d^2\vec r[\phi(\vec r,\tau )]^4,
\label{eq:B23}
\end{eqnarray}
and $S_0$ is given by the same equation, with $\delta_0$ replaced by the
variational parameter $\delta$.  Here $\tilde u=\Pi_4 a^d/4\beta N_0$, with $d=
2$ and $\Pi_4$ evaluated at $\vec q_i=(0,\vec Q,0,-\vec Q)$, $\omega_i=0$, since
higher order corrections are irrelevant.  Then to Gaussian order
\begin{equation}
<S-S_0>_{S_0}={U\over 4}\sum_{\vec q,i\omega_n}(\delta_0-\delta)
<\phi(\vec q,i\omega_n)\phi(-\vec q,-i\omega_n)>_{S_0},
\label{eq:B24}
\end{equation}
\begin{eqnarray}
<\phi(\vec q,i\omega_n)\phi(-\vec q,-i\omega_n)>_{S_0}=
\nonumber \\
Z_0^{-1}\prod_{\vec q,i
\omega_n}\int dRe\phi(\vec q,i\omega_n)dIm\phi(\vec q,i\omega_n)\times
\nonumber \\
\times\phi(\vec q,i\omega_n)\phi(-\vec q,-i\omega_n)e^{-{1\over 2}
\hat D_0^{-1}(\vec q,i\omega_n)\phi(\vec q,i\omega_n)
\phi(-\vec q,-i\omega_n)}
\nonumber \\
={\partial ln(Z_0)\over\partial \hat D_0^{-1}(\vec q,i\omega_n)}
=\hat D_0(\vec q,i\omega_n)
\label{eq:B25}
\end{eqnarray}
with $\hat D_0^{-1}(\vec q,i\omega_n)=(U/2)D_0^{-1}(\vec q,i\omega_n)
=(U/2)[\delta+Aq'^2+C|\omega_n|]$.  Similarly, to Gaussian level
\begin{equation}
F_0=-{1\over\beta}lnZ_0={1\over 2\beta}\sum_{\vec q,i\omega_n}ln\hat D_0^{-1}
(\vec q,i\omega_n),
\label{eq:B26}
\end{equation}
up to a constant.  Writing the quartic term as $S_{int}$, the full partition 
function is approximated by $Z\simeq Z_0[1-<S_{int}>_{Gauss}+{1\over
2}<S_{int}^2>_{Gauss}]$, with
\begin{eqnarray}
<S_{int}>_{Gauss}=\tilde u\int_0^{\beta}d\tau \int d^2\vec r<\phi(\vec r,\tau )
^4>_{Gauss}
\nonumber \\
=3\tilde u\int_0^{\infty}d\tau \int d^2\vec r[<\phi(\vec r,\tau )^2>]^2, 
\label{eq:B27}
\end{eqnarray}
and
\begin{eqnarray} 
<\phi(\vec r,\tau )^2>_{Gauss}={1\over\beta V}\sum_{\vec q,i\omega_n}
<\phi(\vec q,i\omega_n)\phi(-\vec q,-i\omega_n)>_{S_0}
\nonumber \\
={1\over\beta V}\sum_{\vec q,i\omega_n}\hat D_0(\vec q,i\omega_n).
\label{eq:B28}
\end{eqnarray}
[The term in $<S_{int}^2>_{Gauss}$ will not be needed.]

The free energy Eq.~\ref{eq:B22} can thus be rewritten as
\begin{eqnarray}
F={1\over 2\beta}\sum_{\vec q,i\omega_n}lnD_0^{-1}(\vec q,i\omega_n)
\nonumber \\
+{1\over 2\beta}\sum_{\vec q,i\omega_n}(\delta_0-\delta)D_0(\vec q,i\omega_n)
\nonumber \\
+{3u\over\beta^2N_0}[\sum_{\vec q,i\omega_n}D_0(\vec q,i\omega_n)]^2,
\label{eq:B29}
\end{eqnarray}
with $u=\Pi_4/N_0\beta U^2$.

\subsubsection{Stoner Factor}

The variational parameter is found from $\partial F/\partial\delta =0$, or
\begin{equation}
\delta =\delta_0+{12u\over\beta V}\sum_{\vec q,i\omega_n}D_0(\vec q,i\omega_n).
\label{eq:B30}
\end{equation}
The next step is to carry out the sum over Matsubara frequencies and wave 
vectors.  For the former, using
\begin{eqnarray}
{1\over\beta}\sum_{i\omega_n}X(i\omega_n)=-{1\over\beta\pi}\sum_{i\omega_n}\int
_{-\infty}^{\infty}d\epsilon{ImX(\epsilon +i\delta )\over i\omega_n-\epsilon}
\nonumber \\
=-\int_0^{\infty}d{\epsilon\over \pi}coth{\epsilon\over 2T}ImX(
\epsilon +i\delta ),
\label{eq:B31}
\end{eqnarray}
then
\begin{eqnarray}
{1\over\beta V}\sum_{\vec q,i\omega_n}D_0(\vec q,i\omega_n)
\nonumber \\
=\int{d^2\vec qa^2\over(2\pi )^2}\int_0^{\alpha_{\omega}/C}{d\epsilon\over\pi}
coth{\epsilon\over 2T}{C\epsilon\over (\delta +Aq^{'2})^2+(C\epsilon )^2}.
\label{eq:B32}
\end{eqnarray}
Note the sharp energy cutoff in Eq.~\ref{eq:B32}.  This comes about because the
linear-in-$\omega$ dissipation is a result of Landau damping of the spin
waves by electrons near the hot spots, and therefore the dissipation cuts off
when the spin wave spectrum gets out of the electron-hole continuum.  Numerical
calculations (Fig.~\ref{fig:11}) show that the cutoff can be quite sharp,
particularly near the VHS.  

Equations~\ref{eq:B30},~\ref{eq:B32} can easily be solved in the limit $T=0$.  
In this case, there is a transition at
\begin{eqnarray}
\delta_0=-12u\int_0^{q_c^2}{dq'^2a^2\over 4\pi}\int_0^{\alpha_{\omega}/C}{d
\epsilon\over\pi}{C\epsilon\over (Aq'^2)^2+(C\epsilon )^2}
\nonumber \\
=-{3uq_c^2a^2\over \pi^2C}R_0\equiv 1-\eta ,
\label{eq:B35}
\end{eqnarray}
\begin{equation}
R_0={1\over 2}ln[1+a_q^{-2}]+{tan^{-1}(a_q)\over a_q},
\label{eq:B34a}
\end{equation}
with $a_q=Aq_c^2/\alpha_{\omega}$.  Since the right-hand side is finite
and negative, fluctuations reduce but in general do not eliminate the
order at $T=0$.  The quantum corrected Stoner criterion is $U\chi_0=\eta$,
where representative values of $\eta$ are listed in Table I.

However, for finite $T$, there are corrections $\sim ln(\delta )$, so $\delta$
cannot be set to zero, and there is no finite temperature transition (the
Mermin-Wagner theorem is satisfied).  To see this, it is adequate to
approximate $coth(x)$ as $1/x$ for $x\le 1$ and 1 for $x>1$. In this case, 
Eq.~\ref{eq:B30} can be solved exactly, Appendix A3.  However, this exact
solution is not very illuminating, and a simpler approximate solution will
be given here. Since only the term proportional to $T$ is singular, $T$
and $\delta$ can be set to zero in the remaining term.  Defining 
\begin{equation}
\bar\delta_0=\delta_0+\eta -1,
\label{eq:B34g}
\end{equation}
Eq.~\ref{eq:B30} becomes
\begin{eqnarray}
\delta-\bar\delta_0={6uTa^2\over\pi^2A}\int_{\delta}^{\delta+Aq_c^2}{dy\over
y}tan^{-1}({2TC\over y})
\nonumber \\
\simeq{3uTa^2\over\pi A}ln({2CT\over\delta}),
\label{eq:B34h}
\end{eqnarray}
where the second line uses Eq.~\ref{eq:B34bx}, below.
Hence, there is no finite temperature phase transition, and $\delta$ only
approaches zero asymptotically as $T\rightarrow 0$: approximately,
\begin{equation}
\delta =2CTe^{-\pi A|\bar\delta_0|/3uTa^2}.
\label{eq:B34f}
\end{equation}

\subsubsection{Susceptibility}                      

The resulting susceptibility has NAFL form, Eq.~\ref{eq:15}, with  
(Eqs.~\ref{eq:B14},\ref{eq:B18}) the following explicit expressions:
\begin{equation}
\chi_{\vec Q}={\chi_0\over\delta}
\label{eq:B36}
\end{equation}
\begin{equation}
\xi^2={A\over\delta} 
\label{eq:B37}
\end{equation}
\begin{equation}
\Delta^2={\delta\over B} 
\label{eq:B38}
\end{equation}
\begin{equation}
\omega_{sf}={\delta\over C} 
\label{eq:B39}
\end{equation}
If the correlation length $\xi$ is written in the form Eq.~\ref{eq:16}, and 
Eq.~\ref{eq:B34e} is numerically solved for $\delta$, then 
$\rho_s$ is exactly given by 
\begin{equation}
\rho_s={k_BT\over 4\pi}\ln({A\over\xi_0^2\delta}),
\label{eq:B40a}
\end{equation}
with $\xi_0=\sqrt{eA\over 2TC}$.  Using Eq.~\ref{eq:B34f}, an approximate 
$\rho_s$ is:
\begin{equation}
\rho_s^a={A|\bar\delta_0|\over 12ua^2}.
%+{k_BT\over 2\pi}\ln({1\over aq_c}).
\label{eq:B40}
\end{equation}
$\rho_s$ is plotted in Fig.~\ref{fig:11a1}b, with $u^{-1}=0.256eV$, chosen
to give a $\rho_s$ in agreement with the results of Chakravarty, et
al.\cite{CHN} for $x=0$, $T=0$.  

\subsection{`Exact' Solution of Eq.~\protect\ref{eq:B30}}

Approximating $coth(x)=max(1/x,1)$, and introducing the notation $\bar
Aq_c^2=Aq_c^2+\delta$, $\bar a_q=\bar Aq_c^2/\alpha_{\omega}$, and $t=2TC$, 
the solution becomes
\begin{equation}
\delta -\delta_0={3ua^2\over\pi^2AC}\bigl[F_1+F_2\bigr],
\label{eq:B34b}
\end{equation}
with
\begin{eqnarray}
F_1=\int_{\delta}^{\delta+Aq_c^2}dy\int_{t}^{\alpha_{\omega}}dx{x\over
x^2+y^2}=
\nonumber \\
{\bar Aq_c^2\over 2}ln[{1+\bar a_q^2\over\bar a_q^2+(t/\alpha_{\omega})^2}]+
\nonumber \\
\alpha_{\omega}tan^{-1}(\bar a_q)
-{\delta\over 2}ln[{\delta^2+\alpha_{\omega}^2\over \delta^2+t^2}]-\alpha_
{\omega}tan^{-1}({\delta\over \alpha_{\omega}})
\label{eq:B34bi}
\end{eqnarray}
\begin{eqnarray}
F_2=t\int_{\delta}^{\delta+Aq_c^2}dy\int_0^t{dx\over x^2+y^2}=
\nonumber \\
=t\int_{\delta}^{\delta+Aq_c^2}{dy\over y}tan^{-1}({t\over y})=
\nonumber \\
=t[I_1({t\over\bar Aq_c^2})-I_1({t\over\delta})\bigr],
\label{eq:B34bii}
\end{eqnarray}
with 
\begin{equation}
I_1(x)=I_0(tan^{-1}(x))-tan^{-1}(x)ln(x),
\label{eq:B34c}
\end{equation}
\begin{equation}
I_0(x)=\int_0^xln(\tan{\theta})d\theta =L(x)+L({\pi\over 2}-x)-L({\pi\over 2}),
\label{eq:B34c1}
\end{equation}
and $L(x)=-\int_0^xln(\cos{t})dt$ is the Lobachevskiy function\cite{GRyd}.

For most purposes, it can be assumed that $\delta
<<t<<Aq_c^2,\alpha_{\omega}$, in which
case $I_0(tan^{-1}(x))=\theta (\ln{(\theta)}-1)$, with $\theta =
min\{x,1/x\}$, and then $F_2$, Eq.~\ref{eq:B34bii}, simplifies.  
\begin{eqnarray}
F_2=ln({t\over\delta})[\delta +ttan^{-1}({t\over\delta})]+\delta
-{t^2\over Aq_c^2}
\nonumber \\
\simeq {\pi\over 2}t \ln({t\over\delta}).
\label{eq:B34bx}
\end{eqnarray}
Defining 
$Z=1+(3ua^2/\pi^2AC)ln(\alpha_{\omega}/t)$, then 
\begin{equation}
Z\delta -\bar\delta_0={3ua^2T\over\pi A}ln({2CT\over \delta}),
\label{eq:B34e}
\end{equation}
which agrees with Eq.~\ref{eq:B34h} when $Z\rightarrow 1$.

\section{Parameter Evaluations}
                                         
\subsection{Hot Spots}

Hot spots are defined as the intersection of the Fermi surface with the line
$k_x+k_y=\pi /a$.  At these points there is strong scattering, since the vector 
$\vec Q$ connects two hot spots.  The hot spots dominate the
integrals $A$, $B$, Eqs.~\ref{eq:B19},\ref{eq:B20} at $T=0$.  The Fermi 
functions limit the integral to a sum of approximately wedge-shaped areas
centered on the hot spots, Fig.~\ref{fig:44a}.  In the main text, the
susceptibilities were calculated numerically.  Here, analytical
approximations are introduced to clarify the role of the hot spots.

On a single wedge, a typical integral for, e.g., $A$ becomes
\begin{equation}
\int{d^2\vec q\over (2\pi )^2}{1\over (\epsilon_{\vec k+\vec Q}-\epsilon_{\vec k
})^3}={1\over (4t)^3}\int{d^2\vec q\over (2\pi )^2}{1\over (c_x+c_y)^3}.
\label{eq:C2}
\end{equation}
Letting $k_i=k_{i0}-k'_i$, $i=x,y$, then to lowest order the energy difference
becomes 
\begin{eqnarray}
\Delta\epsilon =\epsilon_{\vec k+\vec Q}-\epsilon_{\vec k} =\alpha
_{\theta}k'(1+\beta_{\theta}k')
\nonumber \\
=\alpha_0k_{\perp}(1+\beta_0k_{\parallel}), 
\label{eq:C3a}
\end{eqnarray}
with $\alpha_{\theta}=\alpha_0(\sin{\theta}+\cos{\theta})$, $\beta_{\theta}=
\beta_0(\sin{\theta}-\cos{\theta})$, $\alpha_0=4tas_{x0}$, and $\beta_0=ac_{x0}
/2s_{x0}$; $k_{\parallel}$ and $k_{\perp}$ are the momenta parallel and 
perpendicular to the zone diagonal (magnetic Brillouin zone boundary).
The integral of Eq.~\ref{eq:C2} becomes
\begin{equation}                                                   
{1\over 4\pi^2}\int_{\theta_{min}}^{\theta_{max}}{d\theta
\over \alpha_{\theta}^3}\int{dk'\over k^{'2}},
\label{eq:C3}
\end{equation}
where $\theta_{min}$ and $\theta_{max}$ are the opening angles of the wedge.
Thus, the integral over $k'$ diverges at a hot spot.  In this case, the
expansion Eq.~\ref{eq:B18} must be modified.

\subsection{$A$} 

The apparent singularity of $A$ is an artifact.  In reality, a finite $q'$ 
shifts the location of the hot spots.  To confirm this, the hot spot integrals
can be evaluated as above.  First, the shift in the positions of the hot spots
is found, the points where $\epsilon_{\vec k-\vec q/2}=\epsilon_{\vec k+\vec Q+
\vec q/2}=\mu$.  Denoting the shift by (Eq.~\ref{eq:C1}) $k_{x0}a\rightarrow k_
{x0}a+\phi_x$, etc., then to first order in $q$, the shift is
\begin{eqnarray}
\phi_x+\phi_y={\tau c_{x0}\over 2}(q_y-q_x),
\nonumber \\ 
\phi_y-\phi_x={1\over 2\tau c_{x0}}(q_y+q_x).                       
\label{eq:C8a}
\end{eqnarray}
Expanding about the shifted hot spots (with $k'$ = radial distance from new
hot spots gives $\epsilon_{\vec k+\vec Q+\vec q/2}-\epsilon_{\vec k-\vec
q/2}=4tk'(\alpha_{\theta}(q)+\beta_{\theta}(q)k')$, where $\alpha_{\theta}(q)
\rightarrow\alpha_{\theta}$ and $\beta_{\theta}(q)\rightarrow\beta_{\theta}$ as
$q\rightarrow 0$.  Hence, the integral to be evaluated is
\begin{equation}
\int_0^{k_c}dk'[{1\over \alpha_{\theta}(q)+\beta_{\theta}(q)k'}-
{1\over \alpha_{\theta}+\beta_{\theta}k'}],
\label{eq:C8c}
\end{equation}
which is well behaved as a function of $k'$ and can be expanded as a series in 
$q^n$.  Note that there are terms linear in $q$, which are cancelled in
averaging over the hot spots.

Since there are no singularities, direct numerical evaluation of $\chi (q')
-\chi (q'=0)$ is straightforward, yieldiing $A$, Fig.~\ref{fig:10}.  
[In the numerical calculation, some care is needed due to the terms linear in
$q$.  While they cancel when summed over all eight hot spots, 
to accurately determine the smaller quadratic term, this
summation should be carried out at each $k$-point prior to summing the result
for all $k$.]  

\subsection{$A$ at the C-point}

\begin{figure}
\leavevmode
   \epsfxsize=0.33\textwidth\epsfbox{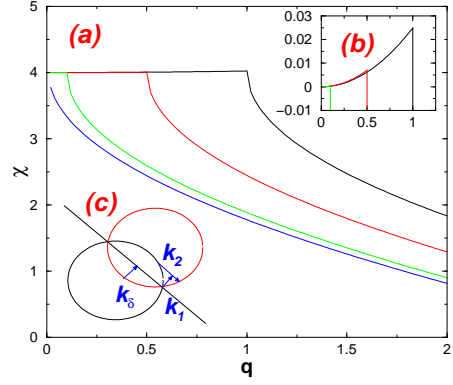}
\vskip0.5cm
\caption{(a) Calculated susceptibility $\chi (q)$ for several values of 
overlap $\delta$.  (b) Blowup of plateau region, for $\chi_q-\chi_{q=0}$. (c) 
Model of Fermi surfaces, defining $\delta$, $k_1$ ($k_{\perp}$) and $k_2$ ($k_
{\parallel}$).}
\label{fig:11b}
\end{figure}

While the above analysis works for small $q$ near the tip of the wedge, it is
hard to extend it to the edge of the $q$-plateau, or in particular to the 
C-point, where the plateau width shrinks to zero.  It is convenient to introduce
a simplified model\cite{GMV}, for which the $q$-dependence of $\chi$ can
be calculated 
{\it analytically}.  It is convenient to recall Fig.~\ref{fig:44a}.  While the
dashed lines in that figure represented an $\omega$ shift, they can equally
well describe the $q$-shift of the energy denominator, Eq.~\ref{eq:C3a}.  The
plateau edge corresponds to the point where the dashed line intersects the
Q-shifted FS (horizontal arrows).  In the simplified model, the energy
denominator is linearized, so $\Delta\epsilon\propto k_{\perp}$, independent of 
$k_{\parallel}$.  Chosing $\vec q$ to point along the $(\pi ,\pi )$ direction, 
the FS can be approximated by two circles of radius $k_F$, centered at
$(\pi ,\pi )$ and $(-\pi ,-\pi )$ (for the choice of $\vec q$ the other
two circles at $(\pi ,- \pi )$ and $(-\pi ,\pi )$ can be ignored).  The
Q-shifted FS is then a circle centered at $\Gamma =(0,0)$.  The FS at
$(\pi ,\pi )$ and the Q-shifted FS are illustrated in Fig.~\ref{fig:11b}c.

Adding the contributions of the overlap of the Q-shifted FS with both the FS
at $(\pi ,\pi )$ and the one at $(-\pi ,-\pi )$, $\chi_q\propto I_{k_\delta+q} 
+I_{k_\delta-q}$, with 
\begin{equation}
I=\int_0^{k_c}{dk_{\perp}dk_{\parallel}\over k_{\perp}},
\label{eq:C6c}
\end{equation}
where the region of integration is over the part of the upper FS in 
Fig.~\ref{fig:11b}c not overlapped by the lower (Q-shifted) FS, and $k_{\perp}$
ranges from zero at the apex of the wedge to the middle of the upper FS, $k_c=
k_F-k_\delta$, where $k_\delta$ is the overlap parameter defined in
Fig.~\ref{fig:11b}c.  To lowest order, for $k_\delta <<k_F$, 
\begin{eqnarray}
I
=2\sqrt{2k_Fk_\delta}[\sqrt{1+\beta}+\ln|{\sqrt{1+\beta}-1\over\sqrt{1+\beta}+1}
|-
\nonumber \\
-\Theta (1-\beta )(\sqrt{1-\beta}+\ln|{\sqrt{1-\beta}-1\over\sqrt{1-\beta}+1}|)
],
\label{eq:C6b}
\end{eqnarray}
with $\beta =k_c/k_\delta$ and $\Theta$ the unit step function.  
The expression for $I_{k_\delta-q}$ must be modified when $q>k_\delta$ and 
the two FSs no longer overlap\cite{GMV}: $I_{k_\delta-q}=2\sqrt{2}k_F[1-\gamma 
\tan^{-1}1/\gamma]$, with $\gamma =\sqrt{(q-k_\delta )/2k_F}$.  
The calculated susceptibilities, Fig.~\ref{fig:11b}a, display the flat topped 
plateaus with weak positive curvature ($A<0$, Fig.~\ref{fig:11b}b).  At the 
plateau edge the susceptibility falls sharply, $\chi\sim 1-\pi\gamma
/2\sim\sqrt{q}$.  The C-point corresponds to $k_\delta =0$.

\subsection{$B$}

The expression for $B$ may be written exactly as the $\omega\rightarrow 0$
limit of
\begin{equation}
B=URe\sum_{\vec k}[{f(\epsilon_{\vec k})-f(\epsilon_{\vec k+\vec Q})\over
(\epsilon_{\vec k+\vec Q}-\epsilon_{\vec k})}]{1\over ((\epsilon_{\vec k+\vec 
Q}-\epsilon_{\vec k})^2-\omega^2}.
\label{eq:C6}
\end{equation}
It can be shown that $B$ has a logarithmic correction due to the hot
spots.  The integral can be approximately evaluated by (a) using symmetry
to reduce the integral to one over an octant of the Brillouin zone
containing one hot spot, (b) splitting the domain of integration into (i)
a circle of radius $k_c$ about the hot spot, and (ii) the remainder of the
domain, and (c) numerically evaluating the integral over domain (ii) while
providing an analytic approximation to that over (i).  
Then the $k$ integral over the hot spot circle can be written approximately as
\begin{eqnarray}
I=\int_0^{k_c}{(1-3\beta_{\theta}k)dk\over \alpha_{\theta}^2k^2-\omega^2}
\nonumber \\
\simeq {1\over\alpha_{\theta}^2}[{1\over k_c}-3\beta_{\theta}\log{\alpha_
{\theta}k_c\over\omega}].
\label{eq:C6a}
\end{eqnarray}
At $T=0$, the integral $I$ must then be integrated in $\theta$ over the wedge
where the difference in Fermi functions does not vanish.  
The integral from outside the hot spot circle will eliminate the 
$k_c$-dependence, but should not affect the $log(\omega )$ term.  The same
logarithmic divergence can be found as a byproduct of the calculation of $C$,
below.

It is difficult to directly evaluate the two-dimensional principal value 
integral for $B$.  Instead, it is much simpler to evaluate $Re(\chi )$ via
Kramers-Kronig transformation of $Im(\chi )$ and find $B$ by numerical
differentiation.  When this is done, it is found that (a) $B$ is numerically
very small due to the plateau in $Re(\chi )$, Fig.~\ref{fig:44b}d, and (b)
the logarithmic correction is too small to determine accurately.

\subsection{$C$}

Parameter $C$ is conveniently found by analytically continuing Eq.~\ref{eq:C6}
back to the Matsubara frequencies,
\begin{equation}
{C\over |\omega_n|}=U\sum_{\vec k}[{f(\epsilon_{\vec k})-f(\epsilon_{\vec k+\vec
Q})\over (\epsilon_{\vec k+\vec Q}-\epsilon_{\vec k})}]{1\over ((\epsilon_{\vec 
k+\vec Q}-\epsilon_{\vec k})^2+\omega_n^2}.
\label{eq:C7}
\end{equation}
The wedge integral can be evaluated as for $B$.  The relevant integral is
\begin{eqnarray}
I\simeq\int_0^{k_c}{dk(1-3\beta_{\theta}k)\over k^2+\hat\omega_n^2}
\nonumber \\
\simeq{1\over\hat\omega_n}tan^{-1}({k_c\over\hat\omega_n})-{3\beta_{\theta}\over
2}ln(1+({k_c\over\hat\omega_n})^2)
\nonumber \\
\simeq{\pi\over 2\hat\omega_n}-{1\over k_c}-3\beta_{\theta}ln({k_c\over\hat
\omega_n}),
\label{eq:C7a}
\end{eqnarray}
($\hat\omega_n=\omega_n/\alpha_{\theta}$) thus giving both the linear in 
$\omega$ dissipation, and confirming the $ln(\omega )$ divergence found above.  
Figure~\ref{fig:11} shows the divergence of $C$ as $\mu$ approaches the VHS.  
Note that it is cut off at increasingly lower frequencies: the arrows 
correspond to $\omega_n=\pi C$.

Alternatively, Eq.~\ref{eq:B21} may be used; this can be integrated to yield
Eq.~\ref{eq:C8} (in agreement with Sachdev, et al.\cite{SCS}), explicitly 
displaying the divergence at the VHS ($s_{x0}\rightarrow 0$).

\subsection{$u$}

Since there is some controversy\cite{Mil,Chu} concerning $u$, it shall be
evaluated in detail.  Millis\cite{Mil} showed that for free electrons
(parabolic bands) this expression is in general well defined, but diverges
when $\vec Q$ is a `spanning' vector of the Fermi surface -- in the
present case, this would correspond to the H- and C-points.  Abanov, et
al.\cite{Chu} found a more severe divergence: $u$ diverges for all $\mu$
in the hot spot regime.  The problem lies in the limit of external frequencies 
$\rightarrow$ 0, momenta $\rightarrow$ 0 or $\vec Q$.  Taking this limit on the
momenta, the expression for $u$ can be written as 
\begin{eqnarray}
u={U^2\over N_0\beta}\sum_{\vec k,i\omega_n}{1
\over (\epsilon_{\vec k}-i\omega_n)(\epsilon_{\vec k}-i\omega_n+i\omega_4)}
\times
\nonumber \\
\times{1\over (\epsilon_{\vec k+\vec Q}-i\omega_n-i\omega_1)(\epsilon_{\vec k+
\vec Q}-i\omega_n-i\omega_1-i\omega_2)}.
\label{eq:C4}
\end{eqnarray}
The sum over Matsubara frequencies yields
\begin{eqnarray}
u=U^2\sum_{\vec k}[{f(\epsilon_{\vec k})\over
i\omega_4}\bigl({1\over (i\omega_3-\Delta\epsilon)(i\omega_3+i\omega_2-
\Delta\epsilon)}-
\nonumber \\
-{1\over (i\omega_1+\Delta\epsilon)(i\omega_1+i\omega_2+\Delta\epsilon)}\bigr)
\nonumber \\
+{f(\epsilon_{\vec k+\vec Q})\over i\omega_2}\bigl({1\over (i\omega_3-\Delta
\epsilon)(i\omega_1+i\omega_2+\Delta\epsilon)}-
\nonumber \\
-{1\over (i\omega_1+\Delta\epsilon )(i\omega_3+i\omega_2-\Delta\epsilon)}\bigr)
],
\label{eq:C4a}
\end{eqnarray}                                                                
where $\Delta\epsilon=\epsilon_{\vec k}-\epsilon_{\vec k+\vec Q}$.  Letting 
$\omega_{i,\pm}=(\omega_i\pm\omega_{i+2})/2$ ($i=1,2$), and noting that $\omega_
{1+}=-\omega_{2+}$, this simplifies to 
\begin{equation}
u=2U^2\sum_{\vec k}{(f(\epsilon_{\vec k+\vec Q})-f(\epsilon_{\vec k}))W_-
\over (W_-^2+\omega_{1+}^2)(W_-^2+\omega_{2-}^2)},
\label{eq:C4b}
\end{equation}
where
\begin{equation}
W_-=(i\omega_{1-}+\Delta\epsilon).
\label{eq:C4c}
\end{equation}
Thus in Matsubara frequency space, $u$ is largest for $\omega_{1+}=
\omega_{2-}=0$, so it should indeed be reasonable to estimate it in that limit:
\begin{equation}
u(i\omega_1,0,0)=U^2{\partial^2\over\partial (i\omega_1)^2} \sum_{\vec k}{f(
\epsilon_{\vec k+\vec Q})-f(\epsilon_{\vec k})\over i\omega_1+\Delta\epsilon}.
\label{eq:C4d}
\end{equation}
In turn, it should be possible to approximate $u$, Eq.~\ref{eq:C4d}, by
its $\omega_1\rightarrow 0$ limit, if this is nonsingular.  From
Eq.~\ref{eq:B18}, $U\chi_0(\vec Q,\omega )=B\omega^2
%ln(\omega_0/\omega )
+iC\omega +1-\delta_0$. 
%where the $B$-term is corrected by the results of subsection 4.  
Thus, the analytic continuation $i\omega_1\rightarrow \omega+i\delta$ yields
\begin{equation}
u(0,0,0)=U^2\lim_{\omega\rightarrow 0}{\partial^2\chi_0(\vec Q,\omega )\over
\partial \omega^2}\simeq 
2BU.
%BU[2ln(\omega_0/\omega )-3],
\label{eq:C4e}
\end{equation}
%which diverges as $\omega\rightarrow 0$.  
Due to the plateau in $\chi (\vec Q,\omega )$, $B$ (Table I) and hence $u$ are
extremely small.  The smallness of $u$ is true only in the limit that all
external frequencies are small, which means that a more complicated expression
should be used to evaluate $u$.  Moreover, there is an additional problem: 
as found above, $B$ has a correction in $\ln{(\omega )}$, which would formally 
be divergent.  Hence, the model is not fully self-consistent, and $u$ will
be treated as an empirical parameter.  The weak logarithmic divergence
will be neglected, and $u$ approximated by a constant.

\section{Interlayer Coupling}

\subsection{Dispersion of $t_z$: Direct and Staggered Stacking}

Andersen, et al.\cite{ALJP} demonstrated that the anomalous form of
interlayer hopping in the cuprates, $t_z=t_{z0}(c_x-c_y)^2$, could be
understood by coupling the Cu$_{d_x^2-d_y^2}$ and O$_p$ orbitals to the
Cu$_{4s}$ orbitals, which have significant interlayer coupling.  Here, I
provide a simplified calculation including only these orbitals, and show
how the dispersion is modified by staggered stacking of the CuO$_2$
layers.  For uniform stacking (Cu above Cu), the hopping matrix becomes
\begin{equation}
H=\left(\matrix{\Delta&-2ts_x&2ts_y&0\cr
                -2ts_x&0&0&-2t_{ps}s_x\cr
                 2ts_y&0&0&-2t_{ps}s_y\cr 
                 0&-2t_{ps}s_x&-2t_{ps}s_y&\Delta_s+E_{sz}\cr},\right)
\label{eq:D1}
\end{equation}
with $s_i=\sin{k_ia/2}$.  
Here the first (last) row is for the Cu$_{d_x^2-d_y^2}$ (Cu$_{4s}$)
orbital, and the middle rows are for the O$_{px}$ and O$_{py}$ orbitals, 
with $E_{sz}=-4t_{sz}\cos{k_zc}$.  In the limit $\Delta_s+E_{sz}>>\Delta>>
t,t_{ps}$, the antibonding band has dispersion
\begin{equation}
E=\Delta -{2t^2\over\Delta}(c_x+c_y-2)-{4t^2t^2_{ps}\over\Delta^2(\Delta_s
+E_{sz})}(c_x-c_y)^2,
\label{eq:D2}
\end{equation}
so if $t_{sz}<<\Delta_s$, the interlayer hopping has the form
$t_{z0}\cos{k_zc}(c_x-c_y)^2$, with $t_{z0}=-16t^2t^2_{ps}t_{sz}/
\Delta^2\Delta_s^2$.  
While this form had been suggested earlier\cite{CAnd} and found
experimentally for the bilayer splitting in BSCCO\cite{Blay}, it should be
noted that it is only
approximate, and that, at least in YBCO, there is considerable splitting
of the bilayer bands along the zone diagonal\cite{ALJP}.  Nevertheless,
this form is adequate for the present purposes.

When successive layers are staggered, the only modification to the hopping
matrix is in the form of $E_s(k_z)$, which now acquires an in-plane
dispersion,
\begin{eqnarray}
E_s(k_z)=-4t_{sz}\cos{k_zc}[\cos{(k_x+k_y)a/2}+\cos{(k_x-k_y)a/2}]
\nonumber \\  
=-8t_{sz}\cos{k_zc}\cos{k_xa/2}\cos{k_ya/2},
\label{eq:D3}
\end{eqnarray}
which leads to Eq.~\ref{eq:n21}.

\subsection{Estimation of $t_z$ from Resistivity Anisotropy}

\begin{figure}
\leavevmode   
   \epsfxsize=0.33\textwidth\epsfbox{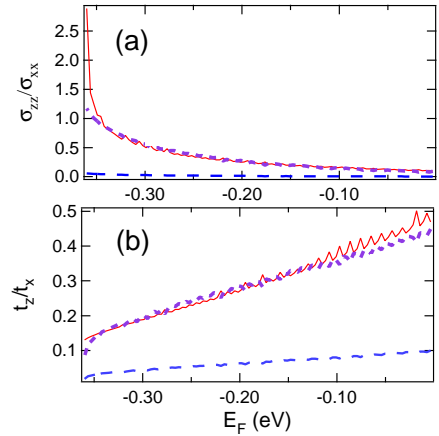}
\vskip0.5cm
\caption{(a) Normalized conductivity ratio,$\hat\sigma_{zz}/
\sigma_{xx}$ vs doping $E_F$, for uniform (solid line) and staggered
stacking (long dashed line and short dashed line, ($\times$20)); and (b) 
resulting normalized interlayer hopping $\hat t_{z0}$ for staggered (solid
line) and uniform stacking (long dashed line and short dashed line,
($\times$4.5).}
\label{fig:nD0}
\end{figure}

The dc conductivity can be estimated 
\begin{equation}
\sigma_{ii}={2e^2\over\Omega}\sum_{\vec k}v_i^2\delta (\epsilon_{\vec
k}-\mu )\tau_{\vec k},
\label{eq:D2a}
\end{equation}
$i=x,y,z$, with $\Omega$ the unit cell volume, $v_i=\hbar^{-1}d 
\epsilon_{\vec k}/dk_i$, and $\tau_{\vec k}$ the scattering rate.  Recent
ARPES data suggest that, when bilayer splitting is resolved, $\tau_{\vec
k}$ is relatively isotropic over the Fermi surface\cite{PashAli}.  Taking
$\tau_{\vec k}$ independent of $\vec k$, the conductivities are given by
integrals over the Fermi surface.  Figure~\ref{fig:nD0}a shows a
normalized conductivity ratio,
\begin{equation}
{\hat\sigma_{zz}\over\sigma_{xx}}={at^2\over ct_{z0}^2}{\sigma_{zz}
\over\sigma_{xx}},
\label{eq:D2b}
\end{equation}
while Fig.~\ref{fig:nD0}b shows the resulting normalized interlayer
hopping $\hat t_{z0}=t_{z0}\sqrt{c/a}$, which would be required to produce
a resistivity anisotropy $\rho_{zz}/\rho_{xx}=1000$.  For simplicity, it
is assumed that $t_{z0}$ is small, and $\hat\sigma_{zz}/\sigma_{xx}$ is
evaluated in the limit $t_{z0}\rightarrow 0$.  It can be seen that
(a) the staggered stacking reduces the conductivity by approximately a
factor of 20, independent of doping (except near the VHS), so (b) assuming
the resistivity anisotropy is 1000 for optimally doped LSCO, it is
estimated that $t_{z0}/t$ = 0.11 (for staggered stacking) or 0.025 (for
uniform stacking).

\subsection{z-Component of Ordering Vector}

\begin{figure}
\leavevmode   
   \epsfxsize=0.33\textwidth\epsfbox{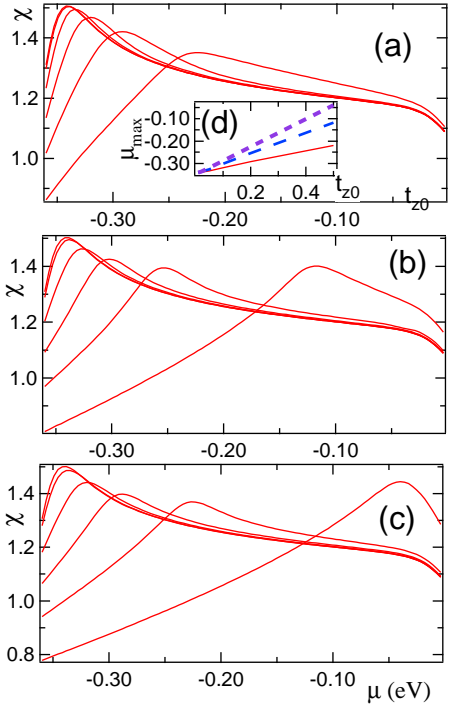}
\vskip0.5cm
\caption{$\chi_0(\vec Q,Q_z)$ at $T=100K$ vs chemical potential $\mu$, for
uniform stacking and $Q_z$ = $\pi$ (a), $\pi /2$ (b), and 0 (c).  The
various curves correspond to $t_{z0}/t$ = 0.01, 0.02, 0.05, 0.1, 0.2, and
0.5, with the peak in $\chi_0$ shifting to the right with increasing
$t_{z0}$.  Inset (d): position of peak, $\mu_{max}$, vs $t_{z0}$ for $Q_z$
= $\pi$ (solid line), $\pi /2$ (long dashed line), and 0 (short dashed
line).}
\label{fig:nD1}
\end{figure}

\begin{figure}
\leavevmode   
   \epsfxsize=0.33\textwidth\epsfbox{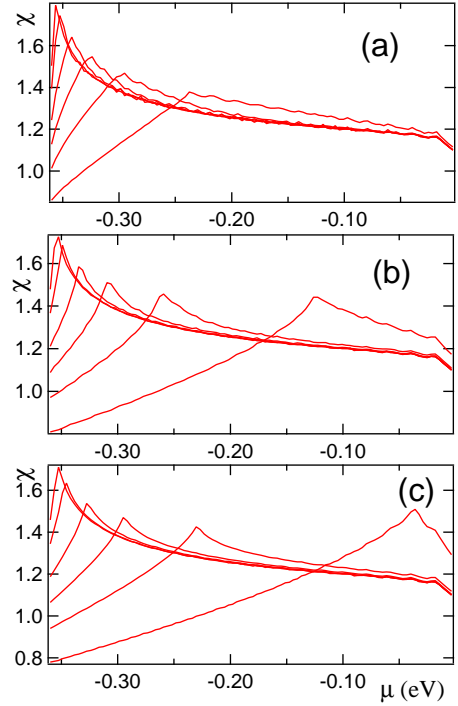}
\vskip0.5cm
\caption{$\chi_0(\vec Q,Q_z)$ vs chemical potential $\mu$, as in Fig.
\protect\ref{fig:nD1}, but at $T=10K$.}
\label{fig:nD2}
\end{figure}

Given a finite interlayer hopping $t_z$, the first issue is to identify
the three-dimensional ordering vector: what $Q_z$ minimizes the
free energy?  At mean field level, the initial magnetic instability will
be associated with the state for which the RPA denominator first diverges,
i.e., the state with the largest value of $Re\chi_0(\vec Q,Q_z)$.  (Note
that these calculations implicitly assume that the two-dimensional ground
state involves commensurate order at $\vec Q$.) For uniform stacking, a 
complicated dependence on doping, temperature, and $t_z$ is found. Figures
\ref{fig:nD1},\ref{fig:nD2} plot $\chi_0$ vs chemical potential for $T$ =
100K, 10K, respectively.  The shift of the susceptibility peak with doping
can readily be understood by comparison with Fig.~\ref{fig:0a}.  Both
temperature and interlayer coupling act to smear out the VHS, and in both
cases cause the susceptibility peak to shift to smaller chemical potential
(lower hole doping), Fig.~\ref{fig:nD1}d.  Note that the peak shifts at
different rates for different $Q_z$-values, showing that the band is
developing a considerable c-axis dispersion.  The fastest shift (short
dashed line in Fig.~\ref{fig:nD1}d, corresponding to $Q_z=0$) can thus be
considered as representing a crossover from quasi-two-dimensional to fully
three dimensional dispersion.


\begin{figure}
\leavevmode   
   \epsfxsize=0.33\textwidth\epsfbox{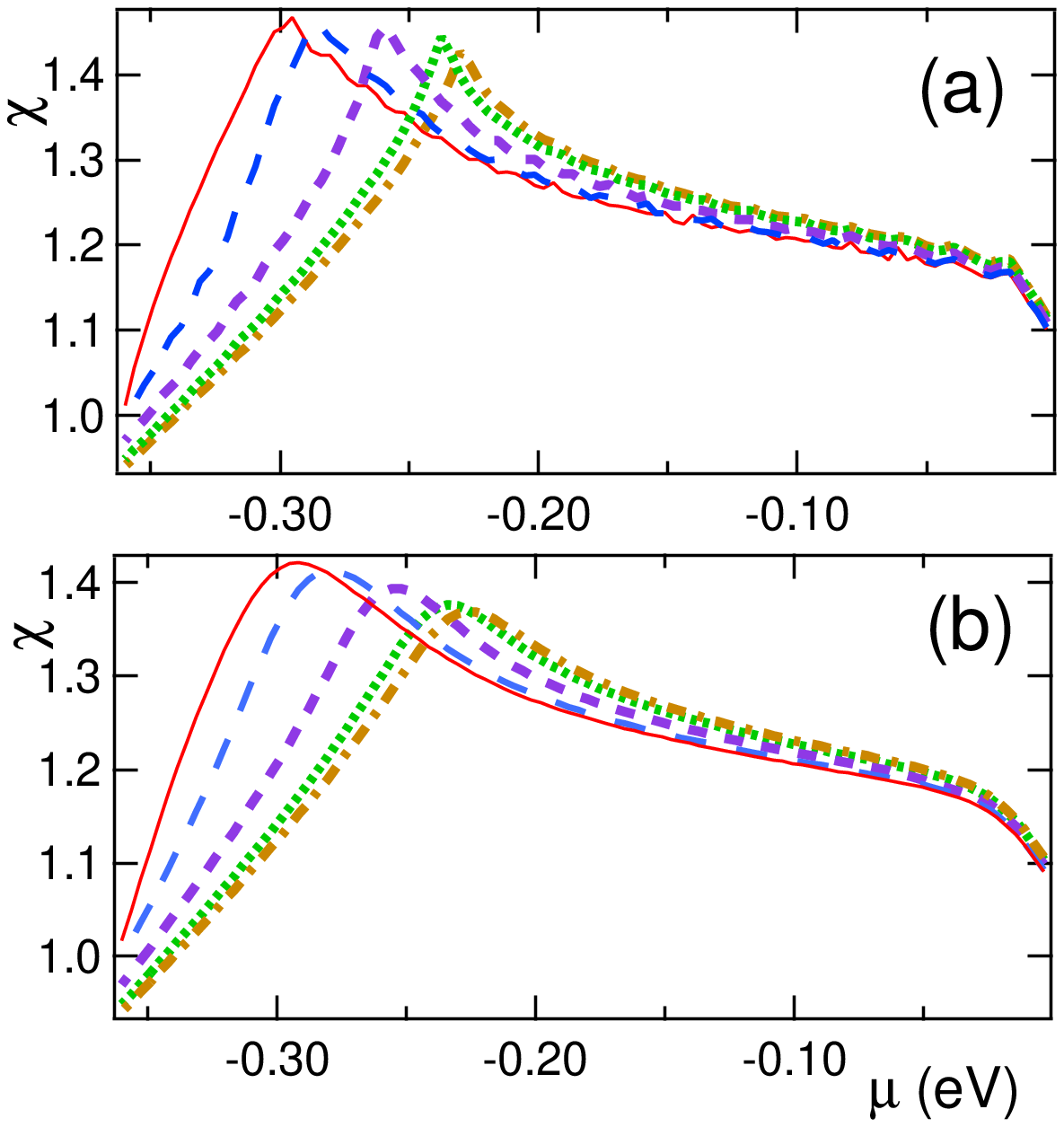}
\vskip0.5cm
\caption{$\chi_0(\vec Q,Q_z)$ vs chemical potential $\mu$, for uniform stacking 
and $t_{z0}=0.2t$, and $T=10K$ (a), or 100K (b), with $Q_z/\pi$ = 1 (solid
line), 0.75 (long dashed line), 0.5 (short dashed line), 0.25 (dotted
line), 0 (dot-dashed line).} 
\label{fig:nD3}
\end{figure}

This dispersive shift of the peak in $\chi_0$ leads to a doping dependence
of the optimal $Q_z$, as illustrated in Fig.~\ref{fig:nD3} for
$t_{z0}=0.2t$.  For large hole doping, near the $t_{z0}=0$ VHS, 
the susceptibility maximum corresponds to $Q_z=\pi /c$, while near the
susceptibility peak, the spin modulation becomes incommensurate
(intermediate values of $Q_z$ have the largest susceptibility).  There is
a rapid evolution of the optimal $Q_z$, and beyond the peak regime, over
essentially the entire electron-doped regime, the optimal $Q_z$ is $0$.
This same pattern is repeated for smaller $t_{z0}$, with only the region
of the susceptibility peak changing.  The results are essentially
independent of the sign of $t_z$.

\subsection{Calculation of $A_z$}

\subsubsection{Uniform Stacking}

Given $t_z$ and $Q_z$, the parameter $A_z$ of Eq.~\ref{eq:0B18} can be
evaluated: $U\chi (\vec Q+q_z\hat z,\omega =0)=U\chi (\vec Q+Q_z\hat z,0)+
A_z(q_z-Q_z)^2$.  The dominant ordering vectors, $Q_z=\pi /c$ and $Q_z=0$,
can be analyzed in more detail.  For the former choice, 
\begin{eqnarray}
A^{\pi}_z={Uc^2\over 4}\sum_{\vec k}\Bigl[{t_zc_z\over \epsilon_{\vec
k}-\epsilon_{\vec k+\vec Q}+i\delta}\bigr(2{f_{\vec k}-f_{\vec k+\vec
Q}\over \epsilon_{\vec k}-\epsilon_{\vec k+\vec Q}+i\delta}
\nonumber \\
-[f'_{\vec k}+f'_{\vec k+\vec Q}]\bigr)
-2t_z^2s_z^2\bigl({f''_{\vec k}-f''_{\vec
k+\vec Q}\over \epsilon_{\vec k}-\epsilon_{\vec k+\vec
Q}+i\delta}\bigr)\Bigr],
\label{eq:D4}
\end{eqnarray}
with $f'_{\vec k}=-f_{\vec k}(1-f_{\vec k})/k_BT$, $f''_{\vec k}=-f'_{\vec
k}(1-2f_{\vec k})/k_BT$, $c_z=\cos{k_zc}$, $s_z=\sin{k_zc}$.  For the
latter case
\begin{eqnarray}
A^0_z={-Uc^2\over 4}\sum_{\vec k}\Bigl[t_zc_z({f'_{\vec k}-f'_{\vec k+\vec
Q}\over \epsilon_{\vec k}-\epsilon_{\vec k+\vec Q}+i\delta})
\nonumber \\
+2t_z^2s_z^2\Bigl[{f''_{\vec k}-f''_{\vec
k+\vec Q}\over \epsilon_{\vec k}-\epsilon_{\vec k+\vec
Q}+i\delta}
\nonumber \\
+8{f_{\vec k}-f_{\vec k+\vec Q}\over (\epsilon_{\vec
k}-\epsilon_{\vec k+\vec Q}+i\delta )^3}-4{f'_{\vec k}+f'_{\vec k+\vec
Q}\over (\epsilon_{\vec k}-\epsilon_{\vec k+\vec Q}+i\delta )^2}\Bigr].
\label{eq:D5}
\end{eqnarray}

\begin{figure}
\leavevmode   
   \epsfxsize=0.33\textwidth\epsfbox{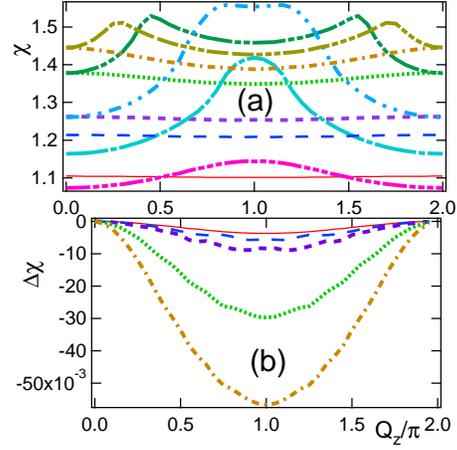}
\vskip0.5cm
\caption{(a)$\chi_0(\vec Q,Q_z)$ vs $Q_z$ for $t_{z0}=0.1t$, and $T=10K$, 
and a variety of chemical potentials $\mu$ = -0.003559 (solid line),
-0.08898 (long dashed line), -0.1779 (short dashed line), -0.2669 (dotted
line), -0.2847 (dot-dashed line), -0.2954 (long-long-short-short-short
dashed line), -0.3025 (long-short-short dashed line), -0.3203
(dash-dot-dot line), -0.3381 (long-short dashed line), and -0.3559 meV
(long-short-short-short dashed line). (b) $\Delta\chi = \chi_0(\vec
Q,Q_z)-\chi_0(\vec Q,Q_z=0)$, where the curves have the same meaning as in
frame (a).} 
\label{fig:nD4}
\end{figure}

\begin{figure}
\leavevmode   
   \epsfxsize=0.33\textwidth\epsfbox{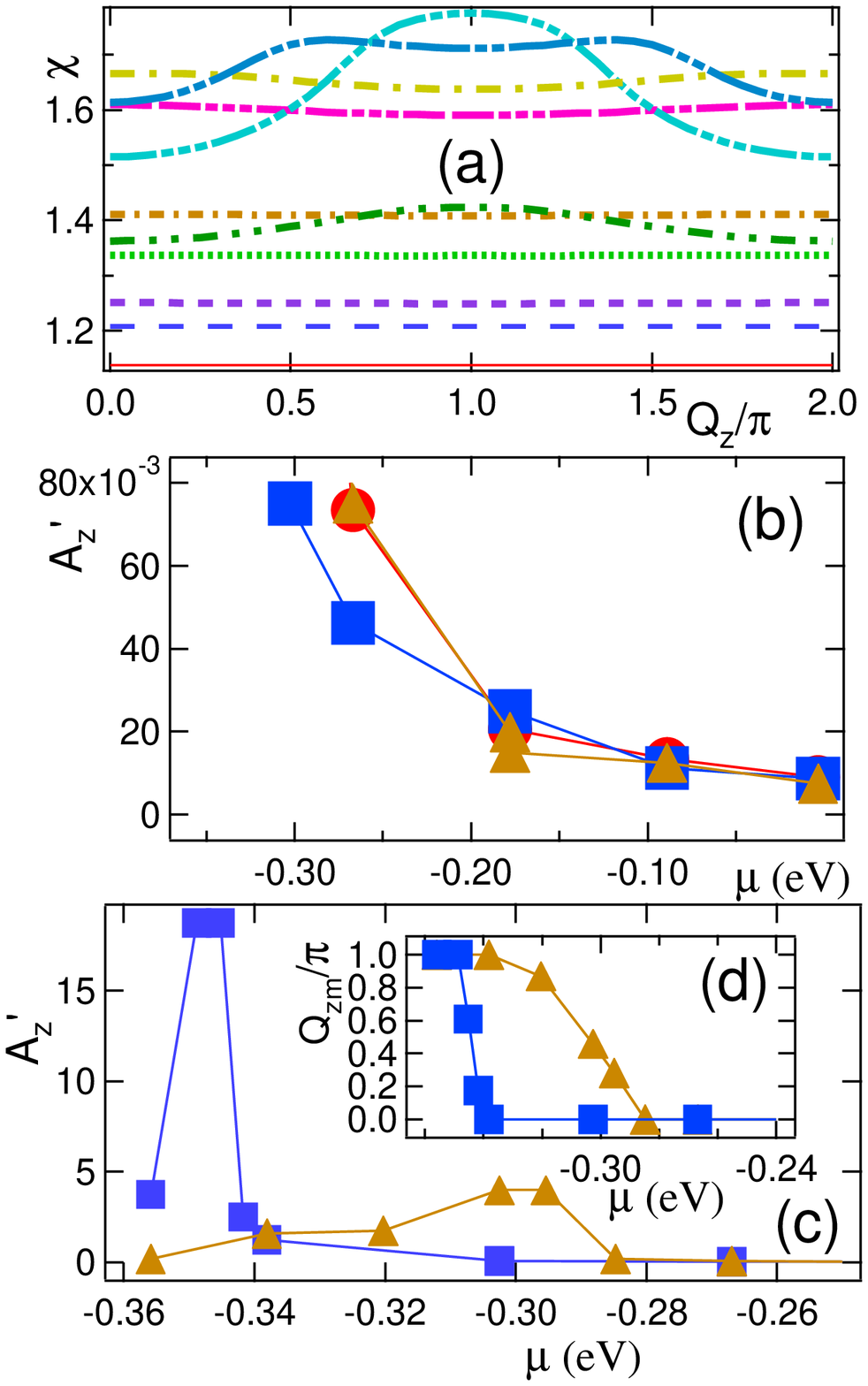}
\vskip0.5cm
\caption{(a)$\chi_0(\vec Q,Q_z)$ vs $Q_z$ for $t_{z0}=0.02t$, and $T=10K$, 
and a variety of chemical potentials $\mu$ = -0.003559 (solid line),
-0.08898 (long dashed line), -0.1779 (short dashed line), -0.2669 (dotted
line), -0.3025 (dot-dashed line), -0.3381 (long-long-short-short-short
dashed line), -0.3417 (long-dashed-dotted line), -0.3452
(long-short-short dashed line), -0.3488 (long-short-short-short dashed
line), and -0.3559 meV (long-dash-dot-dotted line). (b,c) $A_z'=A_z/Uc^2$
vs $\mu$ for $t_{z0}/t$ = 0.02 (squares, $A_z'\times 25$) and 0.1
(triangles,circles). (d) $Q_{zm}$ vs $\mu$ for $t_{z0}/t$ = 0.02 (squares)
and 0.1 (triangles).}
\label{fig:nD5}
\end{figure}

Figure~\ref{fig:nD4} (~\ref{fig:nD5}a) shows how $\chi_0(\vec Q,Q_z)$
varies with $Q_z$ for $t_{z0}=0.1t$ ($0.02t$), for a number of different
dopings. For the entire electron-doped regime, the peak is at $Q_{zm}=0$
(Fig.~\ref{fig:nD4}b,~\ref{fig:nD5}d), crossing over to $Q_{zm}=\pi /c$ in
the hole doped regime.  Away from the peak, the susceptibility varies as
$\hat A_zq_z^2$, with $q_z=Q_z-Q_{zm}$, and in the electron-doped regime
the full variation can be approximated by a cosine.  The amplitude of the
cosine falls to zero as the C-point is approached.  In the 
quasi-two-dimensional regime this amplitude scales with $t_{z0}^2$.
Figure~\ref{fig:nD5}b,c shows plots of the best parabolic fit to 
$A_z'=\hat A_z/c^2$ for $t_{z0}/t$ = 0.02 (squares) and 0.1 (triangles).
For $t_{z0}/t$ = 0.1, an alternative $A_z'$ is shown, found by fitting the
full susceptibility as a cosine in $q_z$ (circles).  The good agreement
between the two techniques shows that this is a reasonable approximation
in the electron-doped regime ($-0.2eV\le\mu\le 0$).  Near the
susceptibility peak, the variation is nonsinusoidal, and the parabolic fit
leads to a large value for $A_z'$.


\subsubsection{Staggered Stacking}

\begin{figure}
\leavevmode   
   \epsfxsize=0.33\textwidth\epsfbox{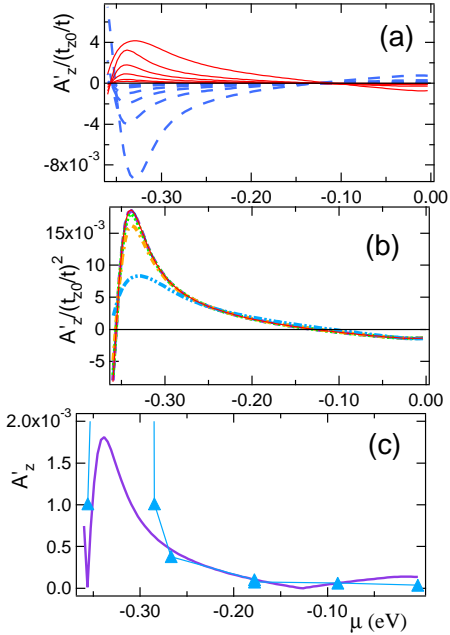}
\vskip0.5cm
\caption{(a)$A'_z=A_z/Uc^2$ vs chemical potential $\mu$ for $Q_z$ = 0
(solid lines) or $\pi$ (dashed lines), for a variety of
values of $t_{z0}$ and $T=100K$.  In order of increasing amplitude, the
values are $t_{z0}/t$ = 0.01, 0.02, 0.05, 0.1, 0.2, and 0.5.  (b) Scaling
of $A_z^{\prime (0)}$ with $(t_{z0}/t)^2$. Curves are $t_{z0}/t$ = 0.01
(solid line), 0.02 (long dashed line), 0.05 (short dashed line), 0.1
(dotted line), 0.2 (dot-dashed line), and 0.5 (dot-dot-dashed line). (c)
Comparison of $max(A_z)$ for staggered stacking (solid line) and uniform
stacking (triangles, $\times$1/20) at $t_{z0}/t$ = 0.1.} 
\label{fig:nD4b}
\end{figure}

The same calculations can be repeated for the $t_z$ of Eq.~\ref{eq:n21},
associated with staggered stacking; Fig.~\ref{fig:nD4b}a shows $A_z$
calculated from Eqs.~\ref{eq:D4},~\ref{eq:D5} at $Q_z$ = 0 (solid lines)
and $\pi$ (dashed lines). The frustration induced by staggering
of the CuO$_2$ layers is reflected in a strong suppression of the
$q_z$-dependence of $\chi$, which leaves a small residual contribution
{\it quadratic} in $t_{z0}$, Fig.~\ref{fig:nD4b}b.  Since $t_z$ vanishes
at $(\pi ,0)$, there is no shift of the susceptibility peak with doping.
Note the symmetry of the $A_z$ values between $0$ and $\pi$.  In fact,
$\chi (Q_z)$ is closely sinusoidal, particularly for small $t_{z0}$, with
maxima either at $\pi$ or $0$.  Thus, near either the H- or C-points, the
maximum of $\chi$ corresponds to $Q_z=\pi$.  For intermediate dopings,
$Q_z=0$ is favored.  At two distinct chemical potentials, the amplitude of
the cosine collapses and changes sign.  At the crossing points, $\chi$ is
independent of $Q_z$, leading formally to $T_N\rightarrow 0$.  Note from
Fig.~\ref{fig:nD4b}c that the suppression of $A_z$ is approximately in the 
same ratio as that of the resistivity, found above.

\subsection{Calculation of $T_N$}

\begin{figure}
\leavevmode   
   \epsfxsize=0.33\textwidth\epsfbox{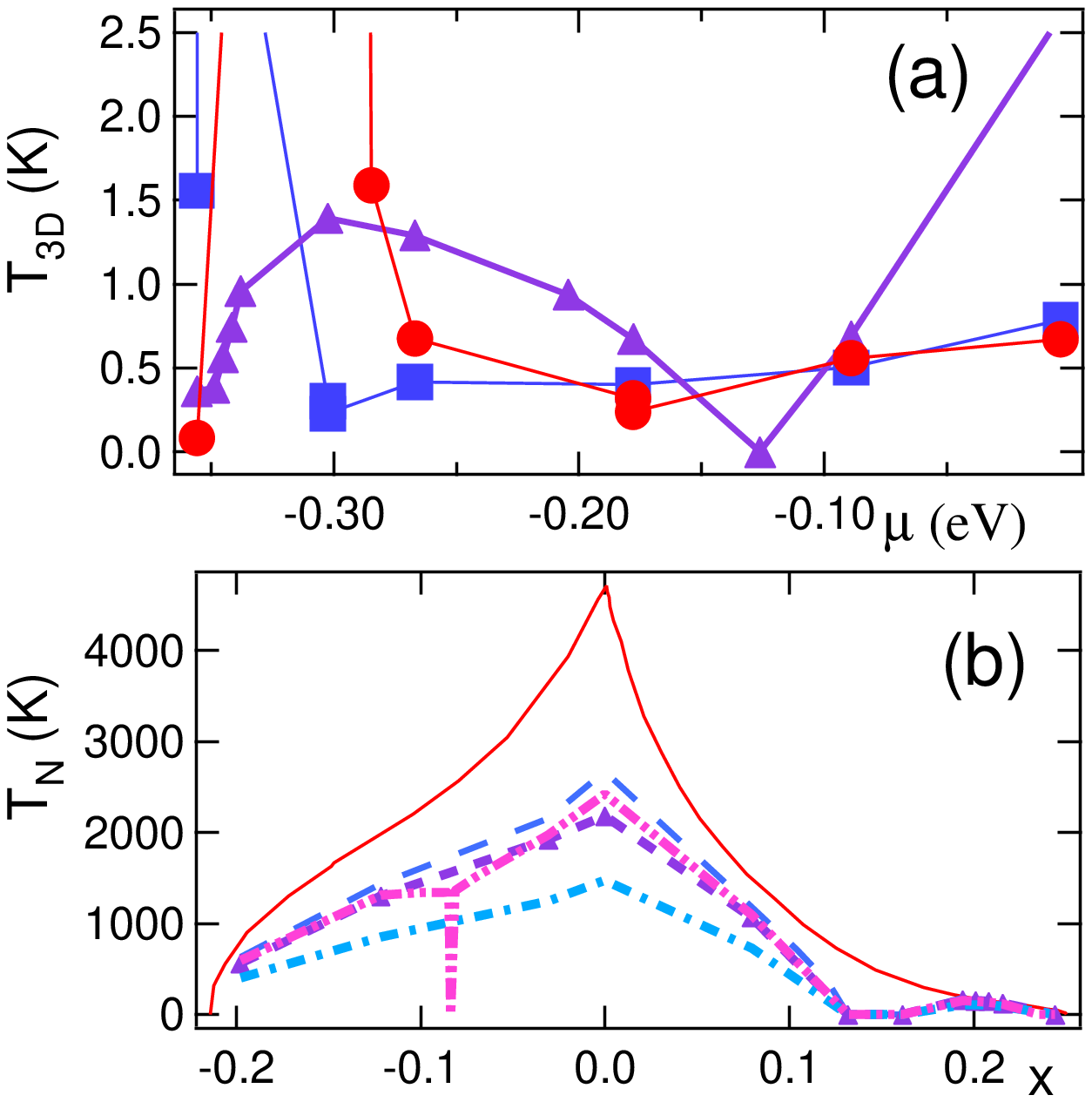}
\vskip0.5cm
\caption{(a)$T_{3D}$ vs $\mu$ for $T=10K$ and uniform stacking with
$t_{z0}=0.1t$ (circles, $\times$1/25) or 0.02 (squares), or staggered
stacking with $t_{z0}$=0.02 (triangles). 
(b) $T_N$ vs $x$, comparing mean field transition (solid line) with
interlayer coupling models (uniform stacking) assuming $t_{z0}/t$ = 0.1
(long dashed line), 0.02 (short dashed line), and 2$\times 10^{-6}$
(dot-dashed line), and the staggered stacking model assuming $t_{z0}/t$ = 
0.1 (dot-dot-dash line).} 
\label{fig:nD6} 
\end{figure}

When there is a finite interlayer hopping $t_z$, Eq.\ref{eq:B34h} becomes
\begin{eqnarray}
\delta-\bar\delta_0={6uTa^2c\over\pi^2A}\int_{0}^{\pi\over 
c}{dq_z\over\pi}\int_{y_0}^{y_0+Aq_c^2}{dy\over y}
tan^{-1}({2TC\over y})
\nonumber \\
\simeq{3uTa^2\over\pi A}\ln ({T\over T_{3D}})],
\label{eq:B34i}
\end{eqnarray}
where $y_0=\delta+A_zq_z'^2$ and $T_{3D}=\pi^2A_z/2Ce^2c^2$.  (A small
correction to $\bar\delta_0$ is neglected.  Treating the $q_z$ dependence
as a cosine rather than a cutoff quadratic
leads to qualitatively similar results.)  Thus a finite $A_z$ always cuts
off the divergence found in Eq.\ref{eq:B34h}, leading to a finite $T_N$
whenever there is a zero-temperature Neel state (e.g., up to a QCP).
It should be noted that the above calculation implicitly assumed that
$T>T_{3D}\sim A_z$: for $T<T_{3D}$ the logarithm is cut off and the
system behaves like an anisotropic three-dimensional magnet.  For
$t_{z0}/t<0.1$, the system is generally in the quasi-two-dimensional
limit, Fig.~\ref{fig:nD6}a.  Figure~\ref{fig:nD6}b compares the mean-field
Neel transition with the Neel transition found assuming uniform
stacking and finite interlayer couplings $t_{z0}/t$ = 0.1, 0.02, and
$2\times 10^{-6}$ [the last found by scaling the $T_{3D}$ for
$t_{z0}/t=0.02$ by the ratio of $t_{z0}^2$'s].  It is seen that
$T_N\rightarrow 0$ as $t_{z0}\rightarrow 0$, albeit exceedingly slowly.

The above calculations are for uniform stacking.  For staggered stacking
$A_z$ is reduced, in approximately the same ratio as the resistivities.
Hence, the staggered stacking with $t_{z0}/t$ = 0.1 should be comparable
to uniform stacking with $t_{z0}/t$ = 0.02, as observed,
Fig.~\ref{fig:nD6}.  While $T_N$ technically goes to zero for
staggered stacking near $x=-0.0838$, the decrease is logarithmic, and in
practice no more than a weak dip is expected to be observed (the point
with $T_N=0K$ is omitted from the plot in Fig.~\ref{fig:nD7}). Hence, if
$t_{z0}$ is estimated from the resistivity, it will be nearly impossible
to distinguish uniform from staggered stacking via measurements of $T_N$.

In the above calculations, a constant value of $A$ was assumed for each
doping, as given in Fig.~\ref{fig:10}.  In fact, for the electron-doped
cuprates, $A\sim 1/T^{1.5}$ for $T>T_A^*$, Fig.~\ref{fig:45}.  This would
cause an enhancement of the logarithmic correction, $\sim T^{2.5}$,
tending to pin $T_N$ close to $T_A^*$.  For the present parameter values,
this could reduce $T_N$ by roughly a factor of two, still larger than the
experimental values.

\begin{figure}
\leavevmode   
   \epsfxsize=0.33\textwidth\epsfbox{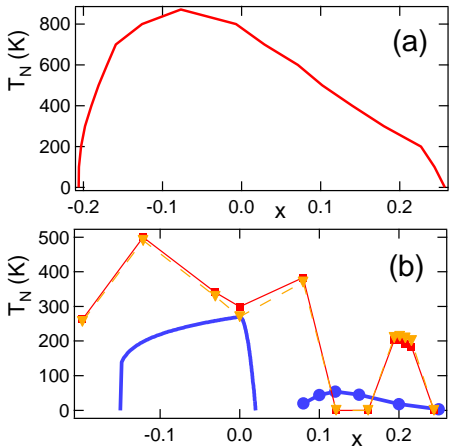}
\vskip0.5cm
\caption{(a) Mean field $T_{N}$ vs $x$ assuming paramagnetic $U_{eff}$
(Appendix B).  
(b) Corresponding $T_N$ vs $x$, calculated using Eq.~\protect\ref{eq:B34i}.
Squares = staggered stacking with $t_{z0}/t$ = 0.1; triangles = uniform
stacking with $t_{z0}/t$ = 0.02; solid line and circles = data, as in 
Fig.~\protect\ref{fig:nD7}.}
\label{fig:nD8} 
\end{figure}

A more likely source of the discrepancy is the possible temperature
dependence of $U_{eff}$, Appendix B.  The large $U_{eff}$ at half filling
arises from lack of screening, in the presence of a Mott gap -- and is
appropriate in analyzing the low-$T$ Fermi surfaces found in ARPES.  For
calculating the onset of the Mott gap, the mean field $T_N$, it is more
appropriate to use the paramagnetic susceptibility, as in
Fig~\ref{fig:111}a.  When this is done, considerably smaller transition
temperatures are found, both at the mean field level, Fig.~\ref{fig:nD8}a,
and when fluctuations and interlayer hopping are included,
Fig.~\ref{fig:nD8}b.  While the latter are closer to the experimental
values, no attempt has been made to correct $U_{eff}$ for the short range
gap.

\end{document}